\documentclass{ws-jai}
\usepackage[flushleft]{threeparttable}
\usepackage{supertabular,lscape,epsfig}
\usepackage{amssymb}
\usepackage{amsmath}
\usepackage{lipsum}
\usepackage{float}
\usepackage[T1]{fontenc}            
\usepackage{newtxtext,newtxmath}    
\usepackage{graphicx}               
\usepackage{amsmath}
\usepackage{subcaption}             
\usepackage{gensymb}                
\usepackage{booktabs}               
\usepackage{anyfontsize}
\usepackage[hidelinks, colorlinks=true, allcolors=blue]{hyperref}
\usepackage{anyfontsize}          
\usepackage{soul}
\usepackage[normalem]{ulem}
\usepackage{subcaption}             

\newcommand{\angstrom}{\textup{\AA} } 

\begin{document}

\catchline{}{}{}{}{} 

\markboth{Hien Phan-Thanh}{SPECTRUMMATE: A Low-cost Spectrometer for Small Telescopes}

\title{SPECTRUMMATE: A Low-cost Spectrometer for Small Telescopes}

\author{
Hien Phan-Thanh$^{1}$, 
Nguyen Nguyen-Duc$^{2,7,*}$, 
Thuy Le-Quang$^{3}$, 
Tobias C. Hinse$^{4}$, 
Quang Nguyen-Luong$^{5,6,2}$
}

\address{
$^1$University of Science and Technology of Hanoi, Vietnam Academy of Science and Technology, Hanoi 100000, Vietnam\\
$^2$TNU Observatory, Tay Nguyen University, 567 Le Duan, Ea Tam, Buon Ma Thuot City, Dak Lak, 630000, Vietnam\\
$^3$Quy Nhon Observatory, ExploraScience Quy Nhon, Quy Nhon City, Vietnam\\
$^4$University of Southern Denmark, Department of Physics, Chemistry and Pharmacy, SDU-Galaxy, Campusvej 55, Odense M, 5230, Denmark\\
$^5$Institute for Astronomy and Space Quantum Communications\\
$^6$Université Paris-Saclay, Université Paris Cité, CEA, CNRS, AIM, Gif-sur-Yvette, 91191, France\\
$^7$Department of Physics, Graduate School of Science, Osaka Metropolitan University, 3-3-138
Sugimoto-cho, Sumiyoshi-ku, Osaka-shi, Osaka, Japan\\
$^*$E-mail: ducnguyen382002@gmail.com (Corresponding author)
}

\maketitle

\corres{$^{*}$Corresponding author.}

\begin{history}
\received{(to be inserted by publisher)};
\revised{(to be inserted by publisher)};
\accepted{(to be inserted by publisher)};
\end{history}

\begin{abstract}
This project presents the development and implementation of a compact spectrometer, named SPECTRUMMATE, tailored for small telescopes. Small telescopes offer several advantages: they are cost-effective, occupy less space, and are simpler to set up than larger instruments. This makes them particularly suitable for amateur astronomers and educational institutions with limited resources. Moreover, small telescopes can effectively observe bright celestial objects, enabling valuable contributions to astronomical projects. Based on the Sol'EX design, SPECTRUMMATE was constructed using optical components available at the Space and Applications Laboratory\footnote{\url{http://remosat.usth.edu.vn/index.php/Main/HomePage}} (SpaceLAB), University of Science and Technology of Hanoi, Vietnam. The instrument is designed to meet the specific needs of astronomers who require detailed analysis within the visible spectrum (SPECTRUMMATE can capture a waveband of $368\ \angstrom/image$, spectral coverage from 4000 to 6600 \angstrom, and resolving power $R=2546$ at 5500 \angstrom). To achieve optimal performance, the design process involved selecting and configuring optical elements, including a collimator, diffraction grating, and objective lens. Experimental setups were tested to minimize spectral dispersion while ensuring the system's compactness and ease of alignment. Spectra obtained by SPECTRUMMATE demonstrated efficient spectral calibration and the capability to capture high-resolution spectra of bright light sources, such as the Sun, making it a valuable tool for specialised spectroscopic observations. The modular design of SPECTRUMMATE also allows users to change its components easily to achieve the desired spectral range and resolution.
\end{abstract}


\section{Introduction}
\noindent Spectroscopy has long been an important element in astronomy, offering crucial insights into the composition, temperature, velocity, and other physical properties of celestial objects. By dispersing light into its constituent wavelengths, astronomers can detect and analyze features beyond the visible spectrum, making spectrometers indispensable tools for both professional and amateur astronomical observations. Developing affordable, high-performance spectrometers is particularly important for amateur astronomers and educational outreach, especially as part of research and education programs. However, commercially available spectrometers are often constrained by high costs, complex setups, or limited customization options.

\begin{figure}[H]
    \centering
    \includegraphics[width=\linewidth]{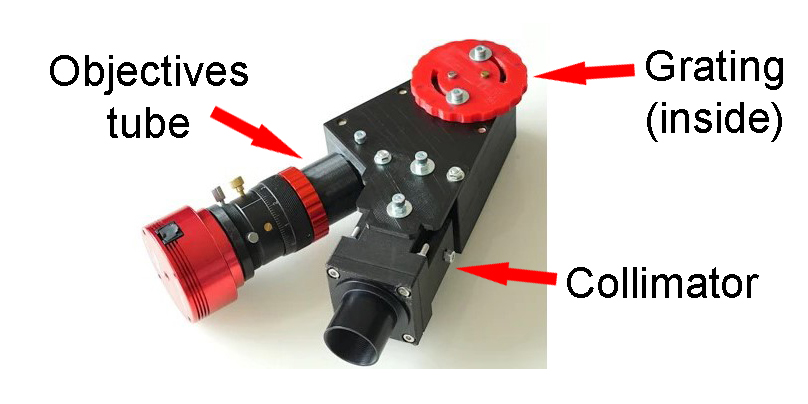}
    \caption{The Sol'EX spectrometer by \citet{builsol} - a good example of an easy-to-build spectrometer.}
    \label{fig:solexspectrometer}
\end{figure}

These challenges inspired the design and construction of a custom-built spectrometer, SPECTRUMMATE. The instrument's optical layout is modeled after the successful Sol'EX project by \citet{builsol}, a documented example of an easy-to-build spectrometer, as shown in \autoref{fig:solexspectrometer}. However, the novelty of the SPECTRUMMATE project lies not in replicating the Sol'EX, but in adapting its principles to a different set of readily available, low-cost components. This work includes the development of a completely new, fully 3D-printable mechanical housing optimized for these specific parts, making the design accessible to educational labs equipped with 3D printers. Furthermore, this paper provides end-to-end documentation of the construction, calibration, and validation process, intended as a guide for educational and amateur use.

A central goal of this project was to create a functional instrument using a limited, pre-existing inventory of parts, a common scenario in educational and amateur settings. SPECTRUMMATE was therefore constructed using optical components available at the SpaceLAB. This constraint dictated the selection of key components, such as the diffraction grating and lenses, shaping the final design and performance of the instrument.

SpectrumMate is designed to be a compact, cost-effective spectrometer suitable for use with small telescopes (aperture $D$ smaller than 10 $inches$). The optical design is optimized for instruments with a focal ratio (f-stop) between 5 and 8 to ensure efficient light throughput. However, it can also be used effectively with telescopes that have higher focal ratios, such as common f/10 Schmidt-Cassegrain models. For such cases, the use of a focal reducer is recommended to match the telescope's output beam to the spectrometer's optimal input f-ratio, preventing the under-filling of the collimator and maximizing system sensitivity. SPECTRUMMATE also incorporates essential optical components such as a slit, collimator, diffraction grating, and objective lenses, carefully chosen for their ability to capture high-resolution spectra from bright light sources. By drawing on the Sol'EX design, SPECTRUMMATE aims to provide a practical solution for users needing a specialized spectroscopic study instrument within a limited spectral range.

\section{Spectrometer design}

\subsection{Optical configuration}
The design of SPECTRUMMATE is based on a classical spectrograph configuration, built around three key optical components: a collimator, a diffraction grating, and an objective lens. Light from the telescope enters through a slit and is collimated into a parallel beam. This beam is then dispersed by a diffraction grating, which separates the light by wavelength. Finally, an objective lens focuses the dispersed spectrum onto the camera's sensor.

\begin{figure}[H]
    \centering
    \includegraphics[width=0.5\linewidth]{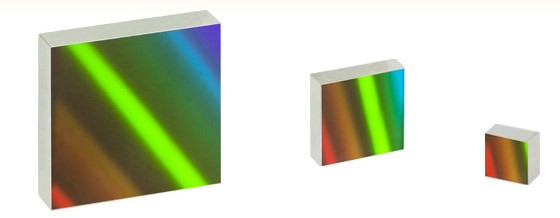}
    \caption{Holographic reflective diffraction gratings of dimension $25 \times 25 \times 6~mm$ by Thorlabs, Inc.}
    \label{fig:threeholreflectgrating}
\end{figure}

For this project, a reflective holographic diffraction grating with 1200 $grooves/mm$ was selected (\autoref{fig:threeholreflectgrating}). This choice was primarily driven by its availability within the SpaceLAB inventory, aligning with the project's goal of using accessible, low-cost components. The reflective design also contributes to a compact and folded optical path.

\subsection{Requirements for SPECTRUMMATE}
\label{subsec:Requirements for SPECTRUMMATE}
SPECTRUMMATE needed all the basic components of a spectrometer, including a collimator tube, a grating, and an objective tube. For product quality, SPECTRUMMATE should be designed to prevent all light leakage, the parts should be tightly connected, and the optical components should fit perfectly. SPECTRUMMATE should also be as compact as possible, since it is intended for use with small telescopes. Since we also aimed to use SPECTRUMMATE for educational purposes, we wanted it to be easy to use and build. For modern off-the-shelf components, these requirements are usually met even for a low-budget setup.

A key quantitative design requirement was to achieve a spectral resolution sufficient to clearly resolve prominent solar Fraunhofer lines, specifically the Sodium doublet ($D1$ at 5896 \angstrom and $D2$ at 5890 \angstrom), which requires a spectral resolution smaller than 6 \angstrom. This level of performance allows for meaningful educational demonstrations of spectral identification and chemical composition analysis.

\subsection{Configuration of SPECTRUMMATE}
To find a suitable configuration, some setups were tested at SpaceLAB following the instructions in the EDU-SPEA1/M Economy Spectrometer Kit Manual and the EDU-SPEB2 Spectrometer Kit Manual by Thorlabs, Inc. The purpose was to find a setup where the optical alignment is not too complex and the spectrum dispersion is small, ensuring the compact design of \textsc{SpectrumMate}. 

\begin{figure}[H]
	\centering	
	\includegraphics[width=0.6\linewidth]{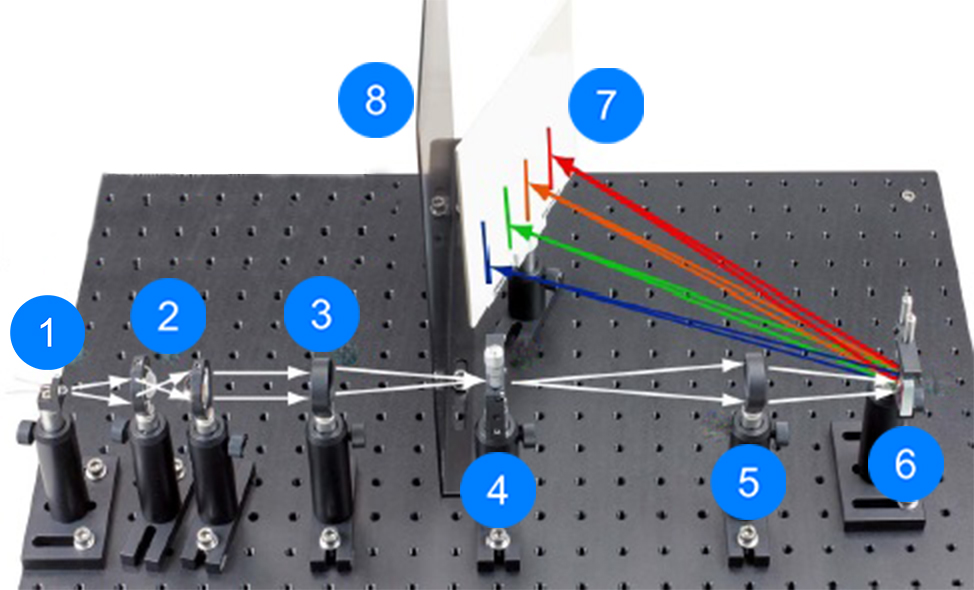}
	\caption{The first configuration by Thorlabs, Inc. 1 - LED light source. 2 - Aspheric Condenser Lenses. 3 - Focusing Lens. 4 - Variable Slit. 5 - Imaging lens. 6 - Reflective Diffraction Grating. 7 - Viewing Screen. 8 - Light Barrier (blocks stray light from source). Image from Thorlabs manual.}
    \label{fig:FirstConfig}
\end{figure}

In the first setup (Fig. \ref{fig:FirstConfig}), components 1, 2, and 3 are used as a light source, while 4, 5, 6, and 7 are for showing a spectrum. This configuration was not chosen since its main purpose is to show smeared colored bands - demonstrate angular diffraction rather than sharp slit images, which mean we can't capture a sharp spectrum image.

\begin{figure}[H]
	\centering
	\includegraphics[width=0.6\linewidth]{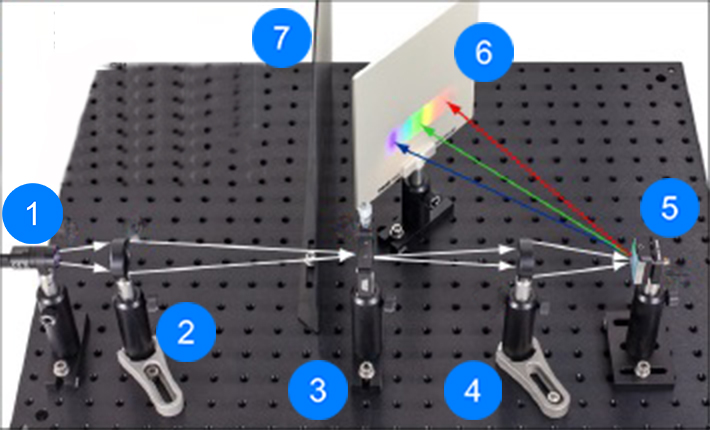}
	\caption{The second configuration by Thorlabs, Inc. 1 - LED light source. 2 - Focusing Lens. 3 - Adjustable Slit. 4 - Imaging lens. 5 - Reflective Diffraction Grating. 6 - Viewing Screen. 7 - Light Barrier (blocks stray light from source). Image from Thorlabs manual.}
    \label{fig:SecondConfig}
\end{figure}

We also couldn't capture a spectrum image with the second setup (\autoref{fig:SecondConfig}). Therefore, we also didn't choose it.

\begin{figure}[H]
	\centering
	\includegraphics[width=0.75\linewidth]{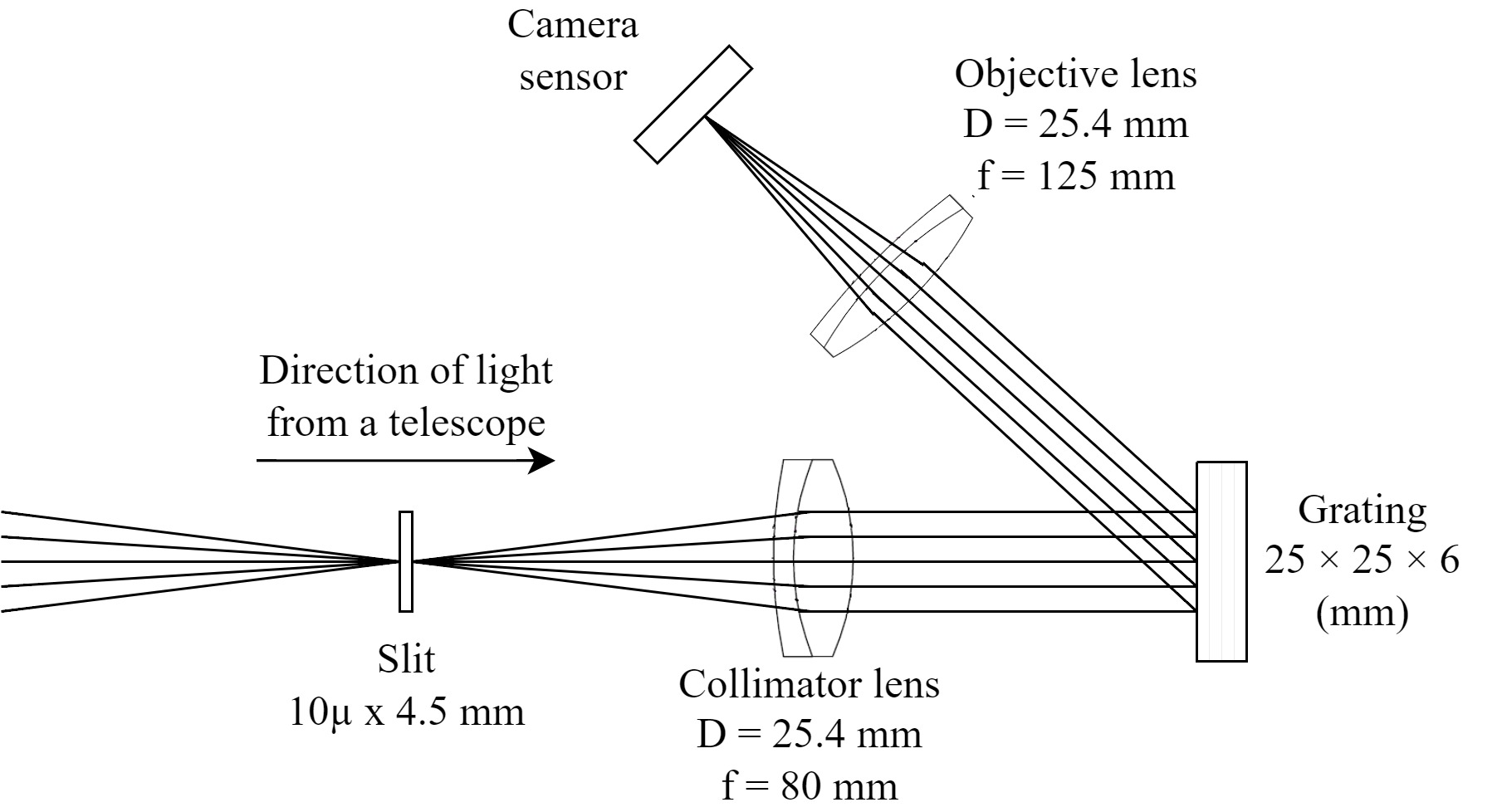}
	\caption{Sol'EX optical configuration.}
    \label{solexconfiguration}
\end{figure}

We then consulted Sol'EX's design. It uses two doublets - one for collimating light from the slit, and the other for converging light onto the camera sensor. This configuration is simple, easy to build, compact, and yet effective since the spectrum is converged onto the camera sensor using an objective lens. \autoref{solexconfiguration} shows Sol'EX optical configuration, reproduced from Sol'EX website\footnote{\url{https://solex.astrosurf.com/solex-theory-en.html}}.

\begin{figure}[H]
	\centering
	\includegraphics[width=0.75\linewidth]{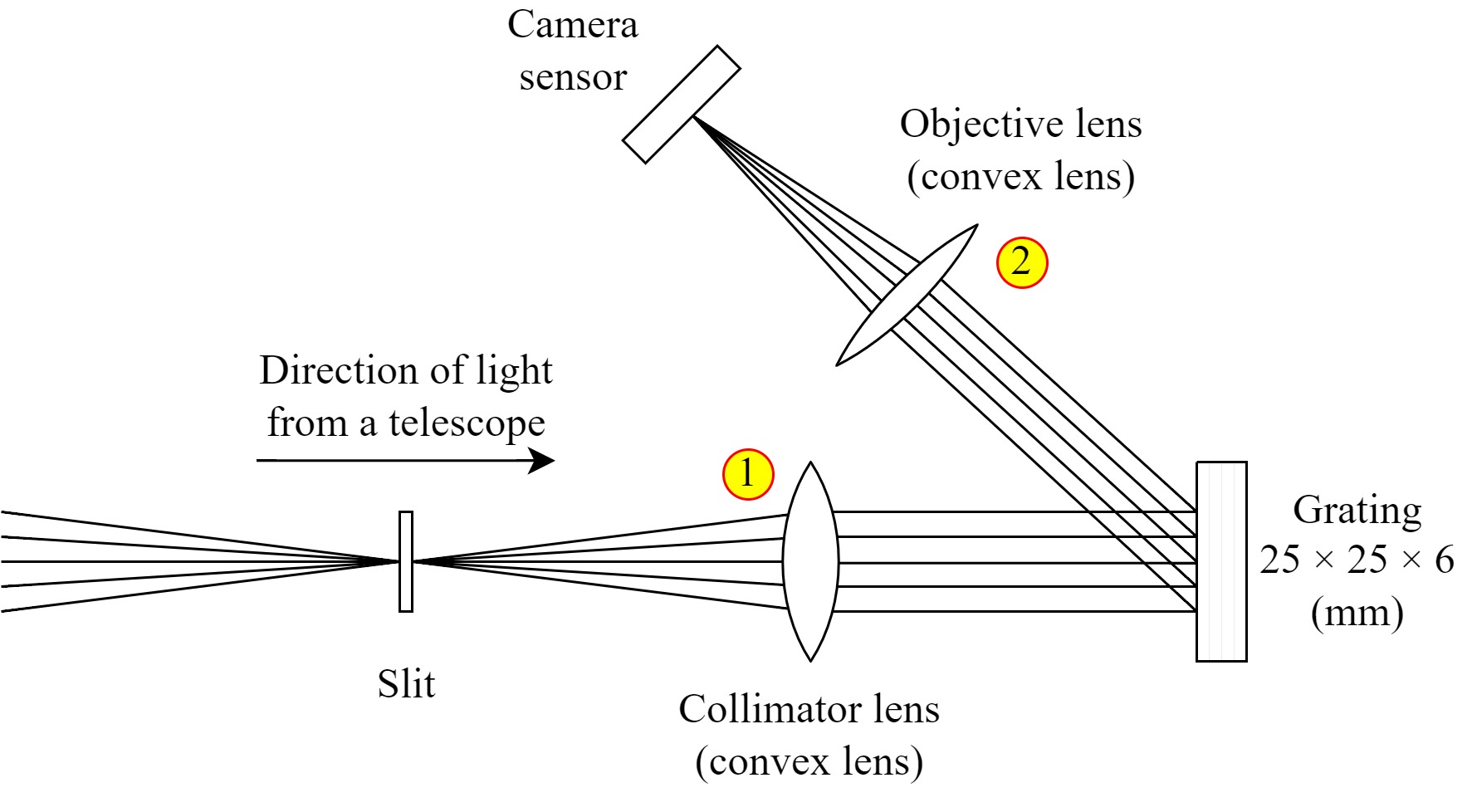}
	\caption{SPECTRUMMATE's configuration.}
        \label{fig:SpectrumMateDrawing}
\end{figure}

The final optical configuration for SPECTRUMMATE, shown in \autoref{fig:SpectrumMateDrawing}, was adopted from the Sol'EX design. This layout was chosen as it best satisfies the primary design requirements for a compact and simple instrument built from available components. It uses a folded light path, with the reflective grating redirecting the collimated beam to an objective lens that focuses the spectrum onto the camera sensor.

\begin{figure}[H]
	\centering
	\includegraphics[width=0.75\linewidth]{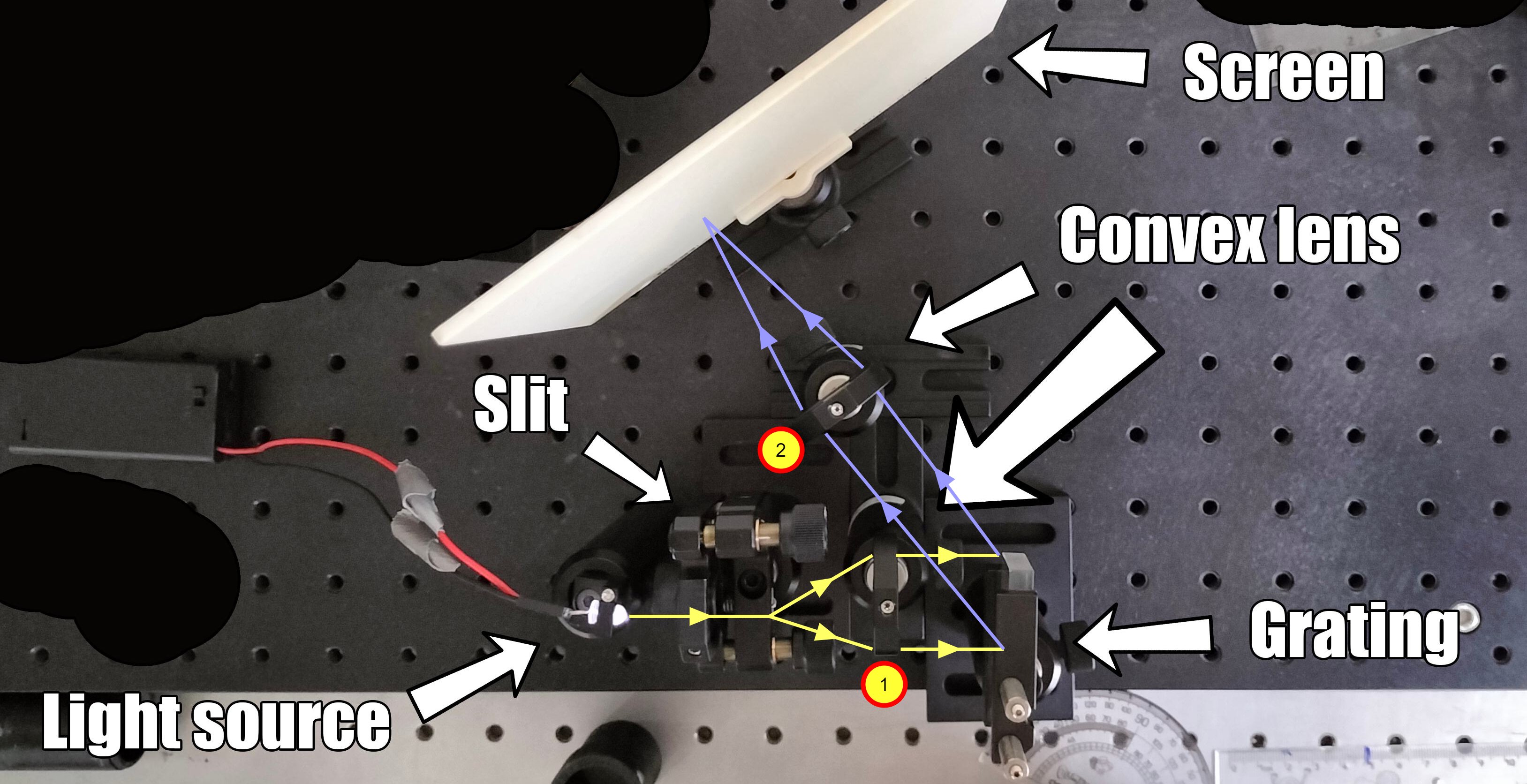}
	\caption{SPECTRUMMATE configuration on the test bench (top view).}
        \label{fig:SpectrumMateTestBench}
\end{figure}

\begin{figure}[H]
	\centering
	\includegraphics[width=0.75\linewidth]{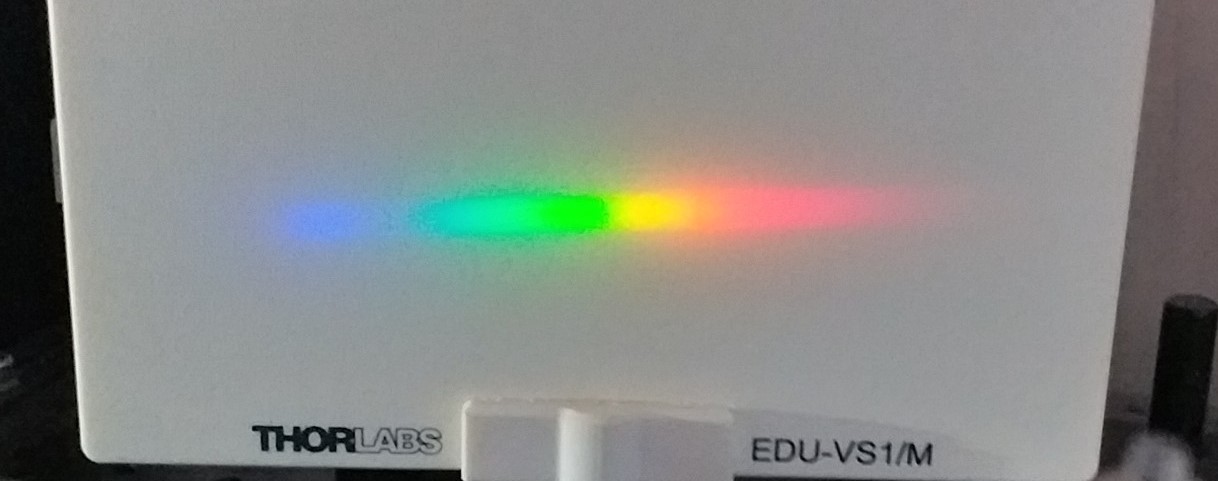}
	\caption{The spectrum of an LED by SPECTRUMMATE.}
    \label{fig:SpectrumMateTestBenchResult}
\end{figure}

A setup of SPECTRUMMATE was also built on a test bench (\autoref{fig:SpectrumMateTestBench}). As the design utilized two singlet lenses for the collimator and objective, chromatic aberration was an expected artifact. This aberration manifests as a slight broadening of spectral lines, as different colors are not brought to a perfect focus across the sensor plane. However, the spectral resolution of the instrument is primarily limited by the projected width of the entrance slit. The additional line broadening from chromatic aberration was found to be a minor contributor to the final measured Full Width at Half Maximum (FWHM) of the spectral lines, and the effect was deemed acceptable for the project's educational goals. The lenses were chosen through basic optical calculations first to make sure that the spectra did not spread too much when put on the test bench. If they do not satisfy further calculations, a different pair of lenses will be chosen. The spectrum of an LED used as the light source for this setup can be seen in Fig. \ref{fig:SpectrumMateTestBenchResult}.

In conclusion, the optical elements specific to SPECTRUMMATE include four components (Appendix \autoref{append:SPECTRUMMATE optical components}): A convex lens of $50\, mm$ focal length and $25.4\, mm$ aperture diameter used as a collimator lens. Another bi-convex lens with a focal length of $175\, mm$ and a diameter of $25.4\, mm$ used as an objective lens. A reflective holographic diffraction grating with a groove density of $1200\ grooves/mm$, $25.0\, mm$ on each side and a thickness of $6\, mm$. A narrow variable slit whose width is adjustable while its height remains constant at $3\, mm$.

\subsection{Accessories}
In addition to the mechanical and optical components, SPECTRUMMATE's functionality was enhanced with several accessories (Appendix \autoref{append:SPECTRUMMATE optical components}). A camera was added, contributing to the overall cost of the spectrometer. SpaceLAB had two cameras available: the ASI178MC (color) and ASI178MM (monochrome), both featuring a Sony IMX178 CMOS image sensor with a resolution of $3096 \times 2080$ pixels. For imaging, a monochrome camera is the preferred choice as the Bayer matrix in color cameras reduces sensitivity and image sharpness. However, if the goal is to display a part of the colorful spectrum in real-time as the grating rotates, a color camera would be more convenient since no spectral calibration is needed. These two cameras also have a small pixel size (2.4 $\mu m$), ensuring a small spectral dispersion for high resolution spectroscopy, since bigger pixel size equals higher spectral dispersion, which means the spectrum image is less detailed.

An SVBONY SV108 Helical Focuser with $0.1\, mm$ precision and a $10\, mm$ focusing stroke was selected for use with the camera to correct any manufacturing and assembly errors that might affect the objective lens to camera sensor distance. Additionally, an SVBONY SV165 guide scope ($D=30\, mm$, $f=120\, mm$) was used for capturing the Sun's spectrum, and a Baader Multi-Purpose Vario Finder $10 \times 60$ ($D=61\, mm$, $f=250\ mm$) was employed for imaging the Sun. Those two "telescopes" were suitable for high-resolution spectra capturing because of their short focal length.

\subsection{Parameter calculations}
\label{paremeterscalculation}

After finalizing the conceptual configuration and selecting the primary optical components from the SpaceLAB inventory, we performed a series of calculations. The purpose of this analysis was to verify that the chosen components would meet the key design requirements established in \autoref{subsec:Requirements for SPECTRUMMATE}, particularly regarding spectral resolution and the absence of vignetting. The key parameters for the SpectrumMate design are listed in \autoref{table:parapara}, alongside those of the original Sol'EX for comparison. The detailed derivations for these parameters are provided in Appendix \autoref{append:Detailed optical design calculations}. These parameters are provided by the component manufacturers.

\begin{wstable}[H]
    \caption{Parameters used for calculation.}
    \label{table:parapara}
    \begin{tabular}{lcc}
        \toprule
        & SPECTRUMMATE & Sol'EX\\
        \midrule
        Collimator lens: focal length $f_1$ and diameter $D_1$ & $50\ mm;\ 25.4\ mm$ & $80\ mm;\ 25.4\ mm$\\
        $F_C=f_1/D_1$ & $1.97$ & $3.15$\\
        Objective lens: focal length $f_2$ and diameter $D_2$ & $175\ mm;\ 25.4\ mm$ & $125\ mm;\ 25.4\ mm$\\
        $F_O=f_2/D_2$ & $6.89$ & $4.92$\\
        Total angle between the incident ray and the diffracted ray $\gamma$ & $40\degree$ & $34\degree$\\
        Distance from the objective lens to the camera & $87\ mm$ & $120\ mm$\\
        Grating groove density $m$ & \multicolumn{2}{c}{$1200\ \textnormal{grooves}/mm$}\\
        Dimensions $H$ & \multicolumn{2}{c}{$25\times25\times6\ mm$}\\
        Telescope's primary aperture diameter $D$ and focal length $f$ * & \multicolumn{2}{c}{$120\ mm;\ 30\ mm$}\\
        Sensor size & \multicolumn{2}{c}{$3096\times2080\ pixel$}\\
        Sensor pixel size $p_{size}$ & \multicolumn{2}{c}{$2.4\times2.4\ \mu m$}\\
        Slit $height\times width$ & \multicolumn{2}{c}{$3\ mm\times25\ \mu m$}\\
        \bottomrule
    \end{tabular}%
\end{wstable}

The calculations confirmed the viability of the design. A vignetting check showed that the collimator lens was sufficiently large for the telescope's f/4 beam. The grating equation was solved to find the optimal angles of incidence ($\alpha=40.49\degree$) and diffraction ($\beta=0.49\degree$) to center the reference wavelength 5500 \angstrom on the sensor.

The theoretical performance was then calculated. The spectral dispersion was determined to be $0.11 \angstrom/pixel$ on the camera sensor, and the spectral resolving power was calculated to be $R\approx1730$. This theoretical resolution of approximately 3.18 \angstrom is well within the requirement to resolve the 6 \angstrom separation of the Sodium D-lines. Finally, a check of the diffracted beam diameter confirmed that the objective lens was large enough to capture the entire dispersed spectrum across the width of the sensor, ensuring no light loss from vignetting.

\subsection{3D model design}
After finalising the underlying optics, we developed a comprehensive 3D model of every mechanical element in SPECTRUMMATE. Each part is dimensioned to mate cleanly with SpaceLAB's optical hardware rather than the Sol'EX standard.

\subsubsection{The main spectrometer body}

\begin{figure}[H]
	\centering
    \begin{subfigure}[b]{0.9\linewidth}
    	\centering
    	\includegraphics[width=0.5\linewidth]{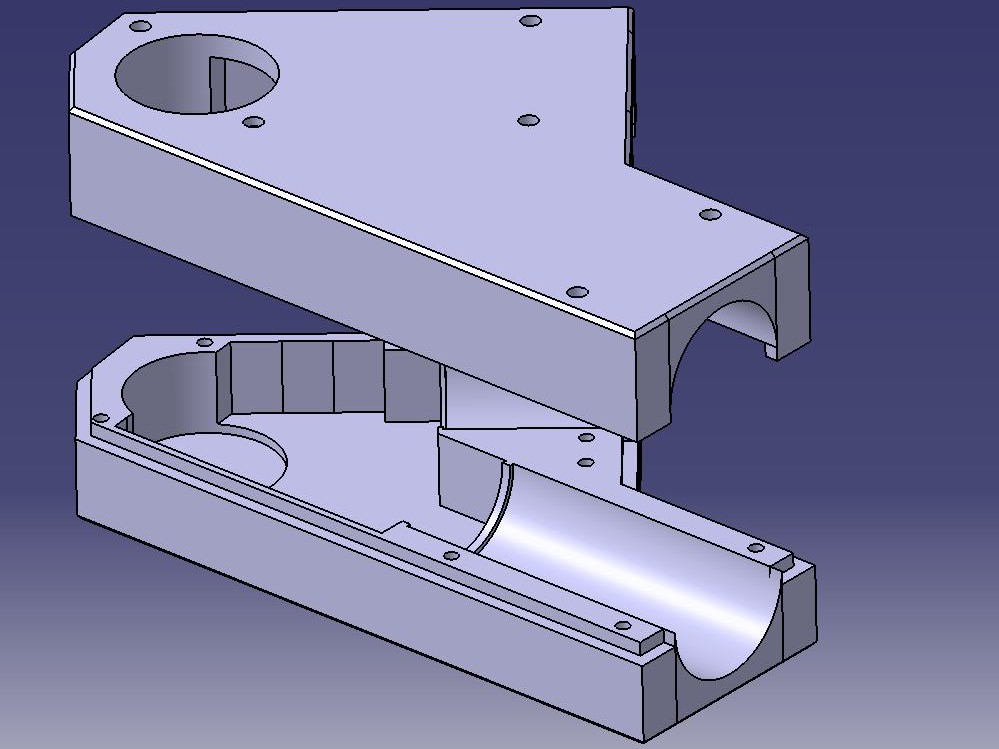}
    	\caption{The top cover (above) and the bottom cover (below).}
        \label{fig:topandbottomcover}
    \end{subfigure}
    
	\begin{subfigure}[b]{0.45\linewidth}
		\includegraphics[width=\linewidth]{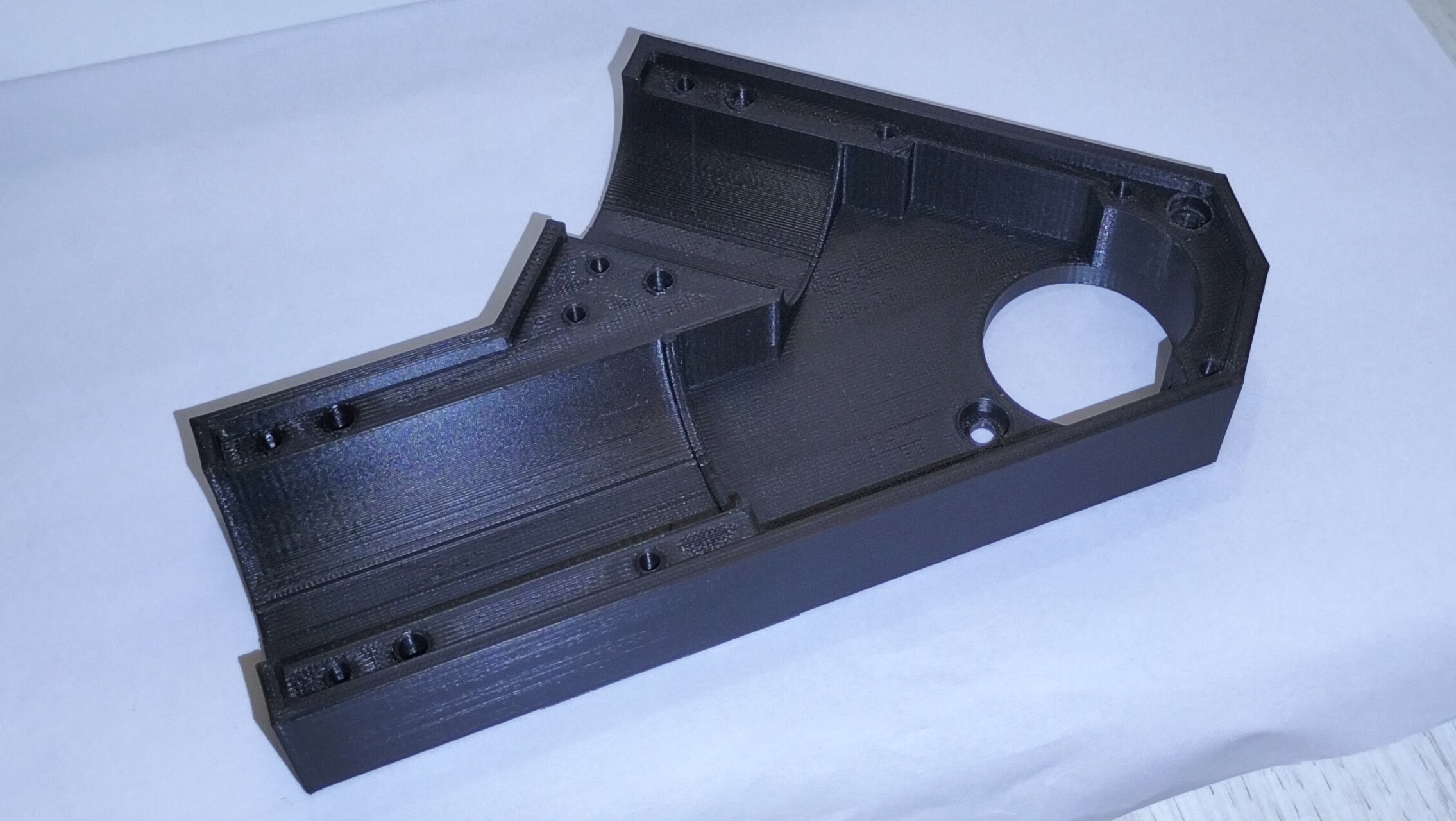}
    	\caption{The top cover.} 
	\end{subfigure}
	\begin{subfigure}[b]{0.45\linewidth}
		\includegraphics[width=\linewidth]{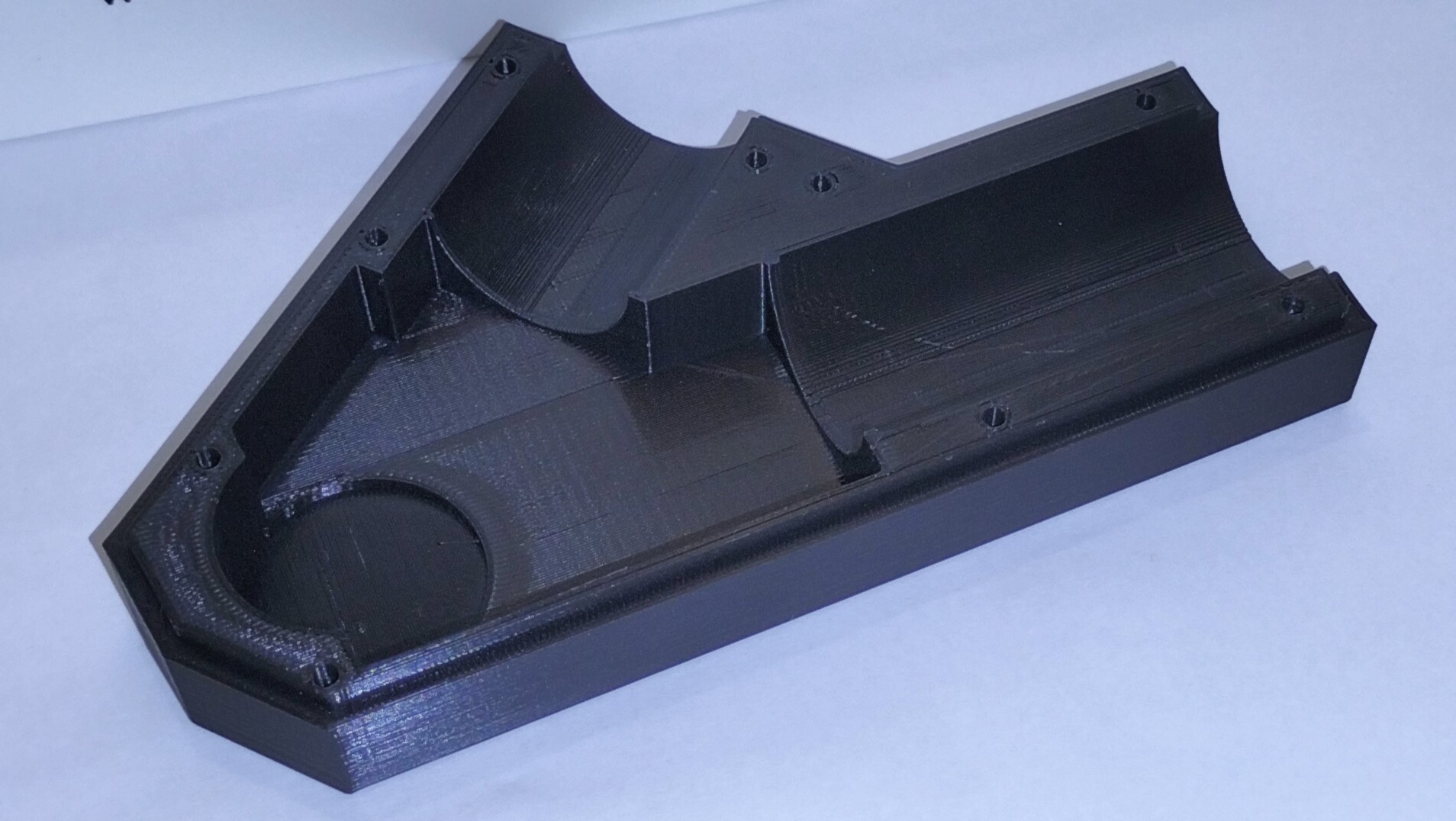}
    	\caption{The bottom cover.}
	\end{subfigure}
	\caption{Two parts of the enclosure.}
\end{figure}

The overall body of SPECTRUMMATE consists of an upper and lower part (\autoref{fig:topandbottomcover}). The two components were joined to form a light-tight enclosure for the internal optics. The critical task at this juncture was to ensure that no stray ambient light could enter the instrument through its cover.

\subsubsection{The grating holder}
\begin{figure}[H]
	\centering
    \begin{subfigure}[b]{0.9\linewidth}
    	\centering
    	\includegraphics[width=0.5\linewidth]{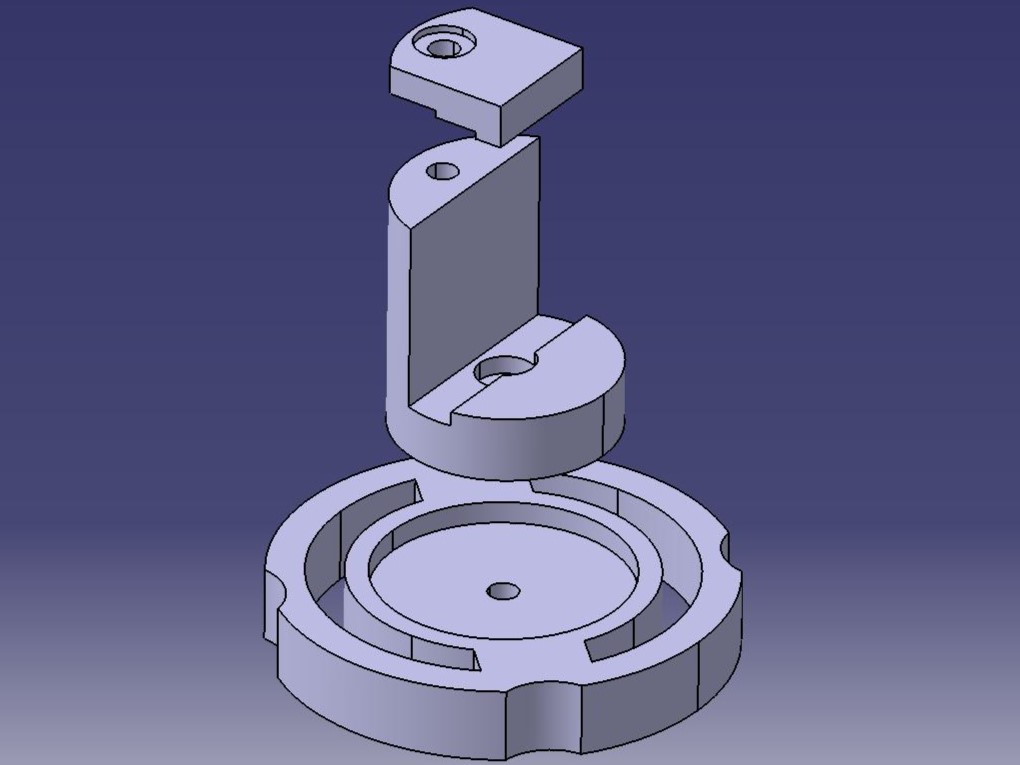}
    	\caption{The grating pat (top), the grating supporter (middle), and the grating rotator (bottom).}
        \label{fig:gratingholder}
    \end{subfigure}

	\begin{subfigure}[c]{0.3\linewidth}
		\includegraphics[width=\linewidth]{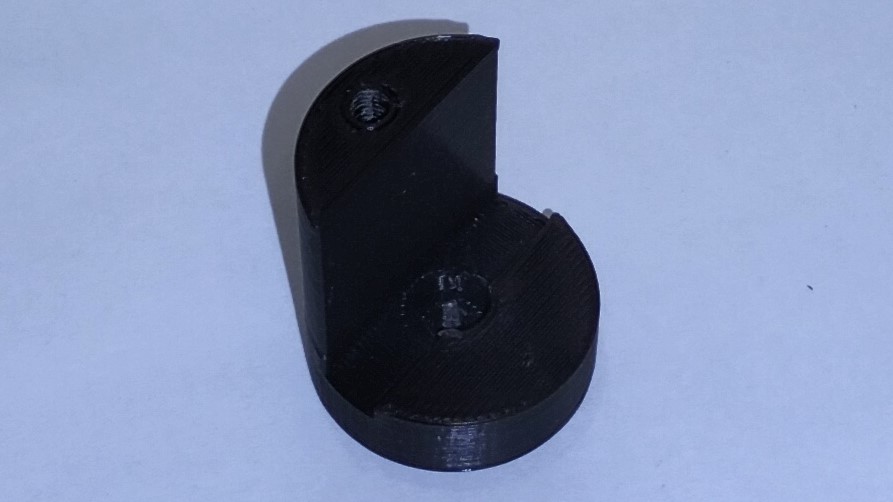}
		\caption{The grating supporter.}
	\end{subfigure}
	\begin{subfigure}[c]{0.3\linewidth}
		\includegraphics[width=\linewidth]{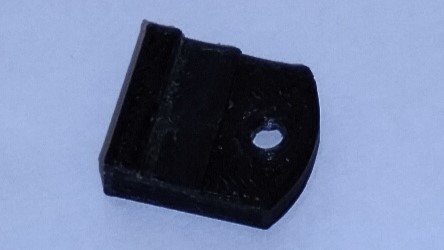}
		\caption{The grating pat.}
	\end{subfigure}
	\begin{subfigure}[c]{0.3\linewidth}
		\includegraphics[width=\linewidth]{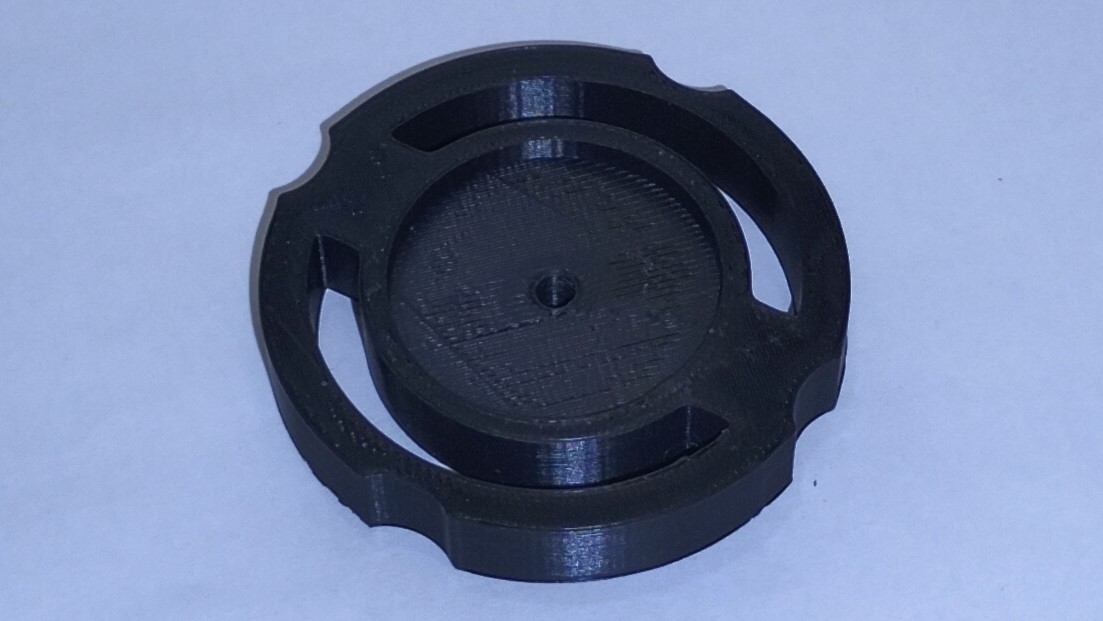}
		\caption{The grating rotator.}
	\end{subfigure}
	\caption{The three components of the grating holder.}
\end{figure}

The grating is placed in the groove of the grating support and is firmly held in place by the grating pad to prevent any movement. The grating rotator features slots for inserting screws and bolts to secure the grating holder to the top cover, allowing for rotation to adjust the incident angle of the incoming light ray.  The part is shown in Fig. \ref{fig:gratingholder}.

\subsubsection{The collimator tube and interface}
The collimator lens is positioned inside the tube and secured with a ring for added stability. A slit can be mounted on the collimator block. To further enhance stability, a stiffener is connected to the top cover and fastened to the tube with a single screw. The interface is designed to be directly connected to a $1.25\ inches$ telescope interface. The component is illustrated in Fig. 11.


\begin{figure}[H]
	\centering
    \begin{subfigure}[b]{0.9\linewidth}
    	\centering
    	\includegraphics[width=0.5\linewidth]{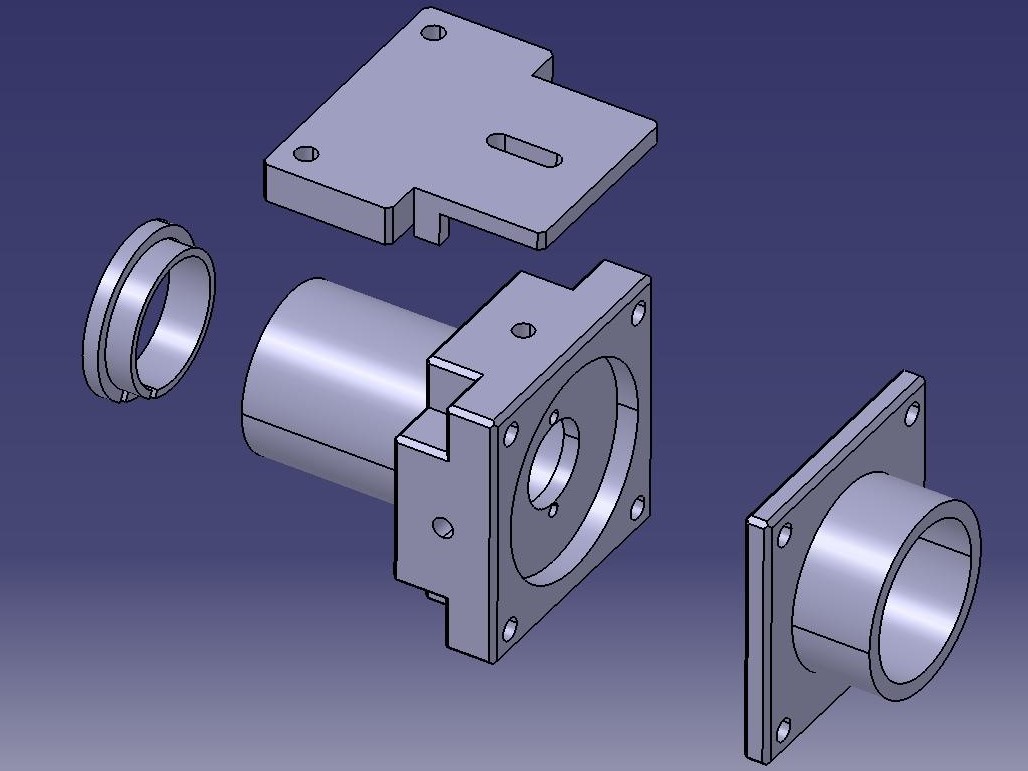}
    	\caption{The collimator stiffener (top), the collimator ring (middle), the collimator tube (top-middle), and the interface (right).}
        \label{fig:collimator}
    \end{subfigure}

	\begin{subfigure}[b]{0.45\linewidth}
		\includegraphics[width=\linewidth]{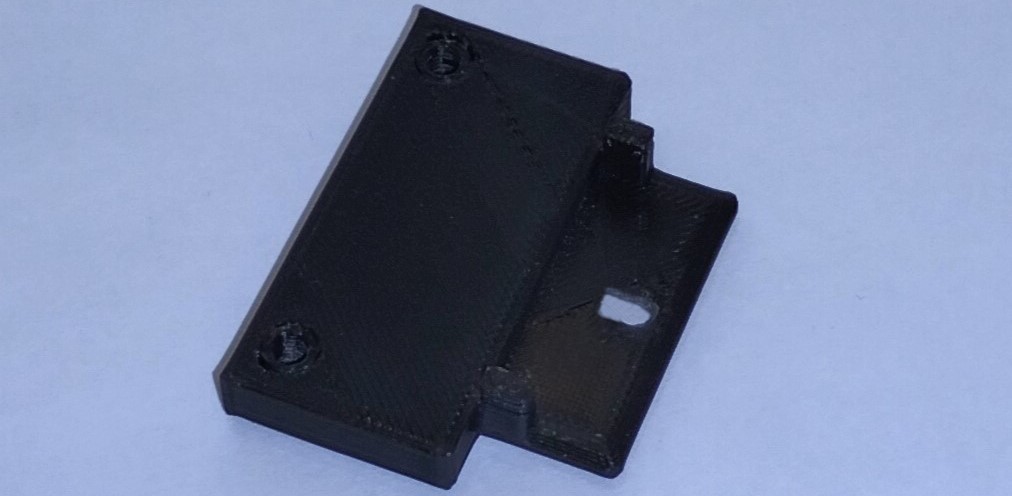}
    	\caption{The collimator stiffener.}
	\end{subfigure}
	\begin{subfigure}[b]{0.45\linewidth}
		\includegraphics[width=\linewidth]{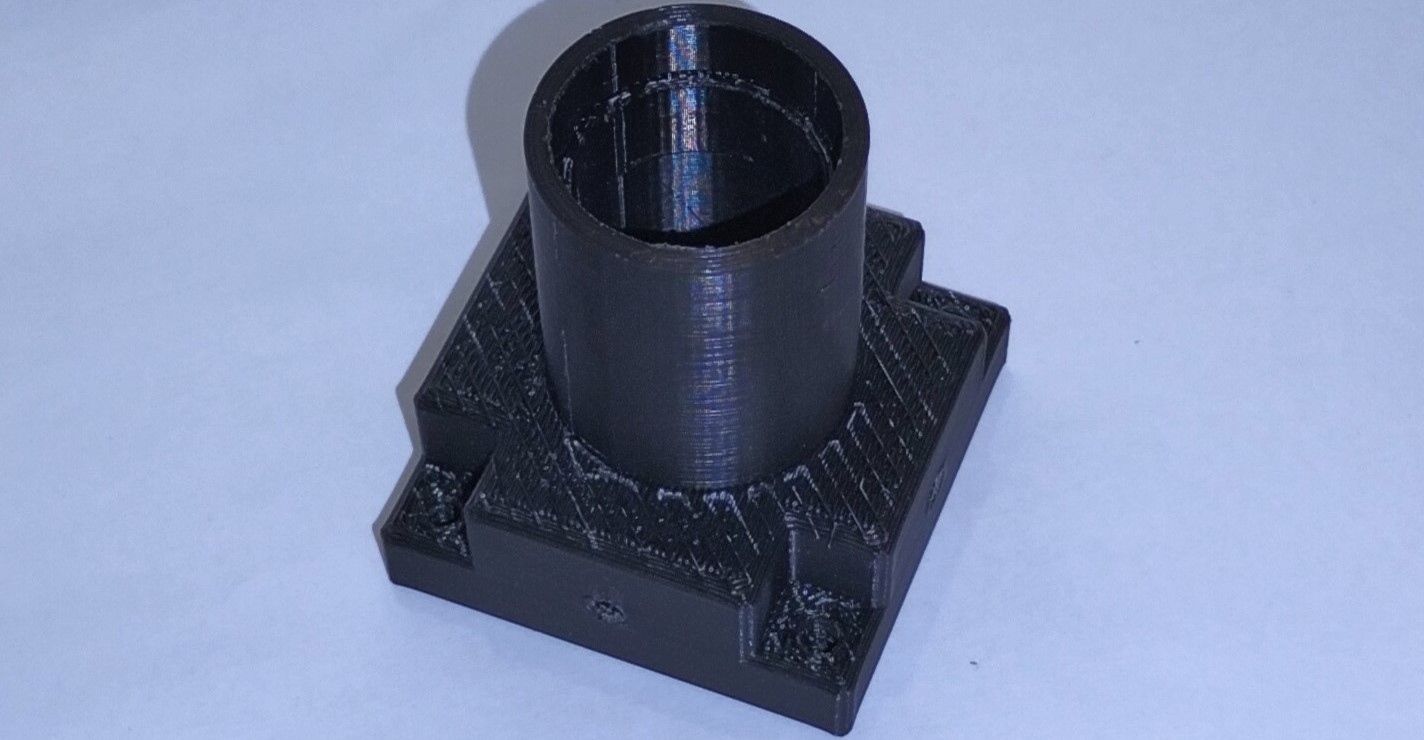}
    	\caption{The collimator tube.}
	\end{subfigure}
	\begin{subfigure}[b]{0.45\linewidth}
		\includegraphics[width=\linewidth]{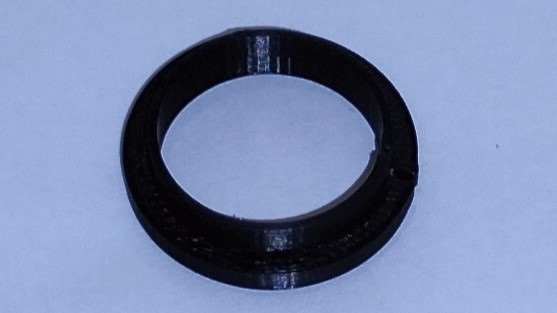}
    	\caption{The collimator ring.}
	\end{subfigure}
	\begin{subfigure}[b]{0.45\linewidth}
		\includegraphics[width=\linewidth]{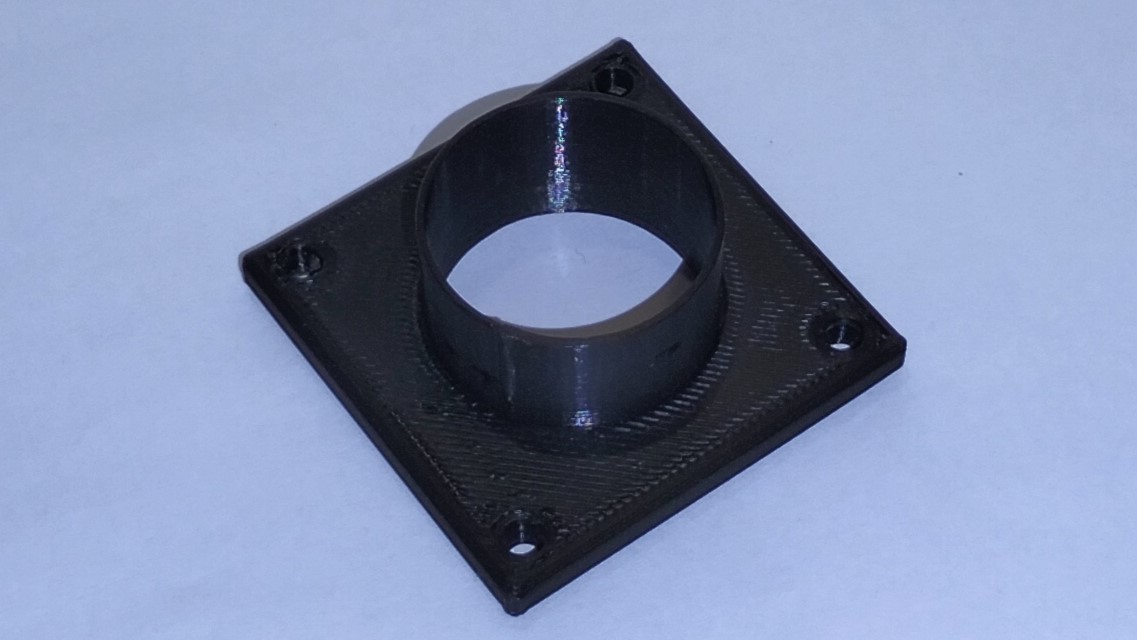}
    	\caption{The interface.}
	\end{subfigure}
	\caption{The collimator tube.}
\end{figure}

\subsubsection{The objective tube}
The objective lens directs the diffracted light rays and focuses the light onto the camera sensor. Also here we introduced a collimator lens ring for enhanced rigidity and stability, securing the lens at a fixed place. The design allows the camera to be attached to the box-shaped structure of the objective tube (Fig. 12). For added support, we chose to pair the objective tube structure with a stiffener for extra rigidity of the enclosure.

\begin{figure}[H]
	\centering
    \begin{subfigure}[b]{0.9\linewidth}
    	\centering
    	\includegraphics[width=0.5\linewidth]{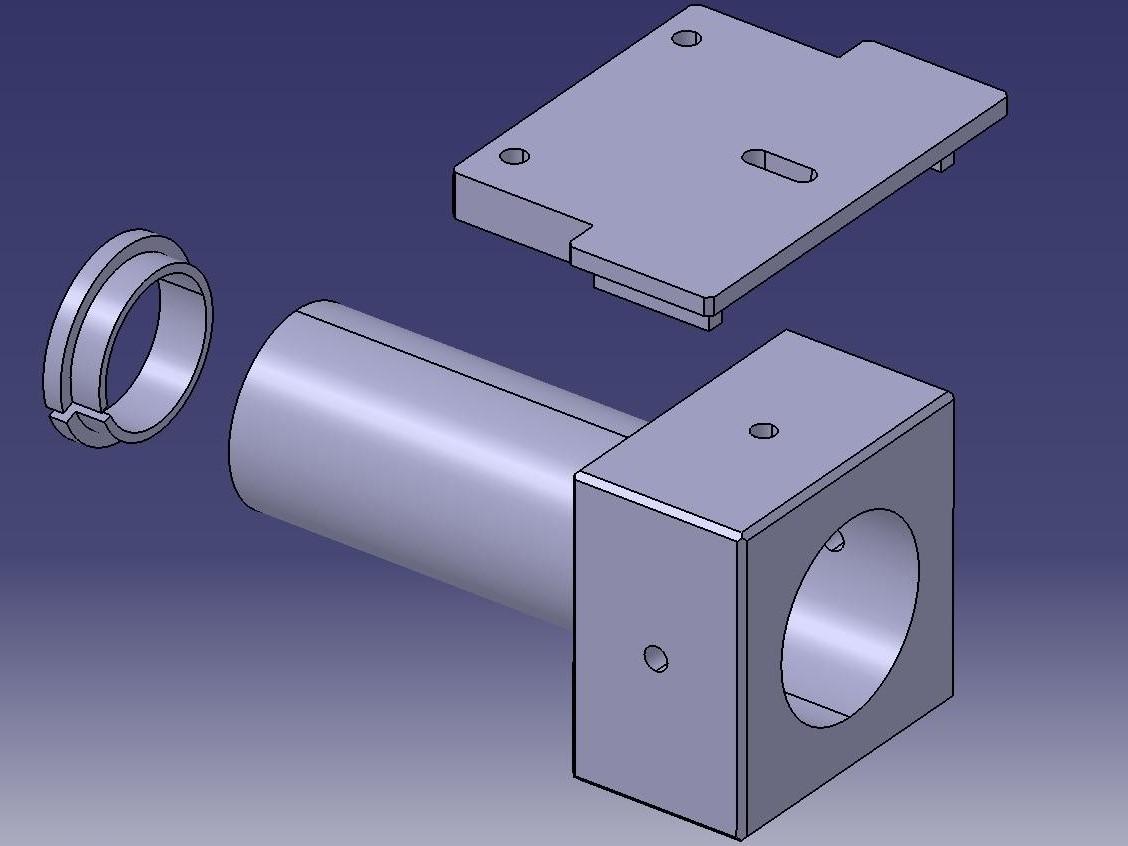}
    	\caption{The objective ring (left), the objective tube (right), and the objective stiffener (top).}
        \label{fig:objectivetube}
    \end{subfigure}
    
	\begin{subfigure}[b]{0.45\linewidth}
		\includegraphics[width=\linewidth]{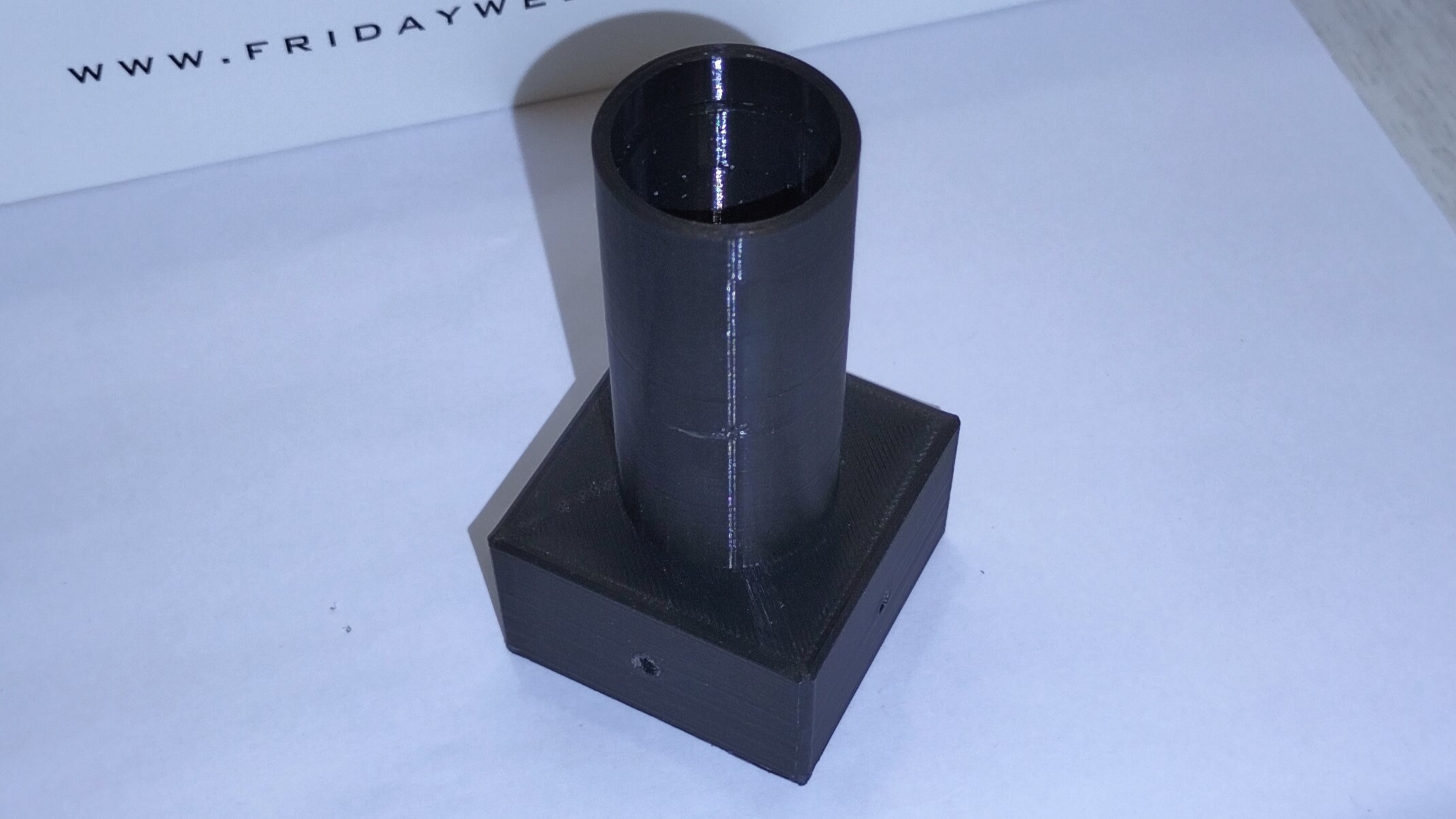}
        \caption{The objective tube.}
	\end{subfigure}
	\begin{subfigure}[b]{0.45\linewidth}
		\includegraphics[width=\linewidth]{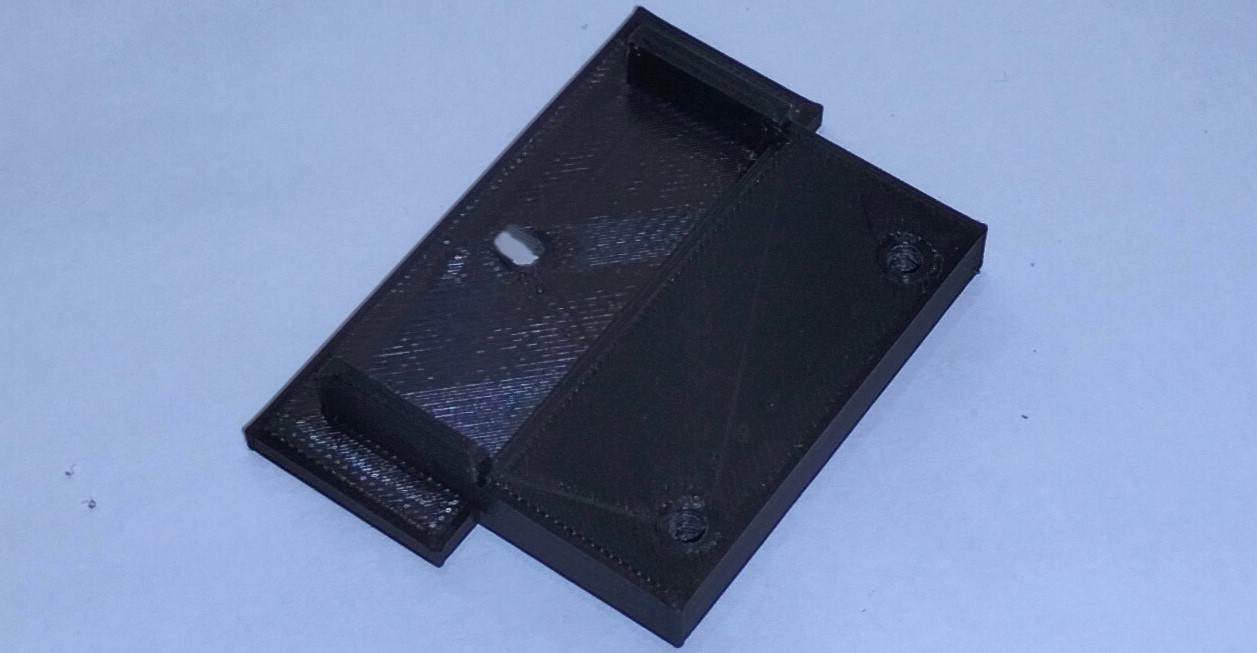}
        \caption{The objective tube stiffener.}
	\end{subfigure}
	\caption{The objective tube.}
\end{figure}

\subsubsection{The complete 3D model}
\begin{figure}[H]
	\centering
	\includegraphics[width=0.6\linewidth]{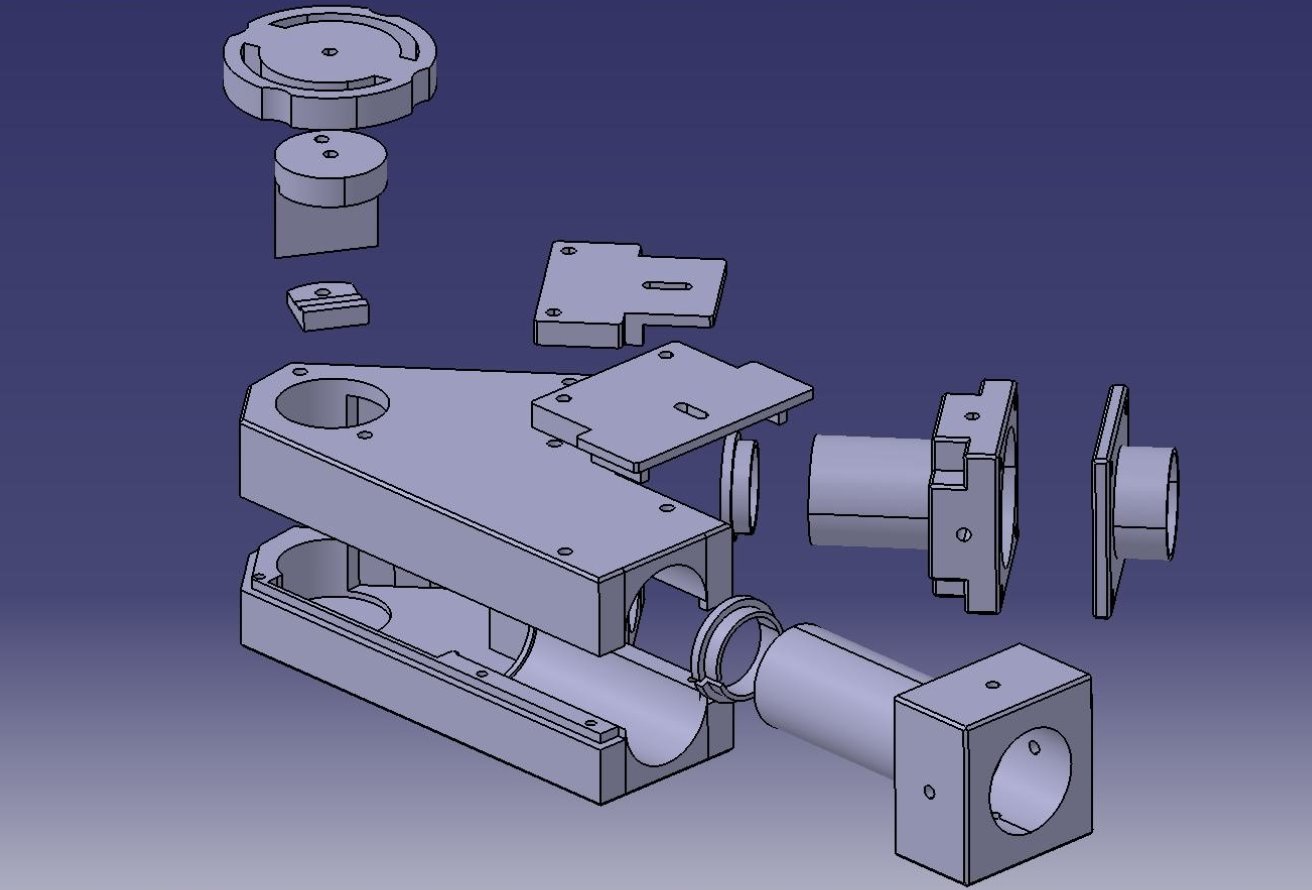}
	\caption{SPECTRUMMATE assembly.}
    \label{fig:12parts}
\end{figure}

Fig. \ref{fig:12parts} shows the assembly of \textsc{SpectrumMate}. Before proceeding with 3D printing, the assembly was simulated on a computer to ensure there was no overlap between parts.

\section{Spectrometer assembly}
After completing the 3D design and assembly on the computer, the files were converted into STL files suitable for 3D printing. Standard PLA filament was selected as the base material for the spectrometer enclosure and related parts. While standard PLA is accessible and easy to print, it is susceptible to deformation when exposed to high temperatures and direct sunlight for extended periods. For the short-duration tests conducted in this work, this did not present an issue. However, for users intending to use SpectrumMate for prolonged solar observations, we strongly recommend printing the components in a more heat-resistant material such as PETG or, ideally, ASA, to ensure long-term mechanical and optical stability.

One important thing to do is to test if the objective tube and camera can focus at infinity. We pointed the system at a distant mountain (about 5 $km$), where the incident rays could be considered parallel, to check if it could form an image.

\begin{figure}[H]
	\centering
	\includegraphics[width=0.8\linewidth]{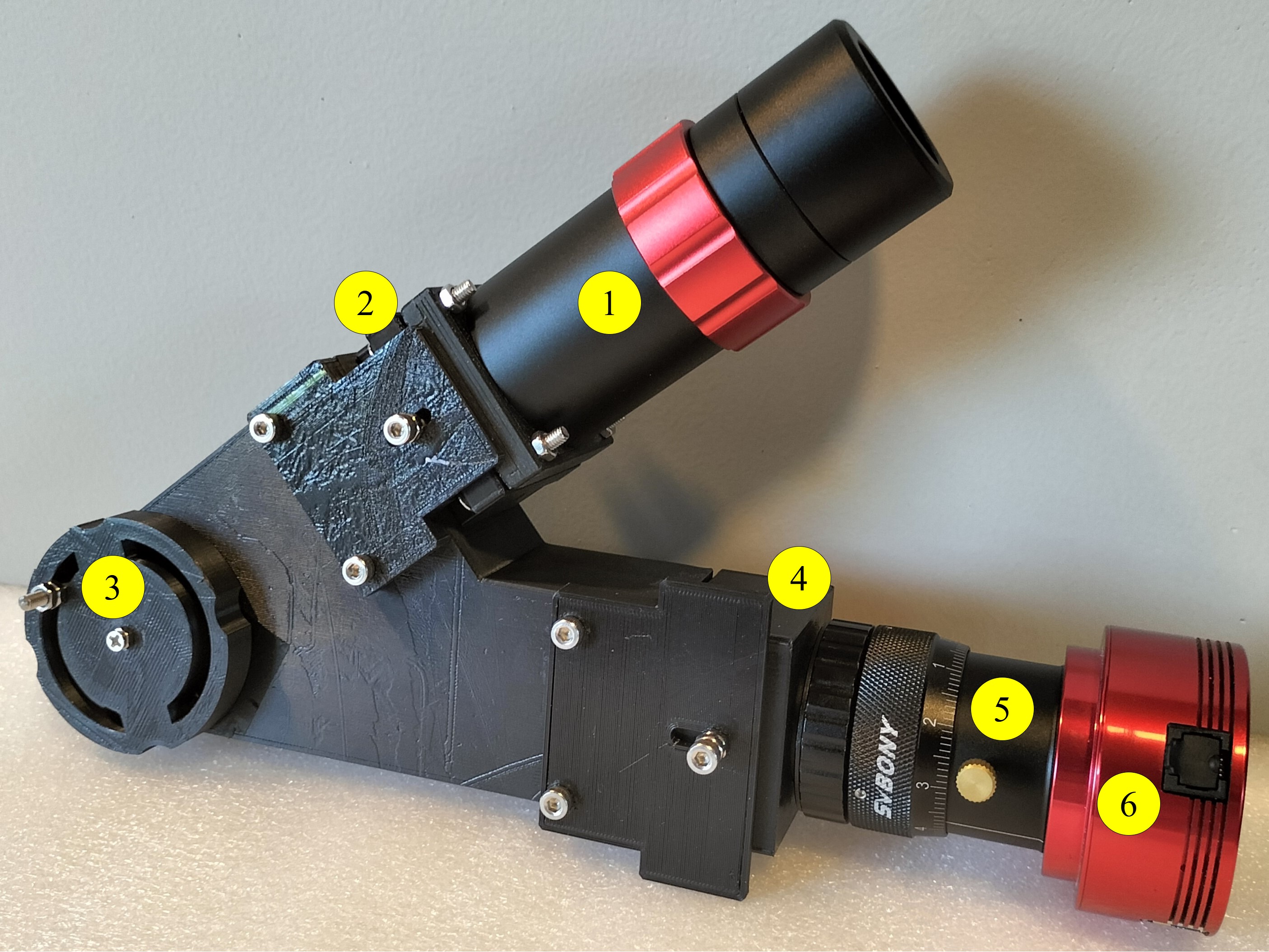}
	\caption{SPECTRUMMATE equipped with accessories. The guide scope (1) is attached to the collimator tube (2), the camera (6) is attached to the focuser (5), and then inserted into the objective tube (4). The grating holder (3) rotates smoothly.}
    \label{fig:IMG20230630090946_1}
\end{figure}

The most important aspect to consider after SPECTRUMMATE was assembled was ensuring incoming light passes through the system along the optical path without obscuring, diminishing, or detecting any light leakage. For the light-blocking check, when pointing a narrow laser in the slit, light could be seen coming out of the other end, so the light ray is not blocked. The laser was also used formerly to check the alignment of optical elements. For light leakage, when blocking the slit and checking the image taken by the camera, there was no light detected in the image. We conclude that SPECTRUMMATE does not suffer from any unintended light leakage. The spectrometer is presented in Fig. \ref{fig:IMG20230630090946_1}.

\section{Observations}
The observations conducted using SPECTRUMMATE have provided valuable insights into the capabilities and limitations of this custom-built spectrometer, particularly in the context of solar and artificial (bright) light source spectroscopy. By capturing the Sun's spectrum and the emission spectrum of a fluorescent lamp, several critical aspects of the instrument's performance were revealed, which will be discussed in detail in the following.

\subsection{Spectrum of the Sun and other light sources}
For our tests, we chose the Sun and a fluorescent lamp. The exposure time and gain of the camera have to be determined carefully so that the brightness of each image does not vary strongly. A long image exposure and low camera gain render the image smoother; a short exposure with a high camera gain setting allows for faster image acquisition, but at the cost of increased noise.

For single-frame image capture, we used SharpCap software. Since the incident light flux from the Sun is high, we can directly point the guide scope toward a blue sky to begin the data retrieval process. If the absorption lines may not be seen or are blurry, the focuser needs to be adjusted so that light will converge on the sensor. Once absorption lines are focused, we adjusted the camera gain and exposure time to prevent image saturation. The photos taken are shown in Appendix \autoref{append:Image taken by SPECTRUMMATE}.

Because of the limited coverage of the wavelength range, we manually stitched single-exposure images using Adobe Photoshop to form a wide-range spectrum. The stitching part is a critical problem, potentially introducing errors. These challenges primarily include maintaining a consistent wavelength scale across the seams and ensuring photometric continuity (i.e., consistent brightness and contrast) between adjacent frames.

\begin{figure}[H]
    \centering
    \includegraphics[width=\linewidth]{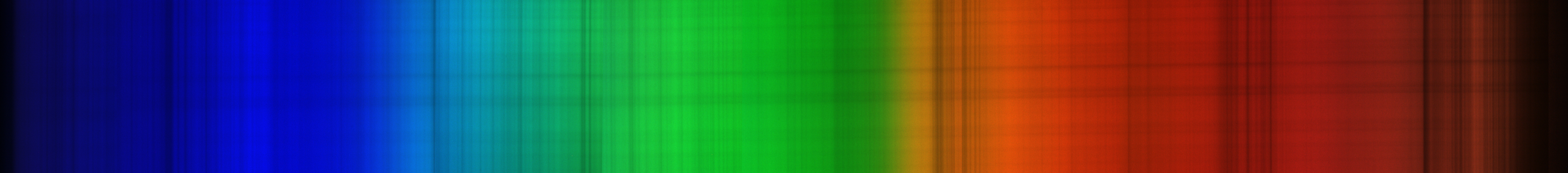}
    \caption{Sun's spectrum taken by SPECTRUMMATE on $120~mm$ telescope, $1200\ grooves/mm$ grating and a $25~\mu m$ wide slit. This spectrum was created by stitching data from the ASI178MC color camera. Prominent absorption lines are qualitatively readily visible. 1D line profile at \autoref{fig:SunSpectrumMate}.}
    \label{fig:sunspectrumbyspectrummate}
\end{figure}

\begin{figure}[H]
    \centering
    \includegraphics[width=\linewidth]{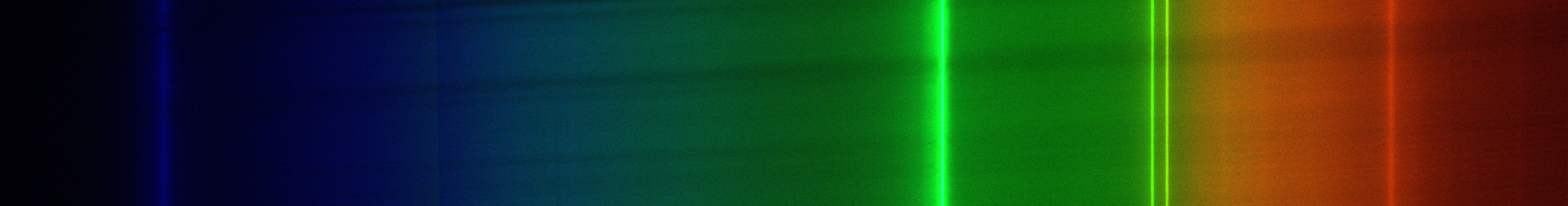}
    \caption{Recorded spectrum of a fluorescent lamp spectrum taken by SPECTRUMMATE on $120~mm$ telescope, $1200~\textnormal{grooves}/mm$ grating and a $25~\mu m$ wide slit. 1D line profile at \autoref{fig:FluorescentLampSpectrumMate}.}
    \label{fig:lampspectrumbyspectrummate}
\end{figure}

To mitigate these issues, a careful procedure was followed. First, a significant spectral overlap of approximately 25\% was intentionally captured between consecutive images. In Photoshop, each image was placed on a separate layer. Prominent Fraunhofer absorption lines present in the overlapping regions of adjacent layers were used as fiduciary markers to precisely align the images, ensuring the wavelength scale remained linear and continuous. Once aligned, the brightness and contrast of each layer were manually adjusted to visually match its neighbors. Finally, layer masks with soft-edged brushes were used to blend the seams, creating a smooth transition. In \autoref{fig:sunspectrumbyspectrummate} we show the result after stitching 14 single-exposure images covering the visible part from approximately 3800 Å in the blue to 6600 Å in the red. The data shown are uncalibrated (no dark subtraction or flat-fielding). The prominent horizontal lines are artifacts caused by dust and physical imperfections on the slit jaws. These imperfections block light before dispersion, casting a shadow across all wavelengths. The specific pattern of these artifacts can change depending on how the slit is illuminated (as they are not shown clearly in \autoref{fig:lampspectrumbyspectrummate}). Such artifacts would typically be removed by dividing the image by a flat-field frame. 

Fig. \ref{fig:lampspectrumbyspectrummate} shows the spectrum of the fluorescent lamp gas when placed directly in front of the telescope objective. Mercury has a well-characterised spectral fingerprint with known (prominent) emission lines and can be used for wavelength calibration.

\subsection{Image of the Sun at H-alpha}
The advantage of SPECTRUMMATE is to capture solar images as seen at different wavelengths. Here, we choose to pay attention to produce an H-alpha image to render the Sun's chromosphere. The H-alpha line, located at $656.28\ nm$, allows us to observe solar features such as prominences, filaments, and active regions, which are otherwise less visible in other parts of the spectrum. This wavelength is crucial for studying solar dynamics and understanding the behaviour of solar flares and other chromospheric phenomena.

\begin{figure}[H]
	\centering
	\includegraphics[width=\linewidth]{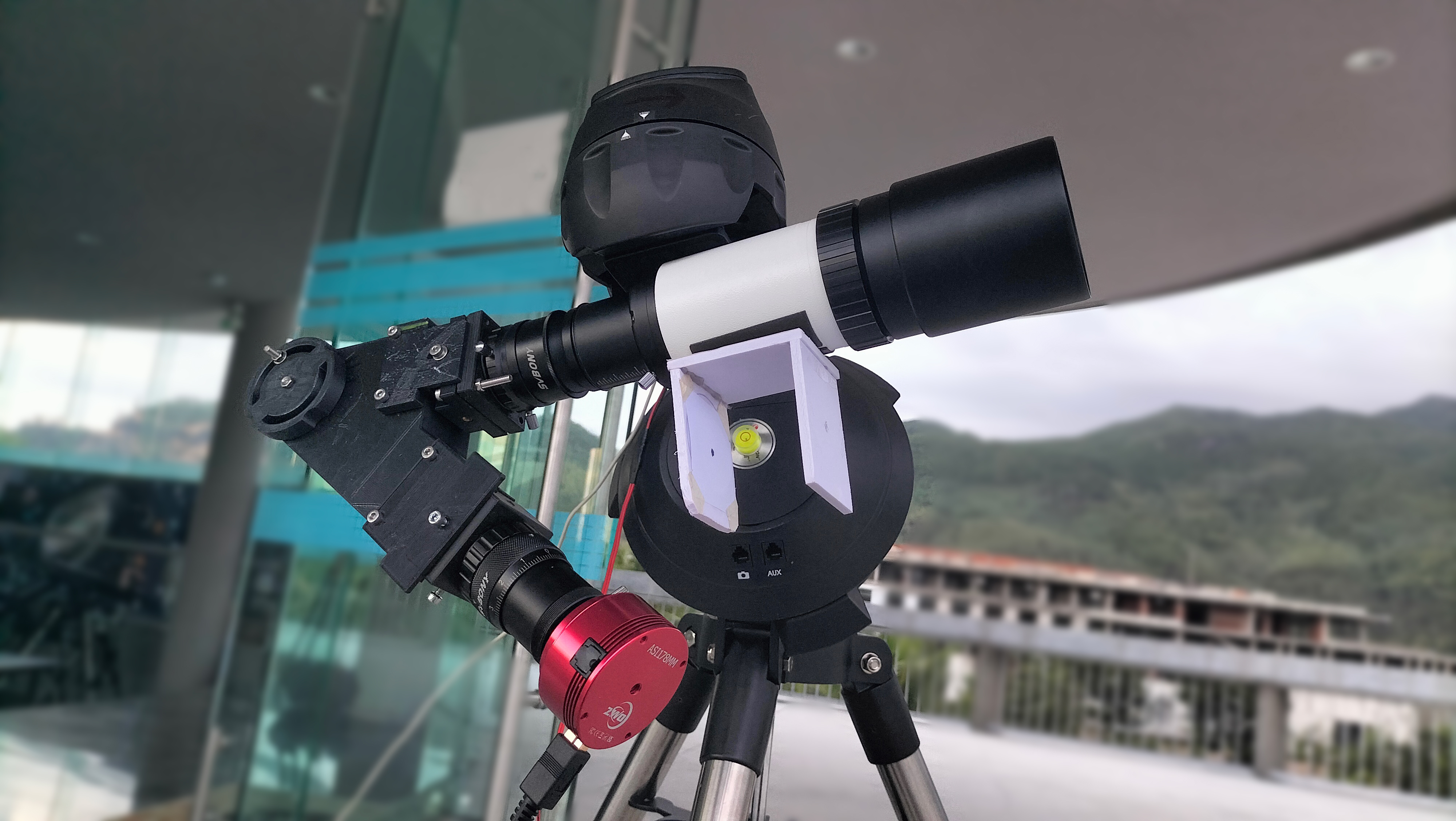}
	\caption{Final assembly of the telescope-spectrograph-camera-mount system to capture for solar observations.}
    \label{fig:CaptureSunSystem}
\end{figure}

We attached SPECTRUMMATE to the NexStar 6 SE Computerized Mount\footnote{\url{https://www.celestron.com/products/nexstar-6se-computerized-telescope}}. For accurate tracking, the mount was polar aligned. With the help of the accompanied Celestron PlaneWave Instrument (CPWI) control software, we directed the telescope towards the solar meridian. Images were captured using the SharpCap software suite. Refer to Fig. \ref{fig:CaptureSunSystem} for a visual of the system.

\begin{figure}[H]
	\centering
	\includegraphics[width=0.5\linewidth]{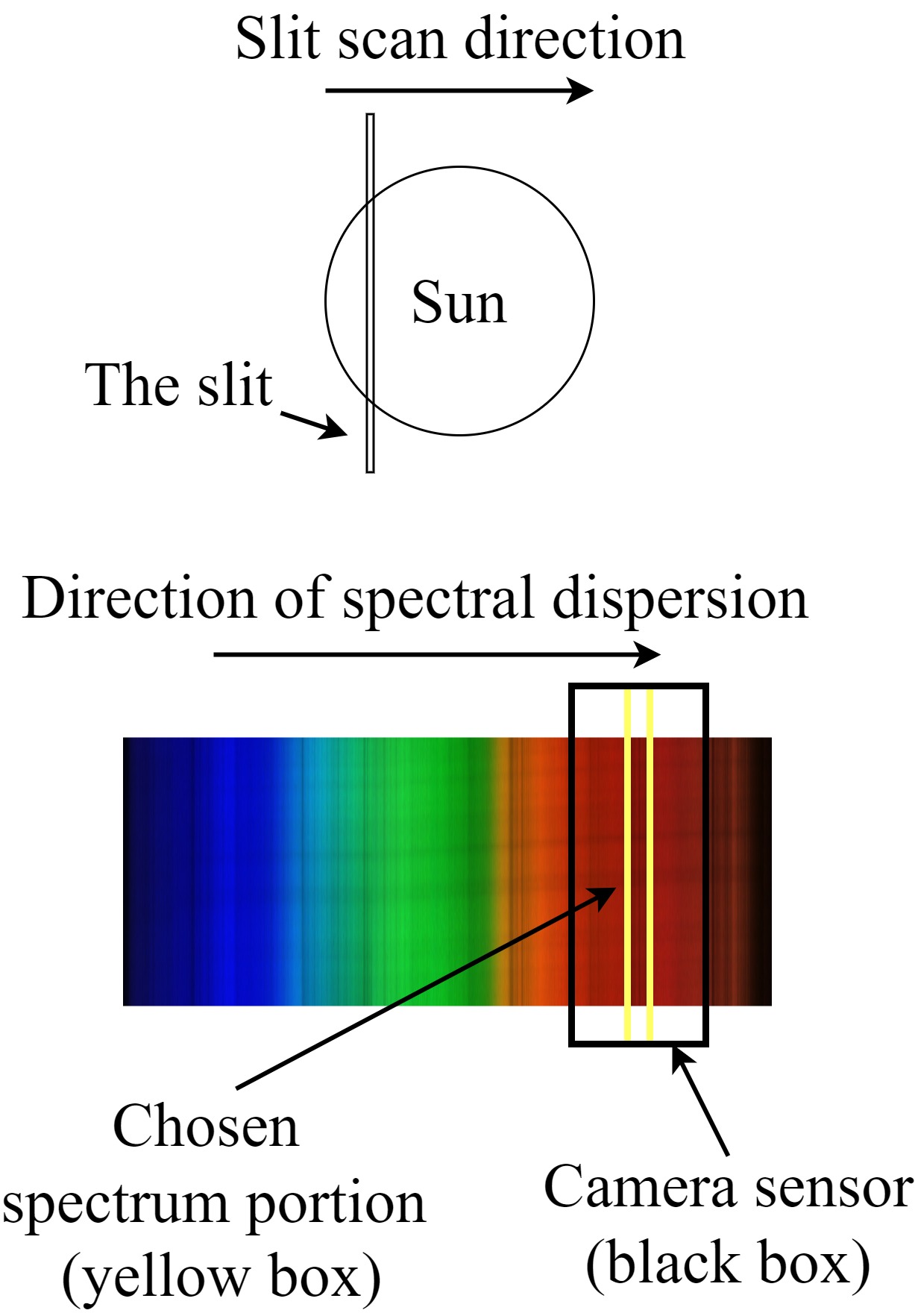}
	\caption{Details of the drift-scan method for imaging of the solar disk at a particular wavelength.}
    \label{Scan}
\end{figure}

The shooting method involved cropping a portion of the spectrum around the desired wavelength. The Sun disk was "scanned" at a slow and steady speed, resulting in multiple slices of the Sun at the chosen wavelength. The camera was positioned to ensure the dispersion direction was perpendicular to the longer image-sensor axis, thereby maximising coverage of the Sun's disk on the sensor plane. Fig. \ref{Scan} displays the method we used.

\begin{figure}[H]
	\centering
	\includegraphics[width=\linewidth]{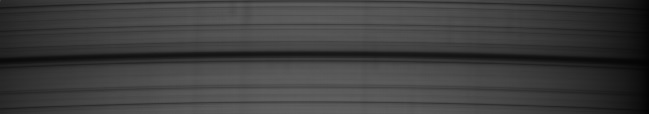}
	\caption{The portion of the spectrum with the H-alpha line (dim black) in the middle, being captured by the camera. The spectral dispersion direction is vertical.}
    \label{fig:HalphaOnCamera}
\end{figure}

Fig. \ref{fig:HalphaOnCamera} demonstrates what is captured by the camera. The data captured is in the form of a video ser file. The software used to process the data was INTI\footnote{\url{http://valerie.desnoux.free.fr/inti/}}.

The Sun's image size within the slit is calculated to be $\tan(0.5\degree) \times f = \tan(0.5\degree) \times 250\ mm = 2.18\ mm$, where $f$ is the focal length of the employed telescope and $0.5\degree$ is the angular size of the Sun on the sky. This means to capture the entire solar disk in a single scan, the slit must be longer than $2.18\ mm$. Therefore, for this specific observation, the 3 $mm$ variable slit used for initial characterization was replaced with a custom reflective slit with dimensions of $10\ \mu m \times 4.5\ mm$, providing ample margin to cover the entire solar image.

\begin{figure}[H]
	\centering
	\includegraphics[width=\linewidth]{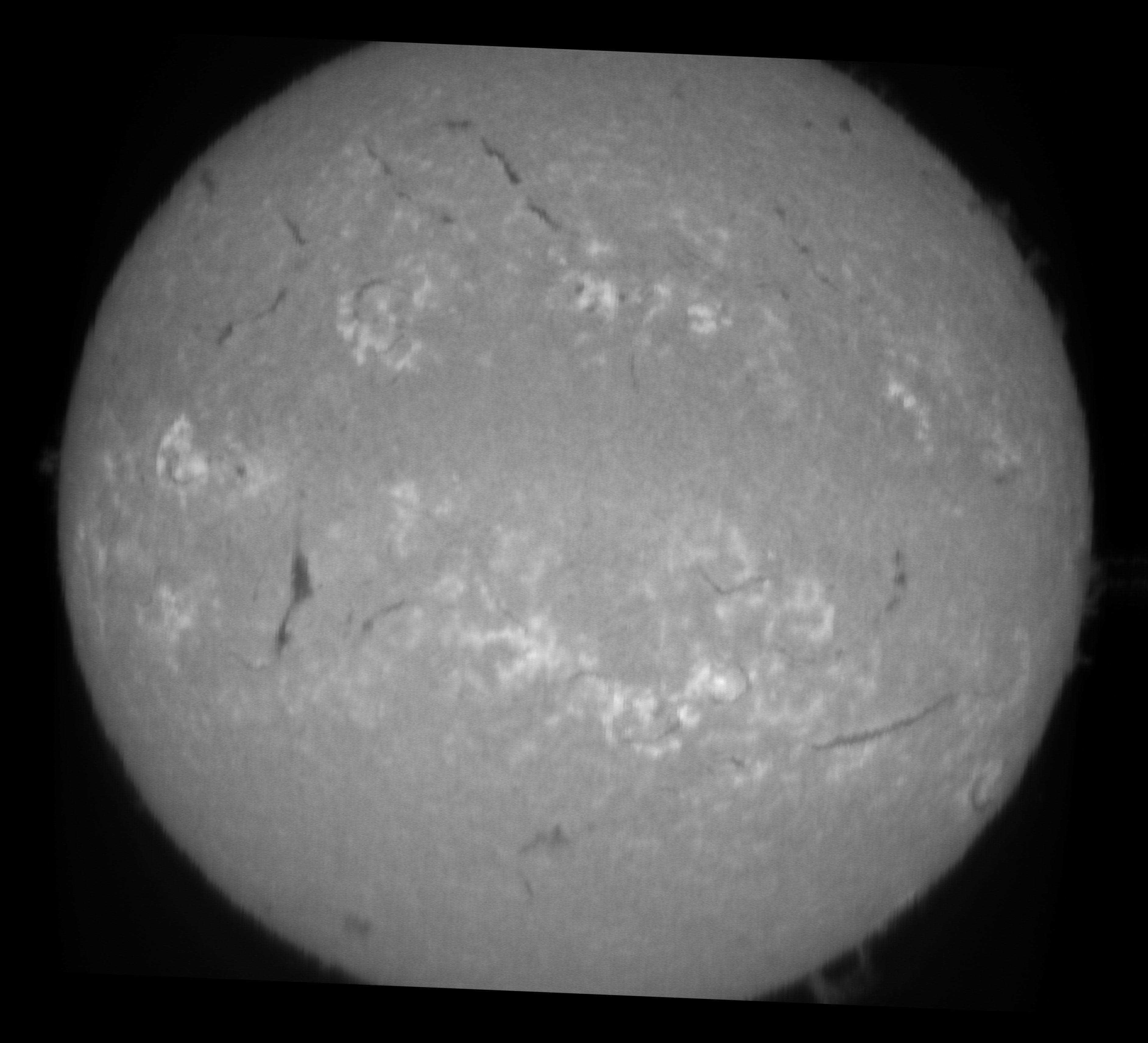}
	\caption{Final result of our observations: image of the Sun in H-alpha.}
    \label{fig:imageofthesun}
\end{figure}

\autoref{fig:imageofthesun} shows the final result displaying the Sun as captured in H-alpha. The apparent clipping at the top and bottom of the solar disk is not caused by the slit being too short. Instead, it is an artifact of the drift-scan method resulting from imperfect alignment between the Sun's drift direction and the long axis of the slit. This misalignment caused the northern and southern limbs of the Sun to move out of the slit's field of view before the full disk had been scanned, resulting in the clipped appearance. The blurring at the solar limb is a combination of this scanning misalignment and minor optical aberrations.

The data for the H-alpha image was captured as a video file in the SER format. The acquisition was performed using the SharpCap software. To achieve a good signal-to-noise ratio while preventing saturation, a short exposure time of approximately 15 $ms$ was used for each frame. The Sun was scanned across the slit at a slow, controlled rate of approximately 45 $arcsec/s$, a process which took about 40 $seconds$ to complete for the full disk. The final image was then reconstructed from the video data using the INTI software.

Though the quality of the image is not perfect, we still resolve and identify several physical features. Most vividly, our observations show extended dark features scattered on the solar face and brighter (higher contrast against a dark background) arc-like structures at the solar limb (affected by being blurred) are identified as solar prominences. A prominence is a phenomenon in which surface plasma material is transported via magnetic field lines to higher altitudes and forms a loop with each end of the prominence anchored to the Sun's surface. The reason they appear dark is that they are rising along the observer's line of sight. In the event of instabilities in the supporting magnetic field, plasma material can escape, thereby entering space at speeds on the order of thousands of kilometres per second.

\section{Data reduction}
This section will focus on reducing the data collected using SPECTRUMMATE, specifically through spectral calibration. The process involves manually calibrating the spectrum images by consulting a spectral atlas. For this project, we used the solar spectrum atlas compiled by \citet{garde2016} and the monograph by \citet{walker2017} as our primary references. By comparing the absorption lines in the Sun's spectrum atlas with those in our images, it is possible to determine the wavelengths of those lines accurately.

\begin{figure}[H]
	\centering
	\includegraphics[width=\linewidth]{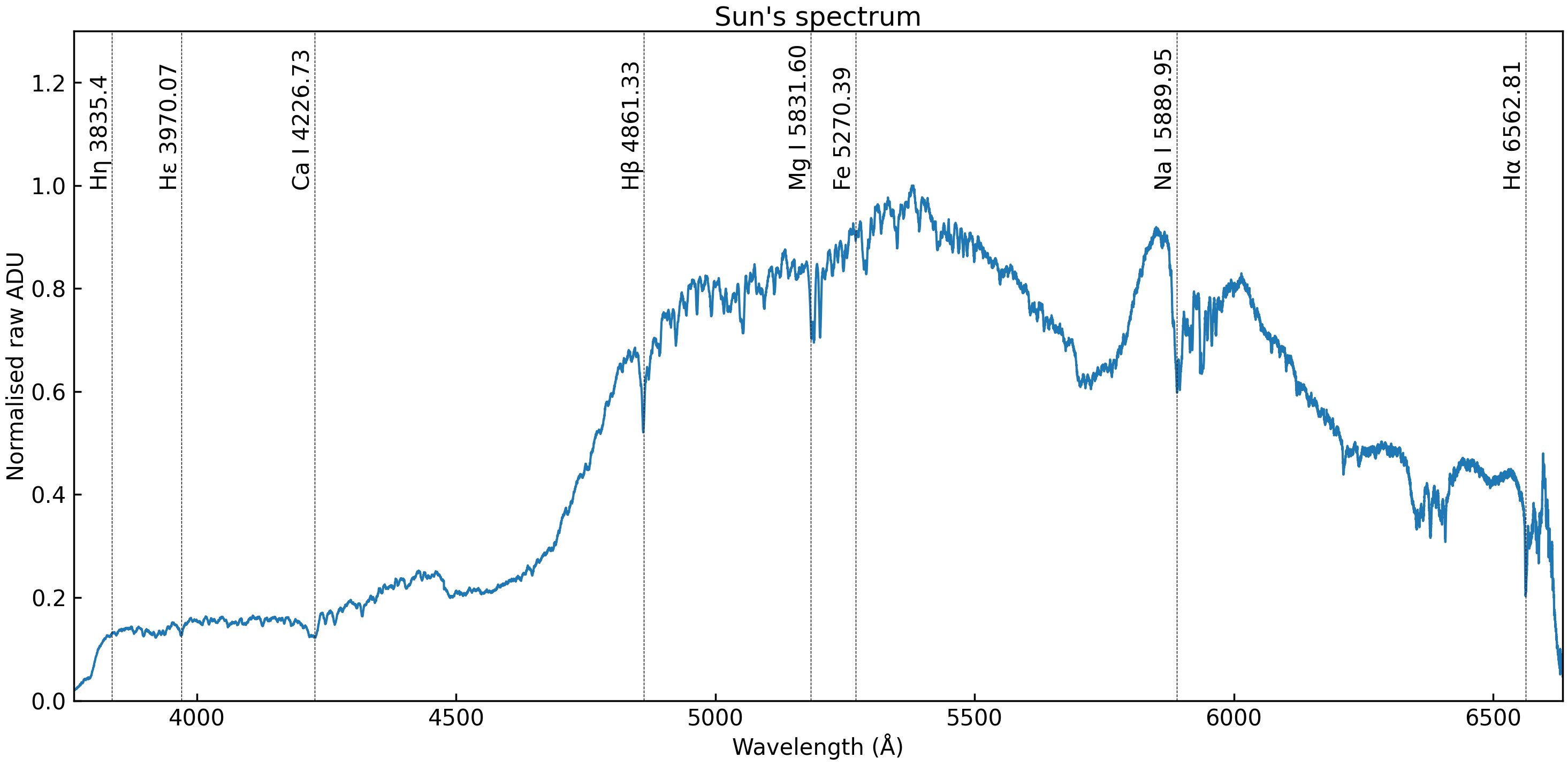}
	\caption{Spectral calibrated Sun's spectrum by SPECTRUMMATE on $120\ mm$ telescope, $1200~\textnormal{grooves}/mm$ grating, and $25\ \mu m$ slit.}
    \label{fig:SunSpectrumMate}
\end{figure}

\begin{figure}[H]
	\centering
	\includegraphics[width=\linewidth]{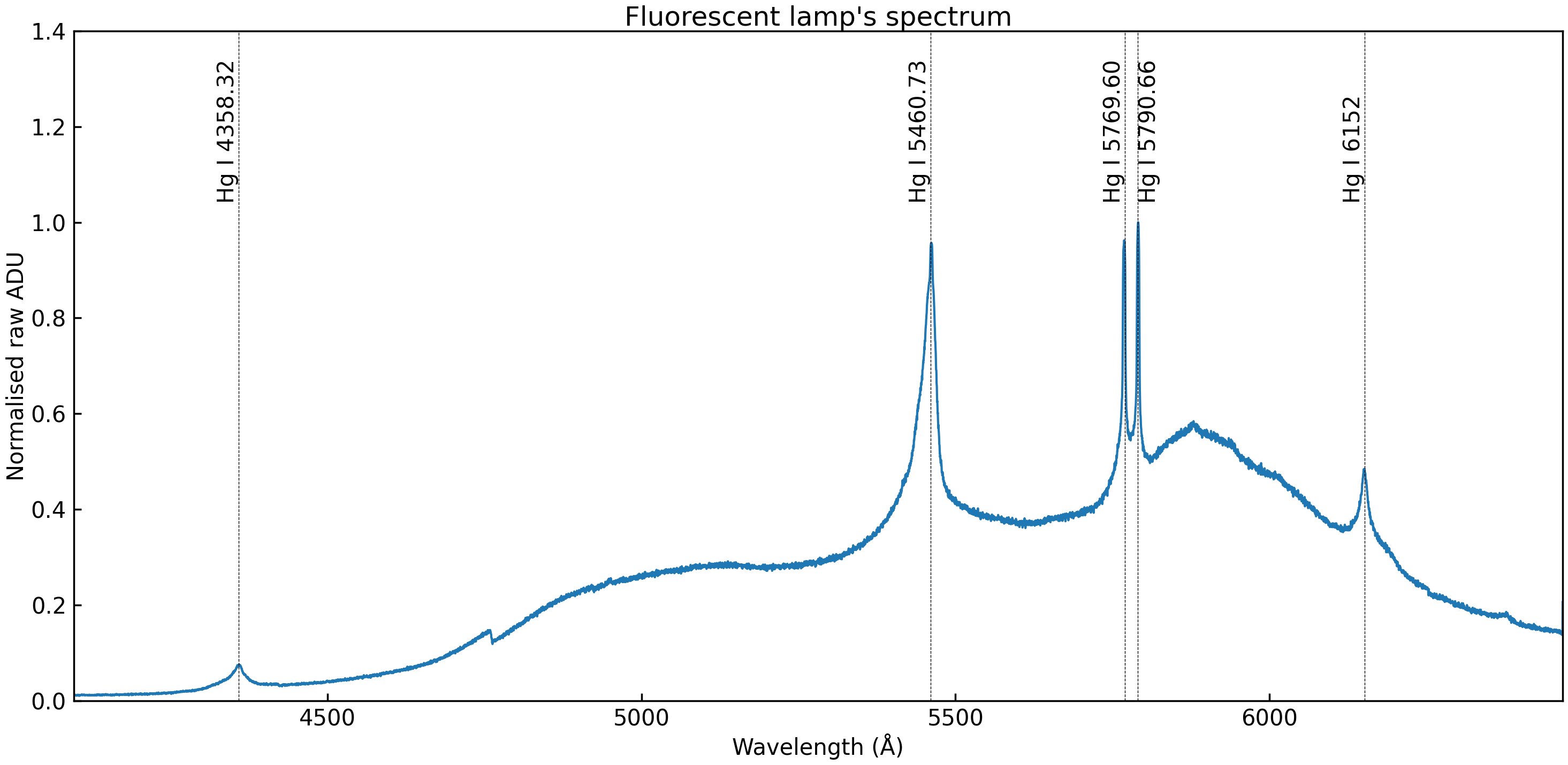}
	\caption{Spectral calibrated fluorescent lamp's spectrum by SPECTRUMMATE on $120\ mm$ telescope, $1200~\textnormal{grooves}/mm$ grating, and $25~\mu m$ slit. The small, sharp step visible at approximately 4750 \angstrom is an artifact resulting from a slight mismatch in the photometric background level between two adjacent frames during the manual stitching process.}
    \label{fig:FluorescentLampSpectrumMate}
\end{figure}

To assist in this process, we used RSpec\footnote{\url{https://rspec-astro.com/}}, a software suite that is capable of extracting a 1-dimensional spectrum from a 2d image, producing instrument spectral flux versus pixel number. The software also provides the option for a wavelength calibration based on known absorption/emission line characteristics. This allows us to determine the (air) wavelength for a given image pixel coordinate. In Fig. \ref{fig:SunSpectrumMate} and Fig. \ref{fig:FluorescentLampSpectrumMate}, we show the relative spectral energy density vs calibrated wavelength for our observation of the blue sky and the fluorescent Mercury lamp.

In Appendix \autoref{append:Spectrum graphs of the Sun by SPECTRUMMATE}, the Sun's visible spectrum graphs are cropped into smaller parts, providing a clearer view of the spectrum. The result showed SPECTRUMMATE can produce the Sun's spectrum image since the absorption lines can be easily calibrated. Based on the calibrated spectrum, the spectral dispersion and spectral range coverage of SPECTRUMMATE were measured using the RSpec calibration wizard feature. The tutorial can be found on the RSpec webpage.

\begin{wstable}[H]
    \caption{Specifications of SPECTRUMMATE.}
    \label{comparison}
    \begin{tabular}{lcc}
        \toprule
        & Theory & Real \\
        \midrule
        Spectral dispersion $\rho\ (\angstrom/\text{pixel})$ & $0.11$ & $0.12$ \\
        Spectral range coverage ($\angstrom$) & $341$ & $368$ \\
        Spectral resolution $\Delta \lambda$ & $3.18$ & $2.16$ \\
        Resolving power R at 5500 \angstrom & $1730$ & $2546$ \\
        \bottomrule
    \end{tabular}
\end{wstable}

The parameters are calculated using data from a single image, which contains the Hg I line. This eliminates the potential for errors introduced during the stitching process. The small discrepancies between the "Theory" and "Real" values are expected and can be attributed to several real-world factors. These include minor manufacturing tolerances in the focal lengths of the off-the-shelf lenses, small inaccuracies in the 3D-printed mechanical components, and any slight misalignment introduced during assembly. Furthermore, the theoretical resolution assumes perfect optics, whereas the real-world measurement is affected by minor line-broadening from the singlet lenses' chromatic and other optical aberrations. The agreement between the theoretical and measured values validates the overall design and construction, with the difference deemed acceptable for an educational instrument.

\section{Conclusion}
This project aimed to design, assemble, and test a portable, lightweight spectrograph for the spectral observation of a bright light source. The project's goal has been achieved since SPECTRUMMATE produced data of sufficient quality suitable for educational, outreach and science communication purposes. The project shows that information on strong light sources could be extracted using a simple-to-build spectrometer and a small telescope. Besides astronomy, the spectrometer is well-suited for educational laboratory settings to demonstrate the principles of chemical composition analysis, such as identifying the primary elements in gas discharge lamps or other bright-line sources. SPECTRUMMATE can be used for educational purposes as an introduction to the wonderful world of light, or as a student's research project aiming at assembling a science-grade instrument as part of an astronomy and astrophysics curriculum. Part of such a project could include low-cost instrumentation, such as a CD or DVD, to demonstrate the principle of a reflection grating. A piece of DVD and differences could replace the high-cost, sensitive grating that could be studied. A low-cost monochrome camera can then be used to capture spectral profiles.

Alongside all the things mentioned, SPECTRUMMATE has a lot of room for improvement. Firstly, the spectrometer itself is made out of PLA (polylactic acid), a widely used plastic in 3D printing, which made it accessible but also caused some problems, such as errors in the size of some parts, the rotation of the grating not smooth, and the threaded 3D printing part was hard to connect with metallic accessories. This can be solved by using the CNC technique to make the parts more durable and accurate in terms of size. Secondly, SPECTRUMMATE was not capable of taking celestial objects (except the Sun) due to the lack of a guiding-responsible part. Without it, we will not know the light from which object is coming into the slit, nor be capable of tracking them. Thirdly, we used singlet lenses for the collimator and objective lens, hence causing chromatic aberration. But SpaceLAB did not have any doublet at the time, so the consequence was to be accepted. SPECTRUMMATE also has a major flaw in achieving a wide-range spectrum because of the lack of a mechanism that allows the grating to rotate in a precise and measurable way, and an automatic image stitch method. Rotating the grating requires an optoelectronic design based on a mechanical mechanism where a motor rotates the grating through an angle at a fixed step. This is a work in progress, and we will leave this part for future reporting.

\section*{Acknowledgments}
We were partly supported by a grant from the Simons Foundation (916424, N.H.) in addition to the enthusiastic support of the IFIRSE/ICISE\footnote{\url{https://www.icisequynhon.com/}} staff.

We would also like to thank the government of Gia Lai Province for providing support to construct the Quy Nhon Observatory\footnote{\url{https://astro.explorascience.vn/}} (QNO), as well as Thierry Montmerle and Jungjoo Sohn for their support during the observatory's construction. We would like to thank the Metaspace company\footnote{\url{http://metaspace.co.kr/wp/?lang=en}} (Cheongju, Korea) for supporting equipment installation at QNO.

We thank Thorlab Inc.\footnote{\url{https://www.thorlabs.com/}}, Mr Olivier Garde\footnote{\url{http://o.garde.free.fr/astro/Spectro1/Bienvenue.html}} for permitting us to reprint their figures.

\appendix{SPECTRUMMATE optical components and accessories}
\label{append:SPECTRUMMATE optical components}

\begin{figure}[H]
    \centering
    \begin{subfigure}[b]{0.45\linewidth}
    	\centering
    	\includegraphics[width=\linewidth]{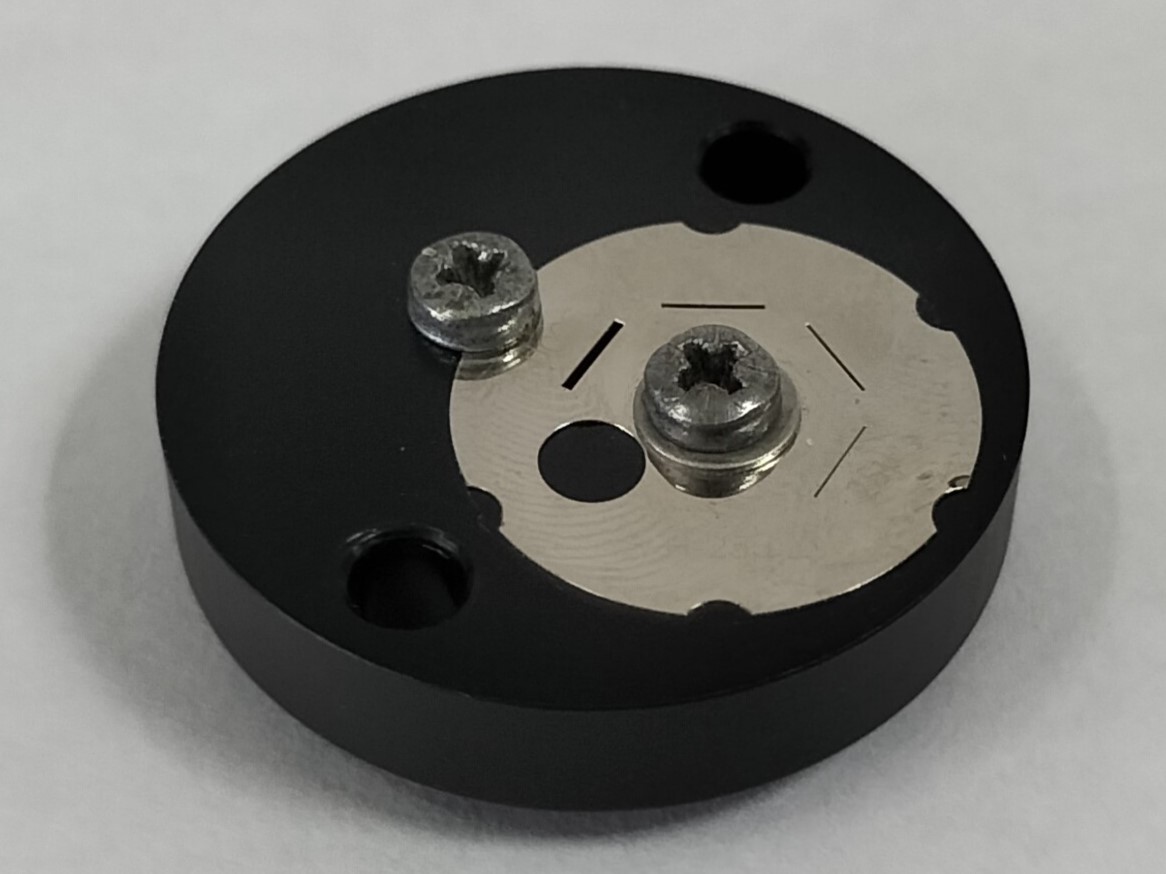}
    	\caption{Variable reflective slit.}
    \end{subfigure}
    \begin{subfigure}[b]{0.45\linewidth}
    	\centering
    	\includegraphics[width=\linewidth]{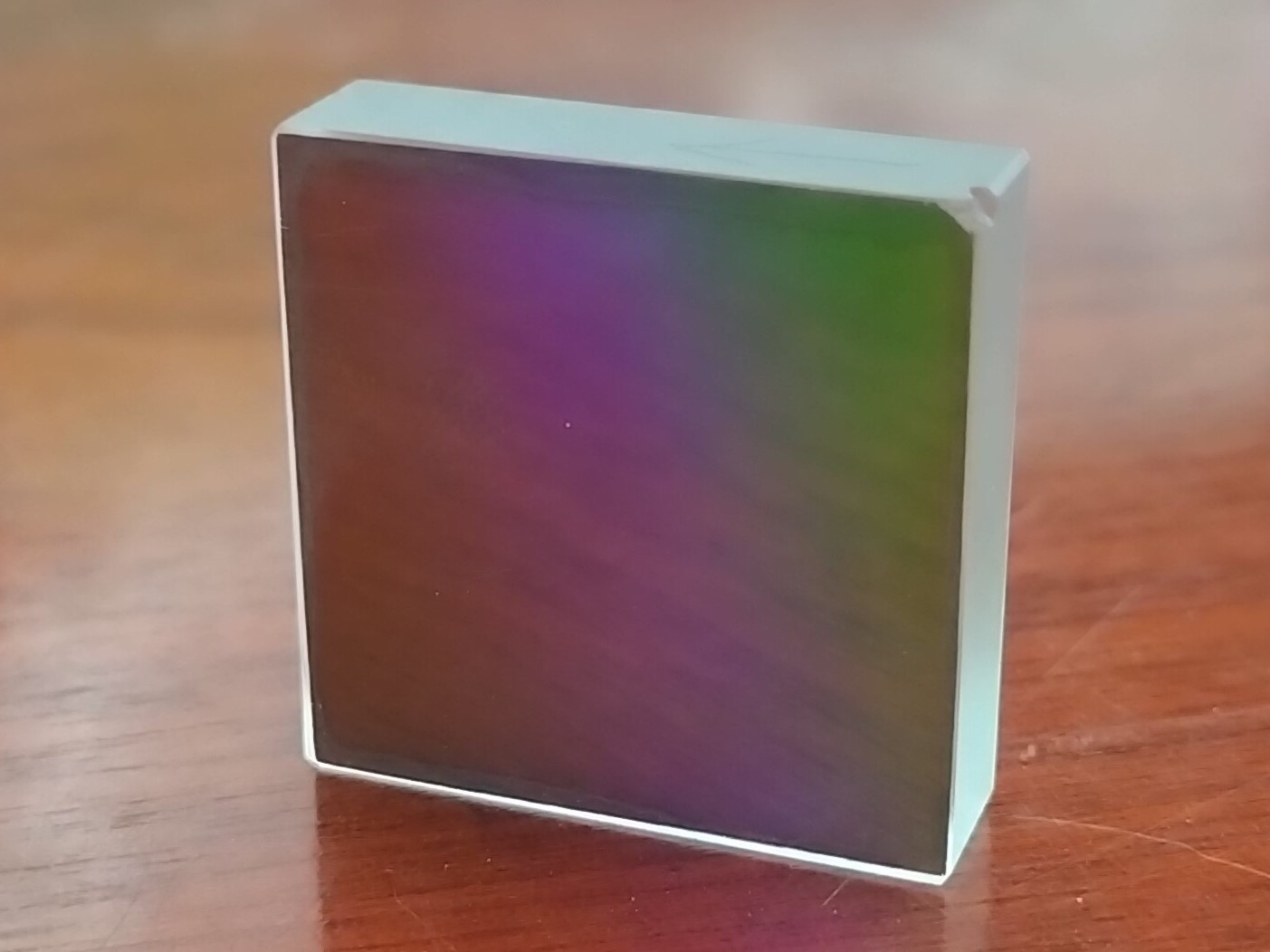}
    	\caption{Holographic reflective grating. Part number: GH25-12V.}
    \end{subfigure}
    
    \begin{subfigure}[b]{0.3\linewidth}
    	\centering
    	\includegraphics[width=\linewidth]{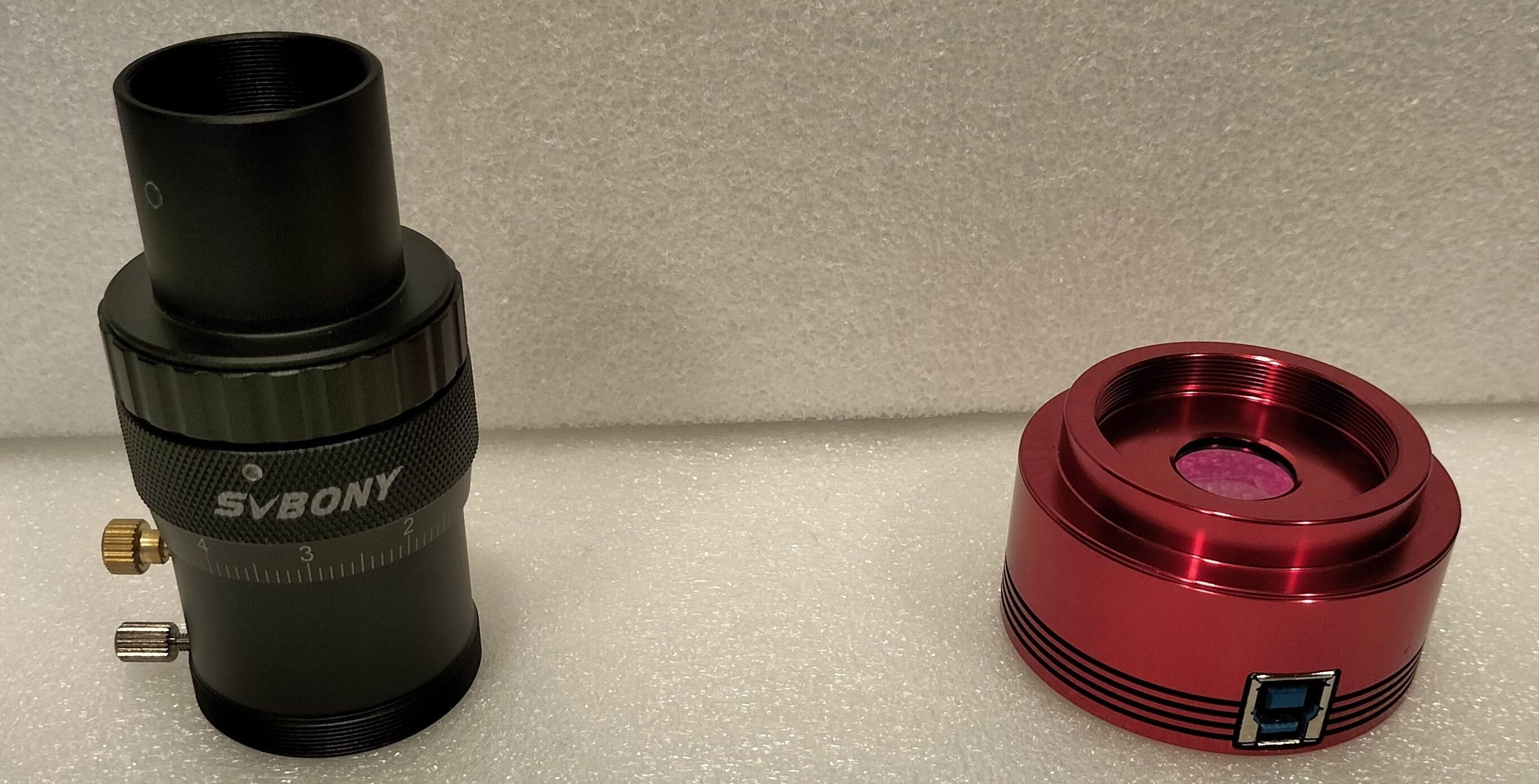}
    	\caption{Helical focuser (left) and ASI178MC/MM camera.}
    \end{subfigure}
    \begin{subfigure}[b]{0.3\linewidth}
    	\centering
    	\includegraphics[width=\linewidth]{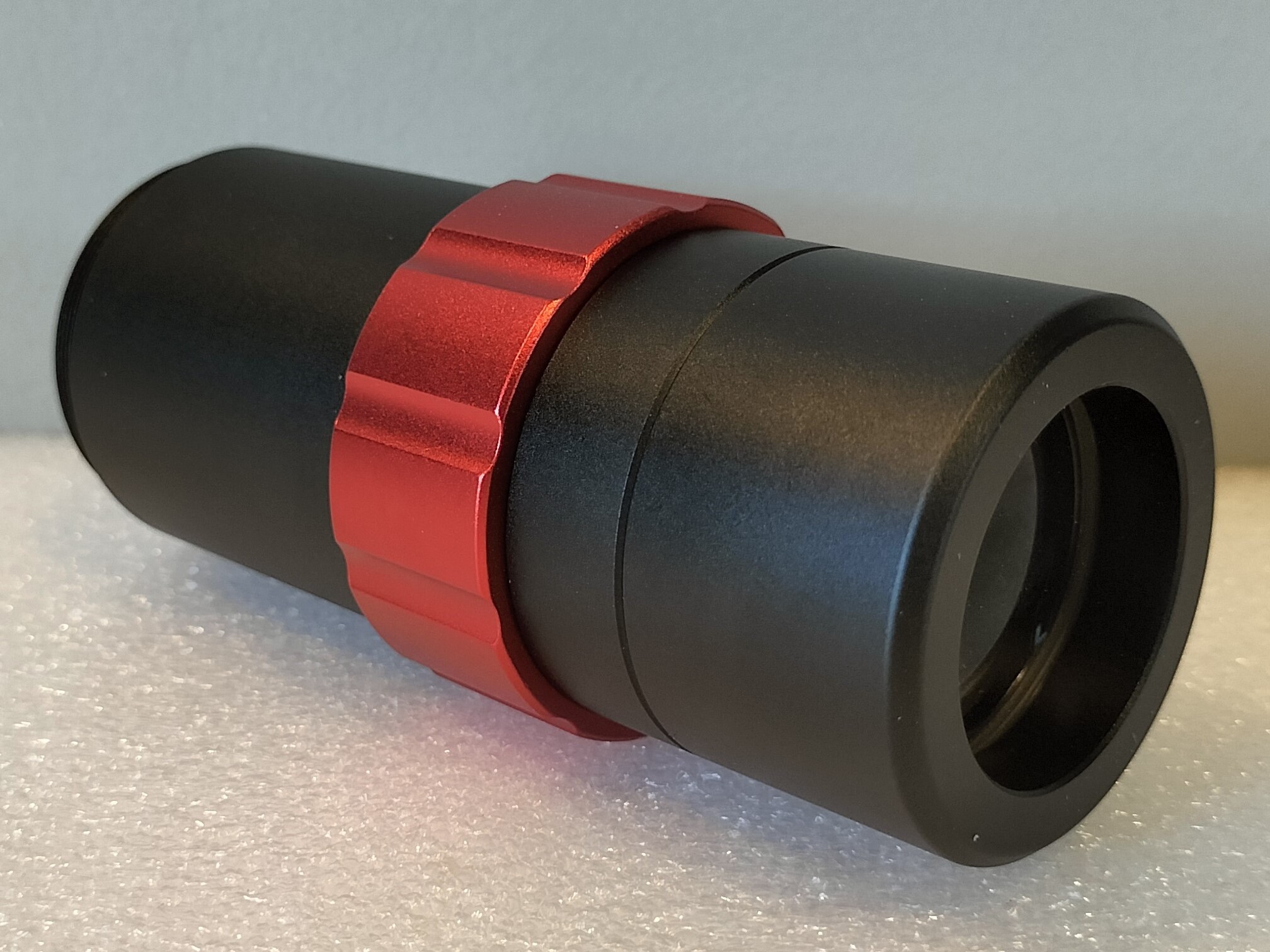}
    	\caption{Guide scope.}
    \end{subfigure}
    \begin{subfigure}[b]{0.3\linewidth}
    	\includegraphics[width=\linewidth]{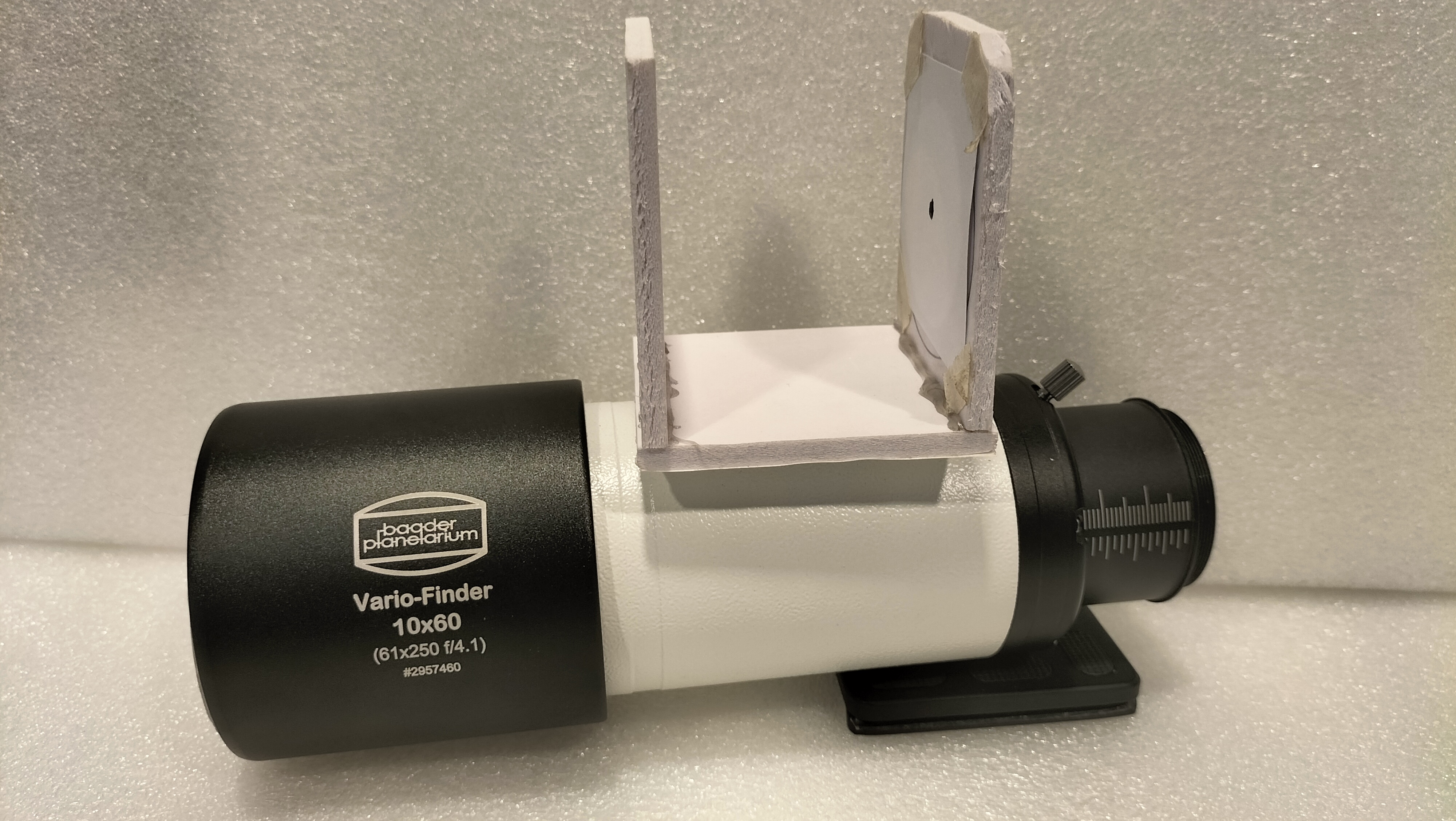}
    	\caption{The Baader Multi-Purpose Vario Finder 10x60. The thing attached on top is a simple finder.}
    \end{subfigure}

    \begin{subfigure}[b]{0.45\linewidth}
    	\centering
    	\includegraphics[width=\linewidth]{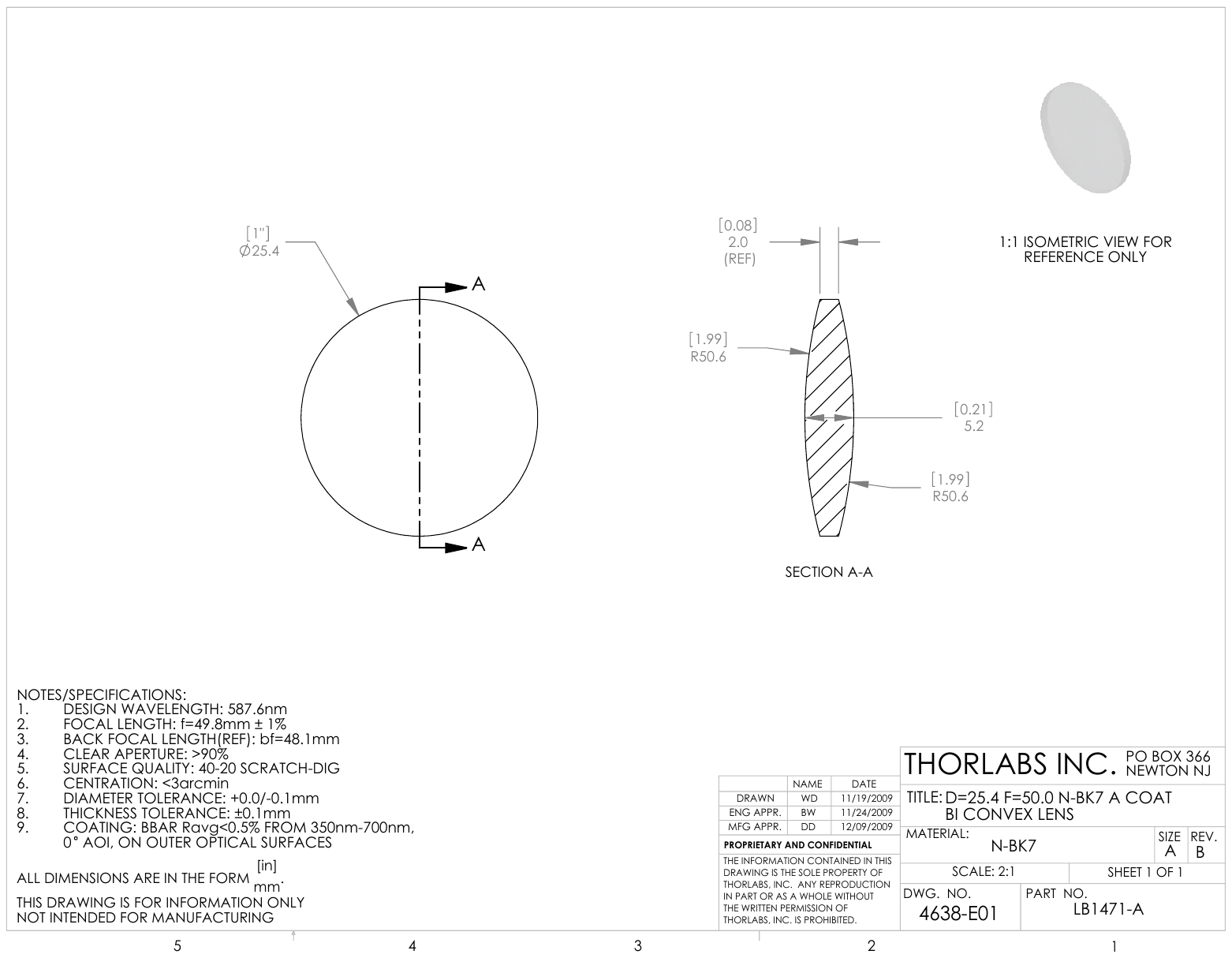}
    	\caption{Collimator lens. Part number: LB1471-A.}
    \end{subfigure}
    \begin{subfigure}[b]{0.45\linewidth}
    	\centering
    	\includegraphics[width=\linewidth]{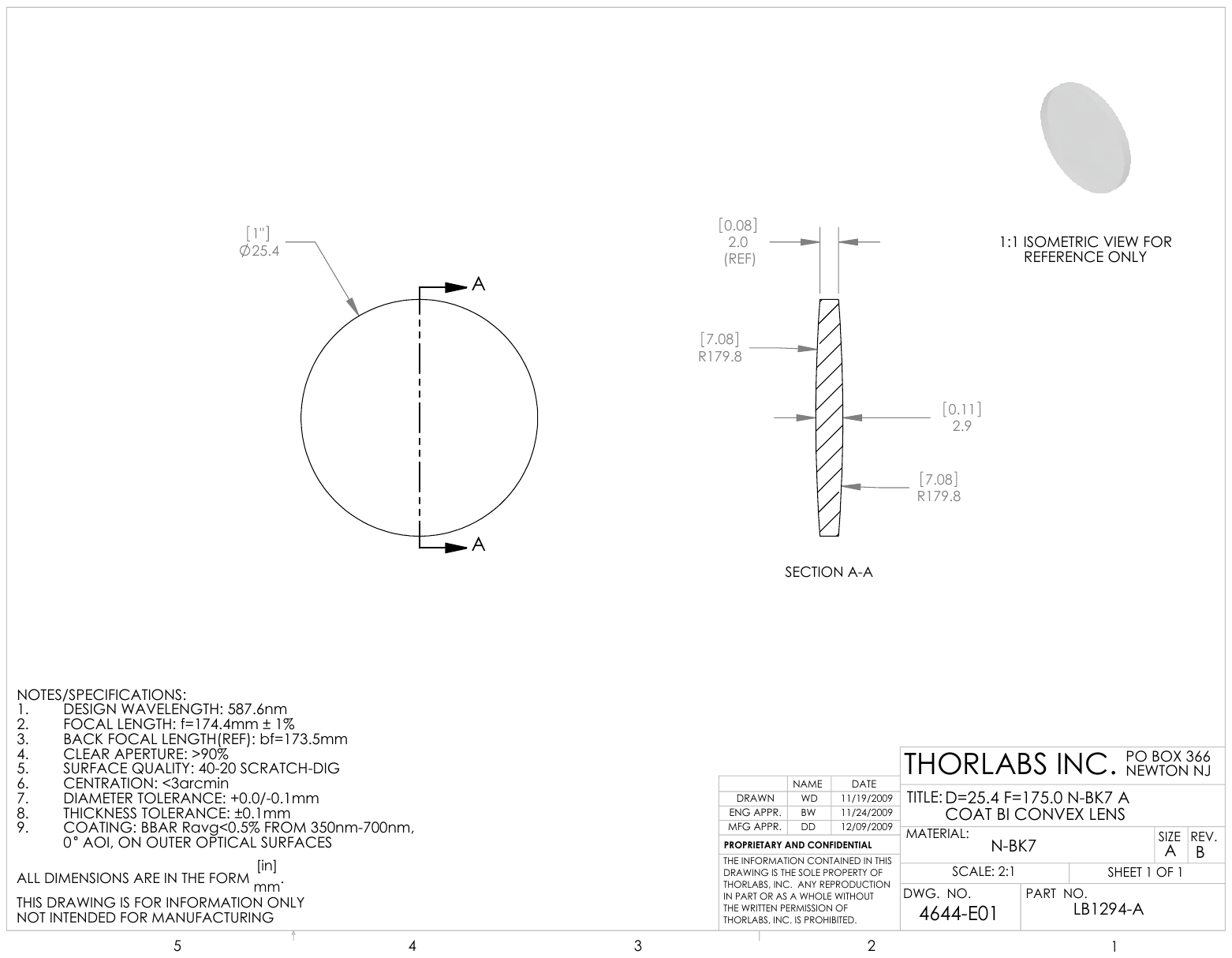}
    	\caption{Objective lens. Part number: LB1294-A.}
    \end{subfigure}

    \caption{SPECTRUMMATE optical components and accessories}
\end{figure}

\appendix{Detailed optical design calculations}
\label{append:Detailed optical design calculations}

\noindent\emph{Vignetting check at the collimator}

The aperture ratio number $F_\#$ of the telescope will be \begin{equation}\label{eqn:Vignetting check at the collimator}F_\#=\dfrac{f}{D}=\dfrac{120}{30}=4\end{equation}

The collimator lens will be large enough to pass all the light from the telescope if $F_C<F_\#$. Since $F_C = f_1 / D_C = 50 / 25.4 \approx 1.969 < 4$, this condition is satisfied.

\noindent\emph{Beam diameter}

We denote the diameter of the beam exiting the collimator as $d_1$. In this case, we find \begin{equation}
d_1=D\times\left(\frac{f_1}{f}\right)=30\times\left(\frac{50}{120}\right)=12.5\, mm
\end{equation}
assuring the correct dimension of the collimator lens diameter.

\noindent\emph{Angles of incidence and diffraction}

The spacing between the grooves in the grating determines the angle at which the different wavelengths of light are diffracted. The grating equation, given by 

\begin{equation}
k\lambda=d(\sin{\alpha}+\sin{\beta})
\end{equation}
relates the wavelength of the light $\lambda$, the distance between the grooves $d$, the angle of incidence $\alpha$, the angle of diffraction $\beta$, and the order of the diffraction $k$. It is sometimes convenient to write the equation as 

\begin{equation}
mk\lambda=\sin{\alpha}+\sin{\beta},
\end{equation}
where $m=1/d$ is the groove density (groove per millimetre).

\begin{figure}[H]
	\centering
	\includegraphics[width=0.5\linewidth]{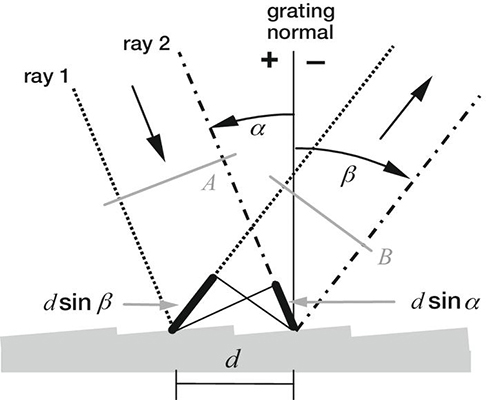}
	\caption{Grating geometry by Palmer (2020).}
    \label{gratinggeometry}
\end{figure}

The equations presented are applicable only when the incident and diffracted rays are situated in a plane that is perpendicular to the grooves at the center of the grating, as shown in Fig. \ref{gratinggeometry}. This condition, called classical or in-plane diffraction, encompasses most grating systems (Palmer 2020).

For the design of SPECTRUMMATE light of wavelength $\lambda=5500\, \angstrom$ is aimed to be located at the centre of the imaging sensor. We have chosen the centre of the visible spectrum (green) as the reference wavelength.

Let $\gamma$ be the total angle between the incident rays on the grating and the direction of diffracted for the reference ray 

\begin{equation}
\gamma = \alpha-\beta.
\end{equation}
However, due to the physical size of the collimator and other components, and we also wanted \textsc{SpectrumMate} to be compact, a value of $40\degree$ is chosen. Replace $\beta$ with $\gamma$, the grating equation becomes

\begin{equation}
\sin\alpha+\sin(\alpha-\gamma)=km\lambda.
\end{equation}
For maximum S/N performance, working with the first order $k=\pm1$ (high light intensity) spectrum is generally recommended. Solving for $\alpha$, the above equation can be rewritten as 

\begin{equation}\sin\left(\alpha-\dfrac{\gamma}{2}\right)=\dfrac{\pm m\lambda}{2\cos(\gamma/2)}=\dfrac{\pm 1200\times0.55\times10^{-3}}{2\cos(40\degree/2)}=\pm0.35.\end{equation}
From this we find

\begin{equation}
\alpha-\frac{40\degree}{2}\approx\pm20.49\degree.
\end{equation}

Thus, the two possible values for $\alpha$ are $40.49\degree$ and $-0.49\degree$ for grating orders $1$ and $-1$ respectively. The corresponding values for $\beta$, using the relation $\gamma = \alpha-\beta$, are $\beta=0.49\degree$ and $\beta=-40.49\degree$. We'll choose $\alpha=40.49\degree$ and $\beta=0.49\degree$. Note that angles of opposite signs mean the incident ray and the diffracted ray lie on opposite sides of the normal to the grating.

\noindent\emph{Minimum grating dimension}

For the required minimum grating dimension $L$ we have 

\begin{equation}
L=\frac{d_1}{\cos\alpha}=\frac{f_1}{F_\#}=\frac{12.5}{\cos40.49\degree}=16.4\, mm.
\end{equation}

For longer wavelengths, for example, $7000\ \angstrom$, the dimension required goes up to $18.2\ mm$. Since $L<H$, the chosen holographic diffraction grating is large enough for no light loss to occur.

\noindent\emph{Spectral dispersion on sensor}

Let $\rho$ be the degree of dispersion in $\angstrom/pixel$, on the camera sensor.  For a sensor of pixel size $p_{size}=2.4\ \mu m$, the dispersion is given by \begin{equation}\rho=10^7\times\frac{p_{size}\cos\beta}{mf_2}\end{equation} which gives \begin{equation}\label{spectraldispersion}\rho=10^7\times\frac{0.0024\times\cos(0.49\degree)}{1200\times175}=0.11\ \angstrom/pixel\end{equation}

\noindent\emph{Diameter $d_{2}$ of the diffracted rays} 

The beam diameter $d_{2}$ of diffracted rays reflected from the grating is
\begin{equation}
d_{2}
=\frac{\cos\beta}{\cos\alpha}\,\frac{f_{1}}{F_{\#}}
+\frac{X\,p_{\rm size}\,N}{f_{2}},
\end{equation}
where $X$ is the grating–to–objective distance and $N$ is the number of pixels across the sensor in the dispersion direction. Due to mechanical constraints, $X=75~\mathrm{mm}$. For the chosen sensor (Sony IMX178) $N=3096$ pixels. 

The first term is the \emph{anamorphic enlargement} of the monochromatic pupil, set solely by the grating geometry. The second term budgets the extra angular span needed to place a spectrum of physical length $S$ on the detector: with small–angle imaging $x\simeq f_{2}\beta$, the span is $\Delta\beta\simeq S/f_{2}$ and the added footprint over the distance $X$ is $X\,\Delta\beta=X\,S/f_{2}$. Writing $S=N\,p_{\rm size}$ merely converts pixels to millimeters; if $S$ is specified directly in mm, the explicit pixel-size dependence vanishes.

Numerically,
\[
d_{2}
=\frac{\cos 0.49^{\circ}}{\cos 40.49^{\circ}}\times\frac{50}{4}
+\frac{75\times 2.4\times10^{-3}\times 3096}{175}
=19.62~\mathrm{mm}.
\]
For total absence of vignetting, the condition $F_{0}<f_{2}/d_{2}$ is met: $6.89<8.92$.

\noindent\emph{Spectral range coverage}

The extreme wavelength limits from one side of the sensor to the other are given by the relation

\begin{equation}
\label{spectralrange}
\lambda_{2,1}=\lambda\pm\frac{N\rho}{2}=5500\pm\frac{3096\times0.11}{2}.
\end{equation}

Substituting for numbers we have $\lambda_1=5329\ \angstrom,\ \lambda_2=5670\ \angstrom$. Thus, the spectrometer in this configuration covers a visible region of $341\ \angstrom$ with an image size of $0.11\ \angstrom/pixel$. The grating must be rotated if the other regions are to be examined. Since the spectrometer can only cover a range of wavelengths of $341\ \angstrom$, about 9 images need to be taken so that an image of the visible spectrum of the objects can be formed. However, to ensure a smooth transition of the spectrum between images, subsequent images (in wavelength) preferably need to overlap, resulting in an increase in the number of images to record a complete spectrum in the visible.

\noindent\emph{Spectral resolution and resolving power $R$}

The spectral resolving power $R$ is defined as \begin{equation}R=\frac{\lambda}{\Delta\lambda}=\frac{f_1}{slit \ width}\left(tan(\alpha)+\frac{sin(\beta)}{cos(\alpha)}\right)\end{equation} where $\lambda$ is the considered wavelength and $\Delta\lambda$ is the smallest resolvable spectral element (the measured FWHM in the recorded data). SPECTRUMMATE is of medium resolution with a value of $R = 1730$, which tells us that SPECTRUMMATE allows one to see details of $\Delta\lambda = 5500/1730 = 3.18\ \angstrom$ in the range near $5500\ \angstrom$.

\appendix{Image taken by SPECTRUMMATE}
\label{append:Image taken by SPECTRUMMATE}

\begin{figure}[H]
	\centering
        \caption{Images taken by {\sc SpectrumMate}.}
	\begin{subfigure}[c]{0.3\linewidth}
		\includegraphics[width=\linewidth]{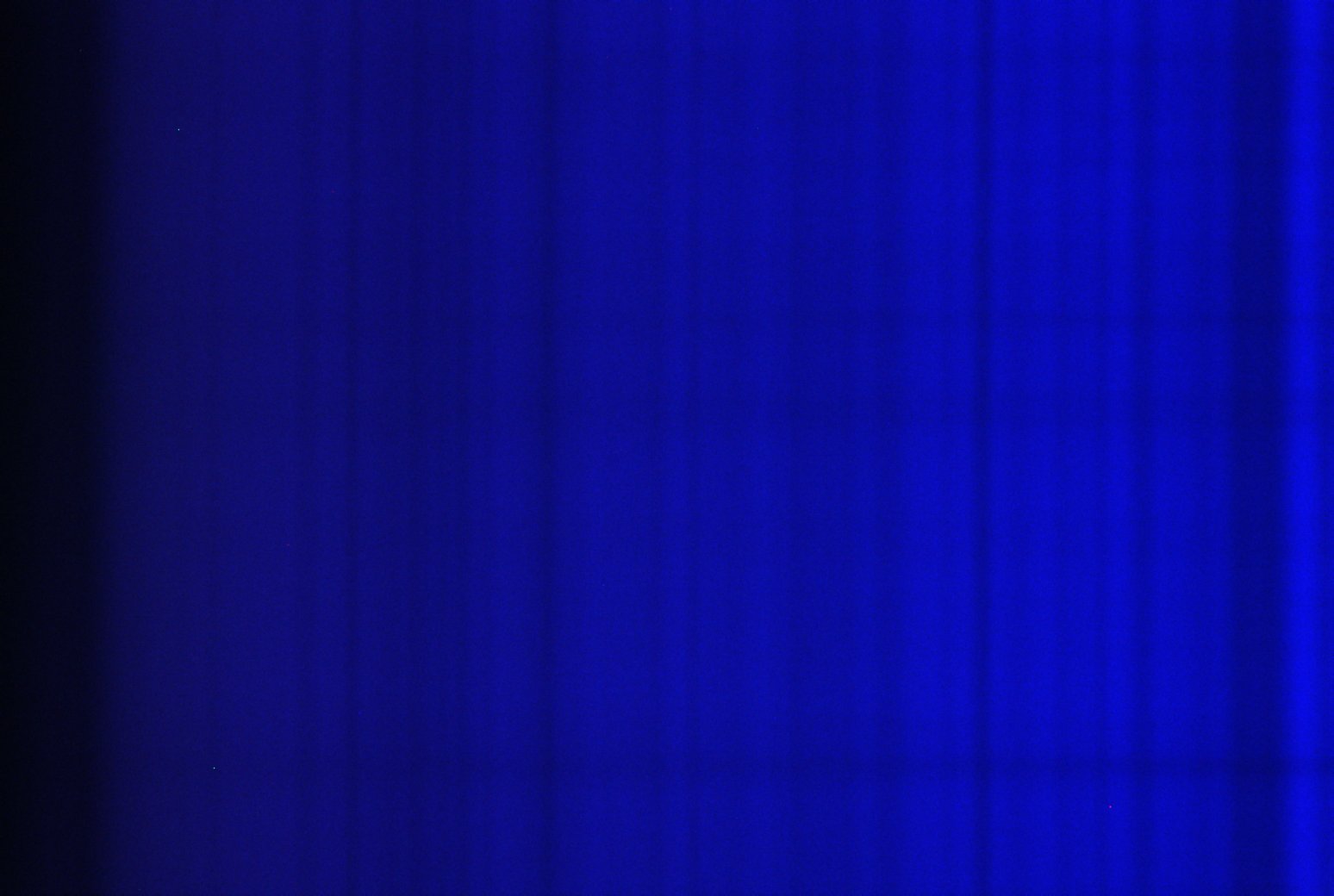}
	\end{subfigure}
	\begin{subfigure}[c]{0.3\linewidth}
		\includegraphics[width=\linewidth]{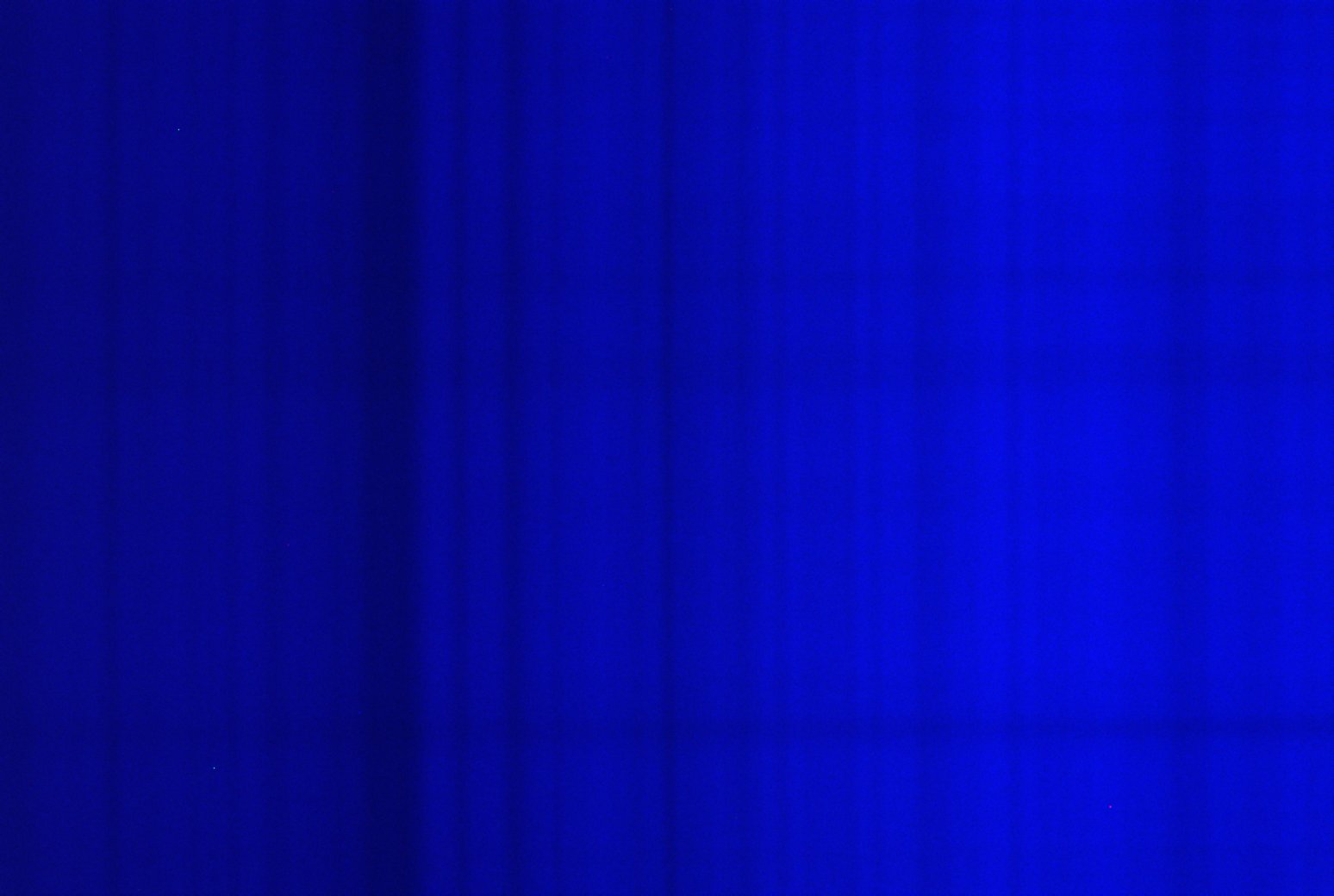}
	\end{subfigure}
	\begin{subfigure}[c]{0.3\linewidth}
		\includegraphics[width=\linewidth]{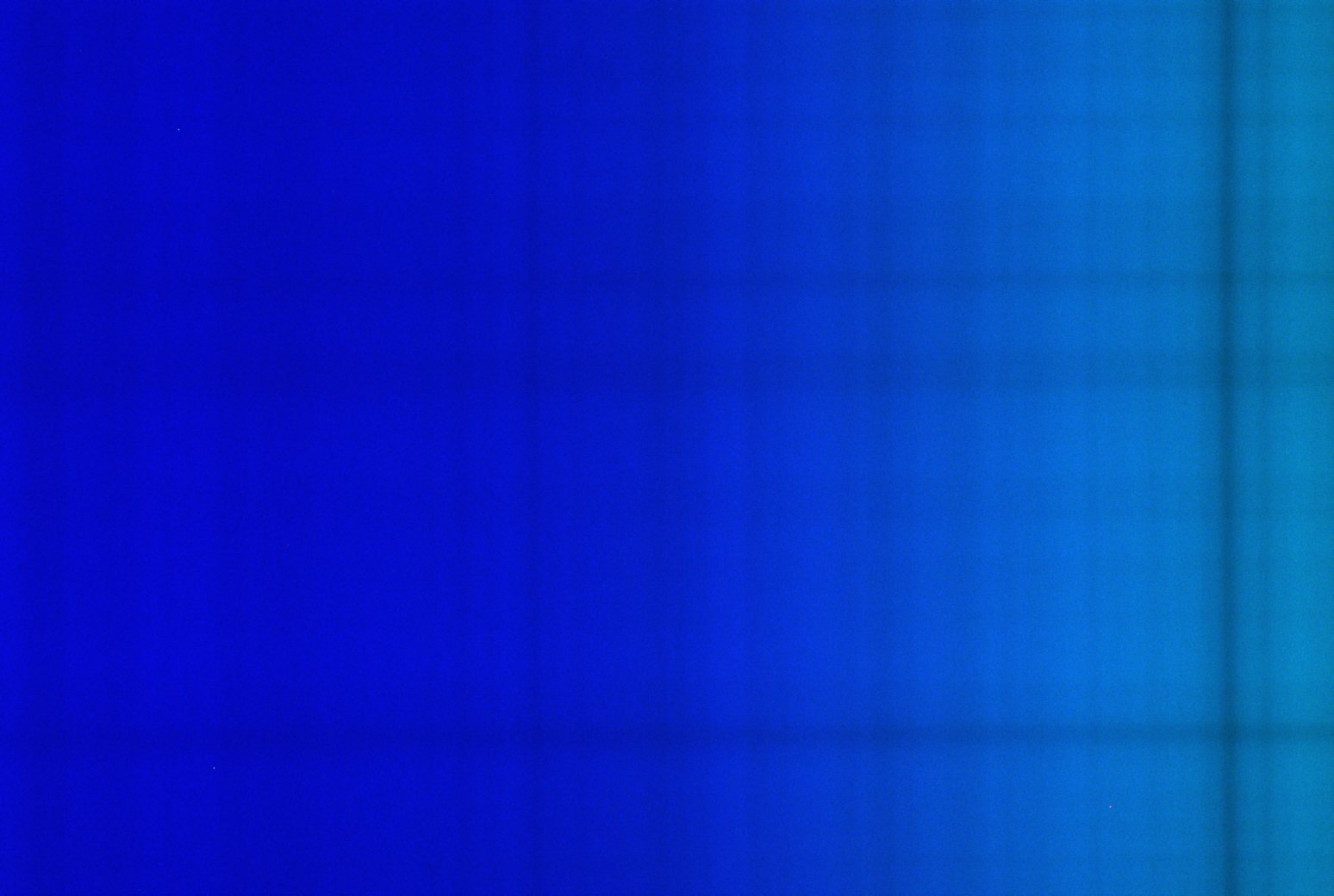}
	\end{subfigure}
	
	\begin{subfigure}[c]{0.3\linewidth}
		\includegraphics[width=\linewidth]{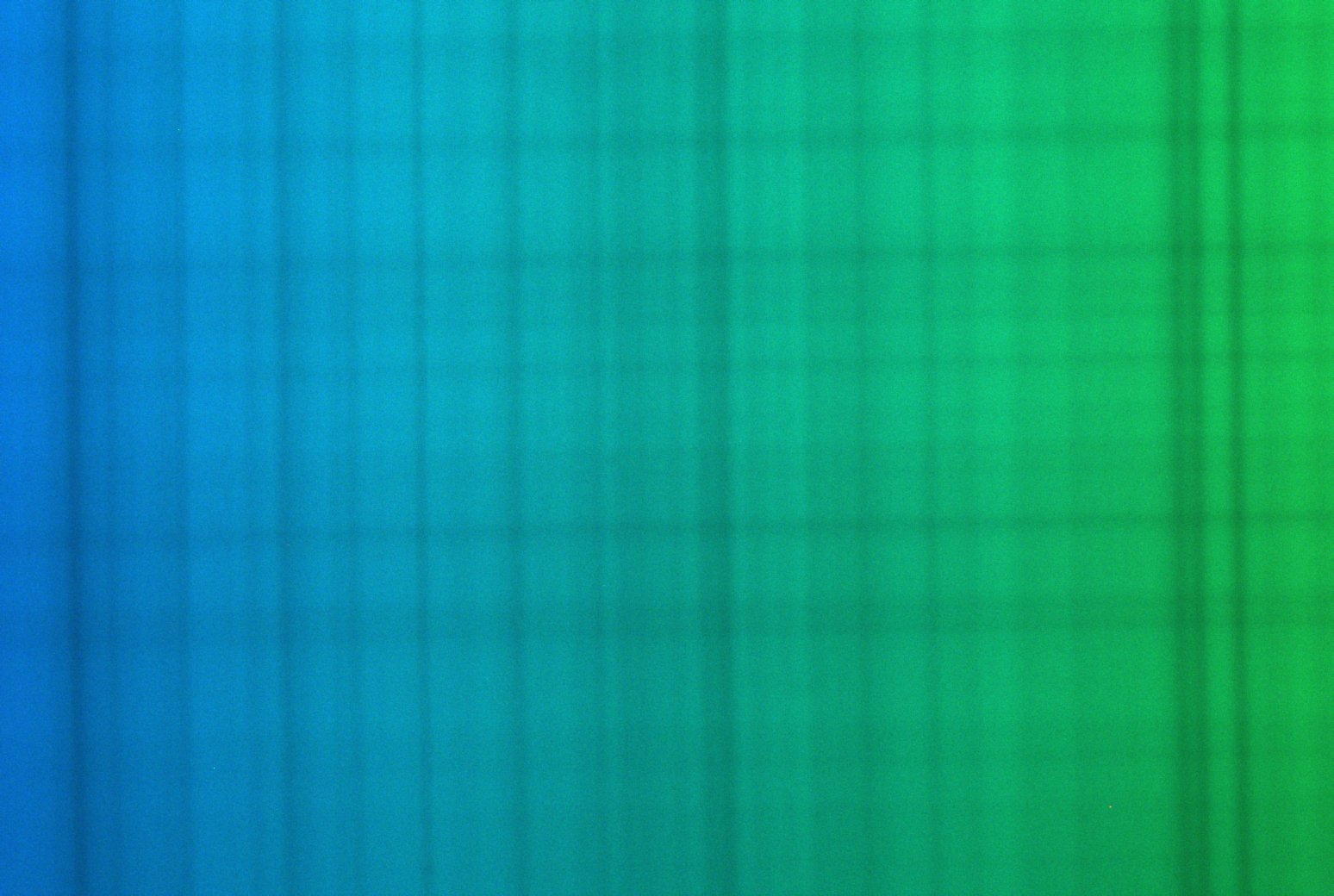}
	\end{subfigure}
	\begin{subfigure}[c]{0.3\linewidth}
		\includegraphics[width=\linewidth]{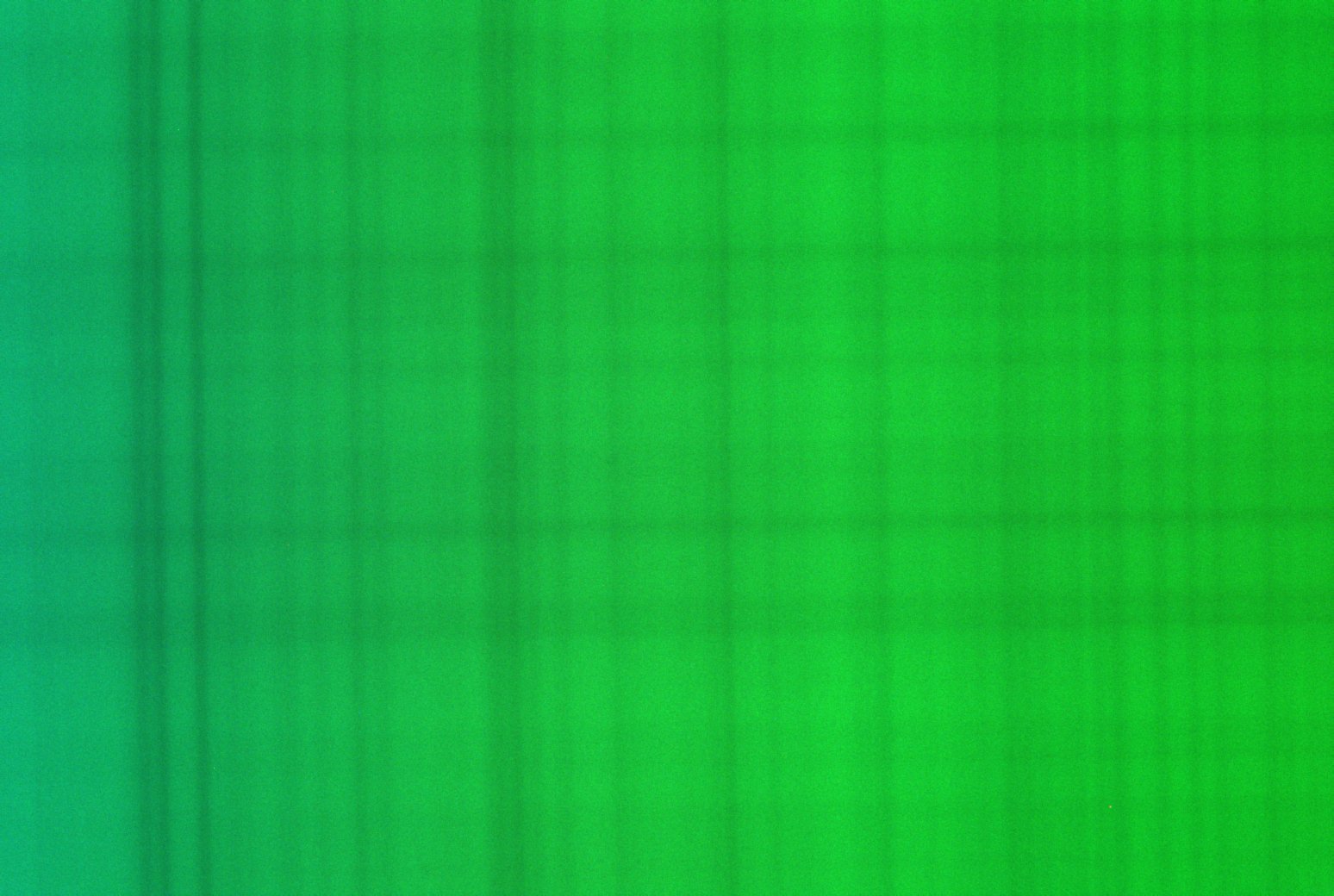}
	\end{subfigure}
	\begin{subfigure}[c]{0.3\linewidth}
		\includegraphics[width=\linewidth]{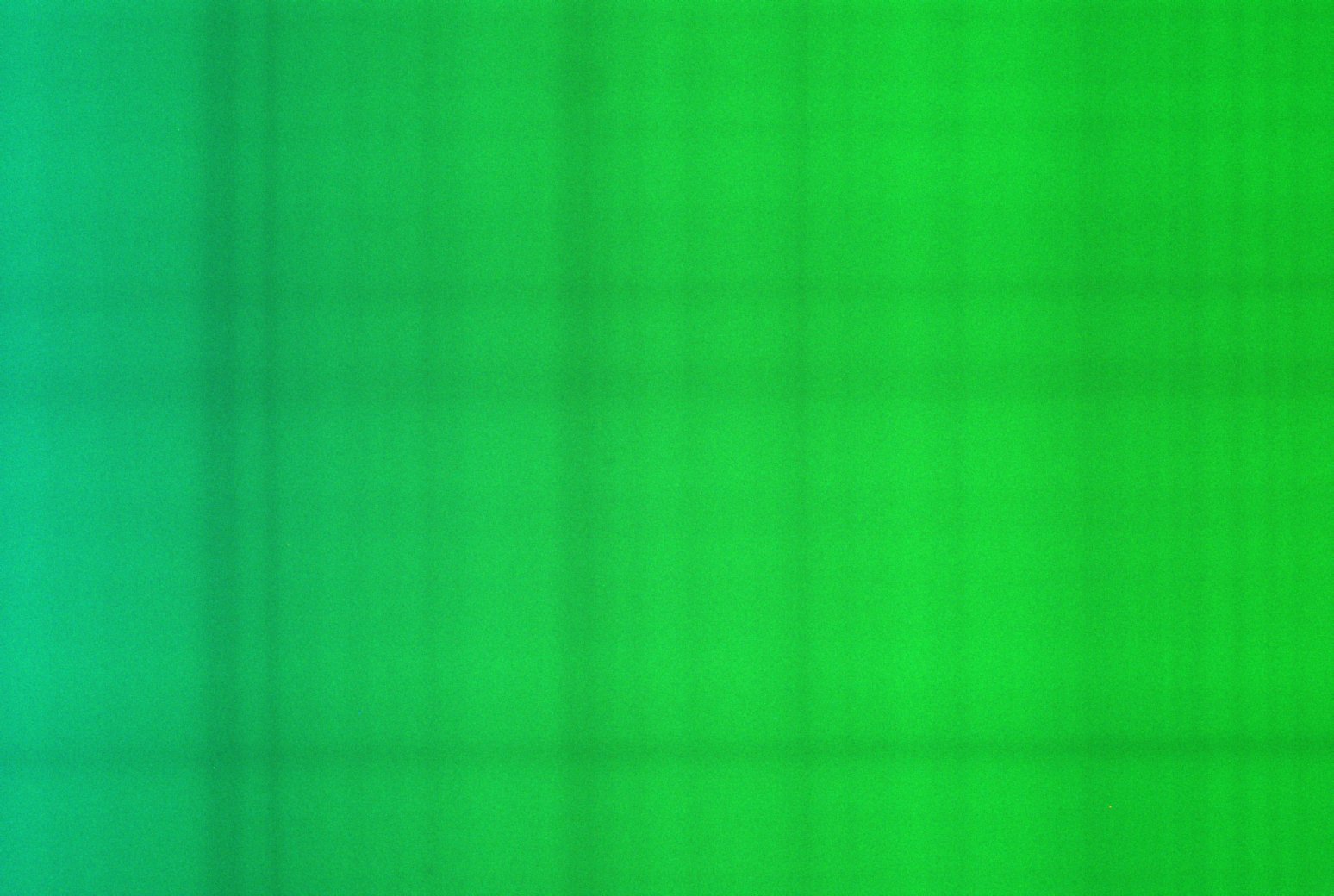}
	\end{subfigure}

	\begin{subfigure}[c]{0.3\linewidth}
		\includegraphics[width=\linewidth]{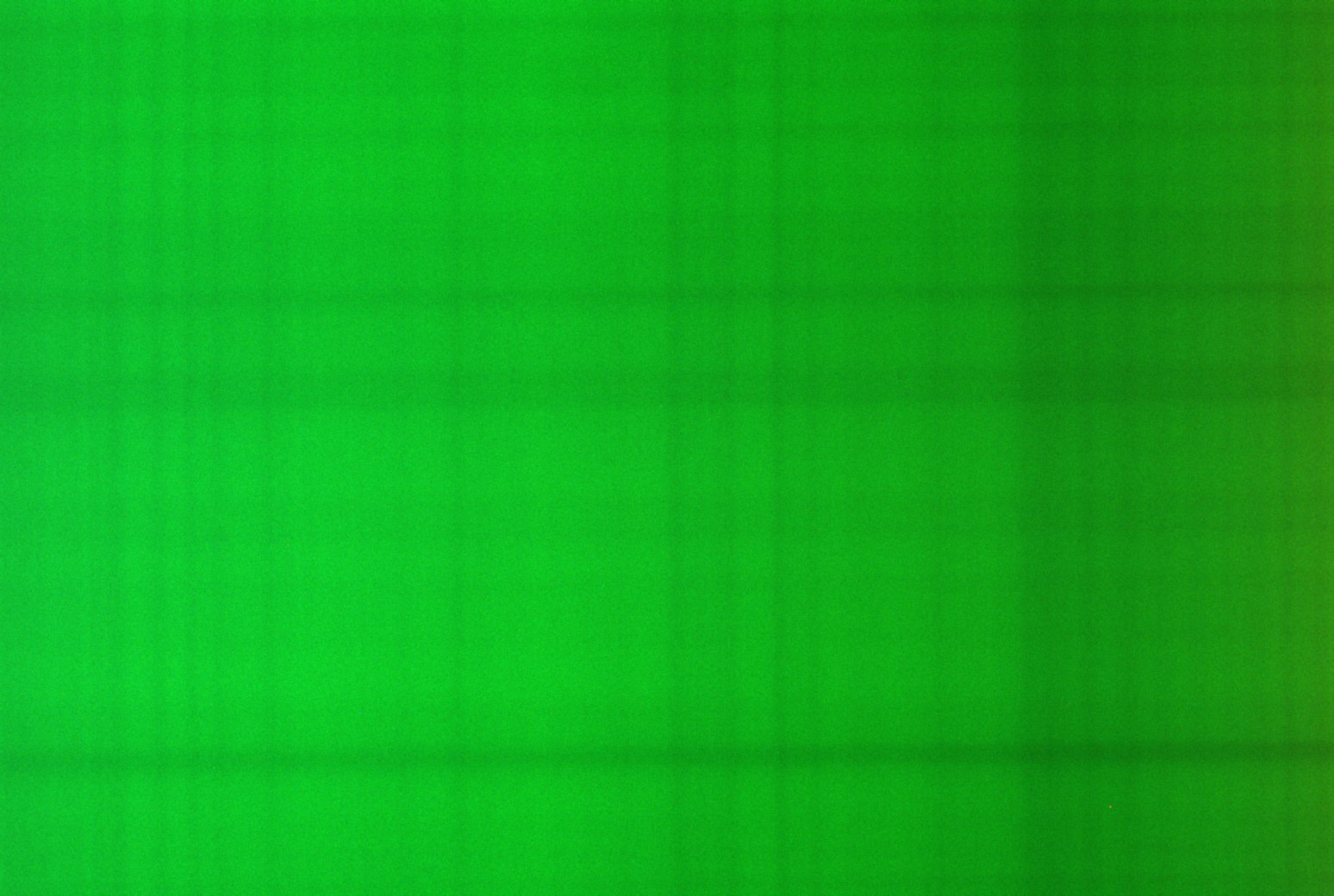}
	\end{subfigure}
	\begin{subfigure}[c]{0.3\linewidth}
		\includegraphics[width=\linewidth]{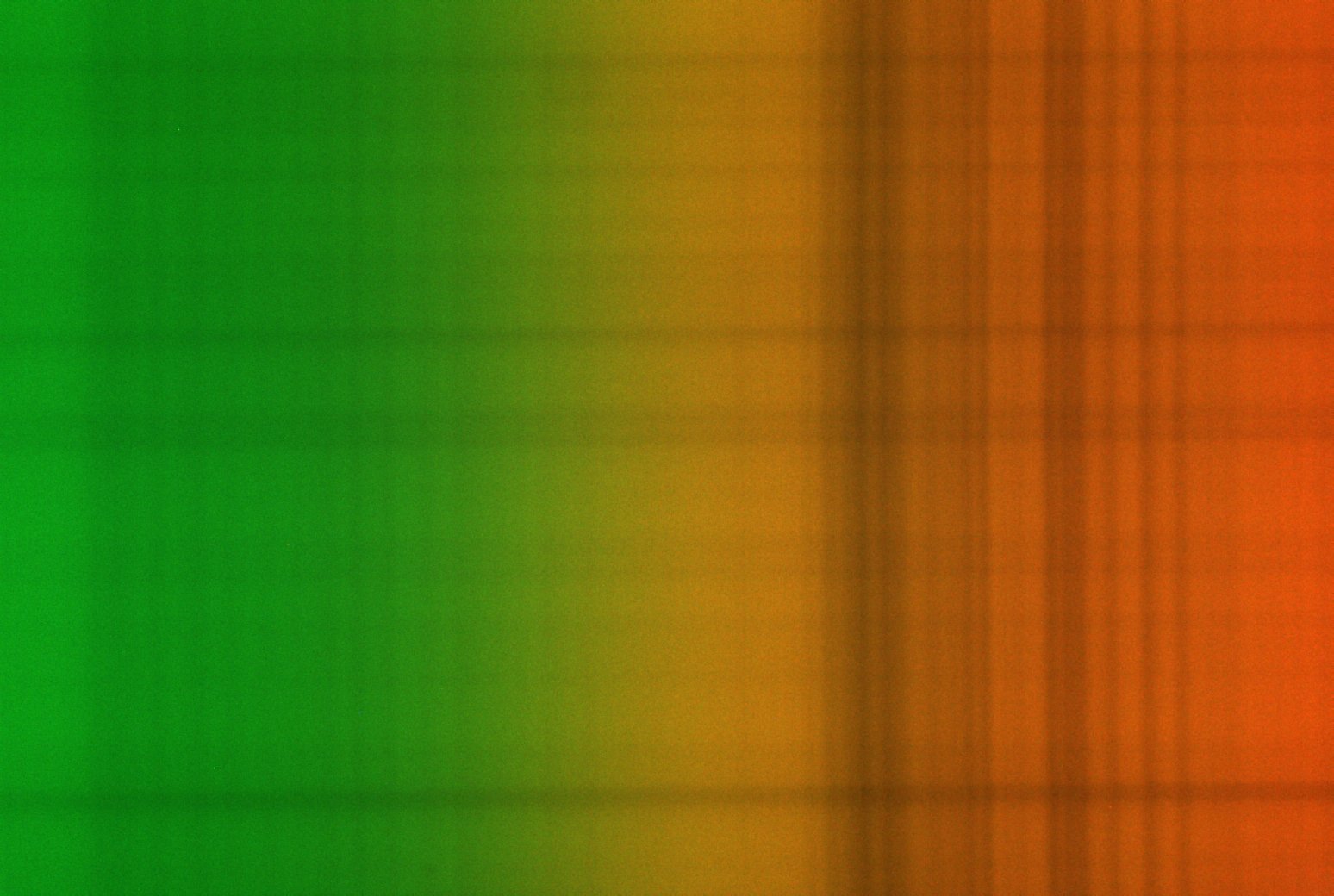}
	\end{subfigure}
	\begin{subfigure}[c]{0.3\linewidth}
		\includegraphics[width=\linewidth]{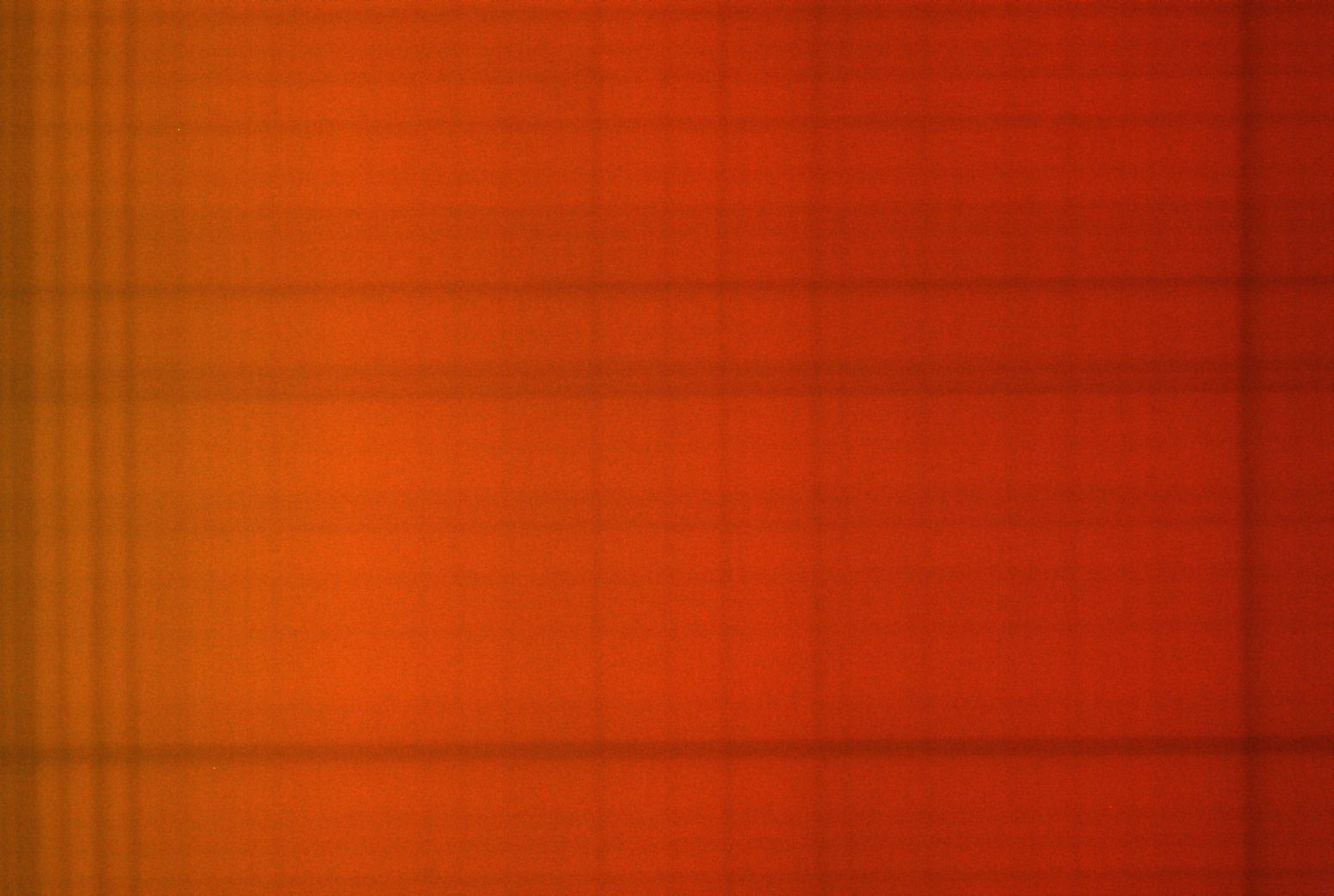}
	\end{subfigure}
	
	\begin{subfigure}[c]{0.3\linewidth}
		\includegraphics[width=\linewidth]{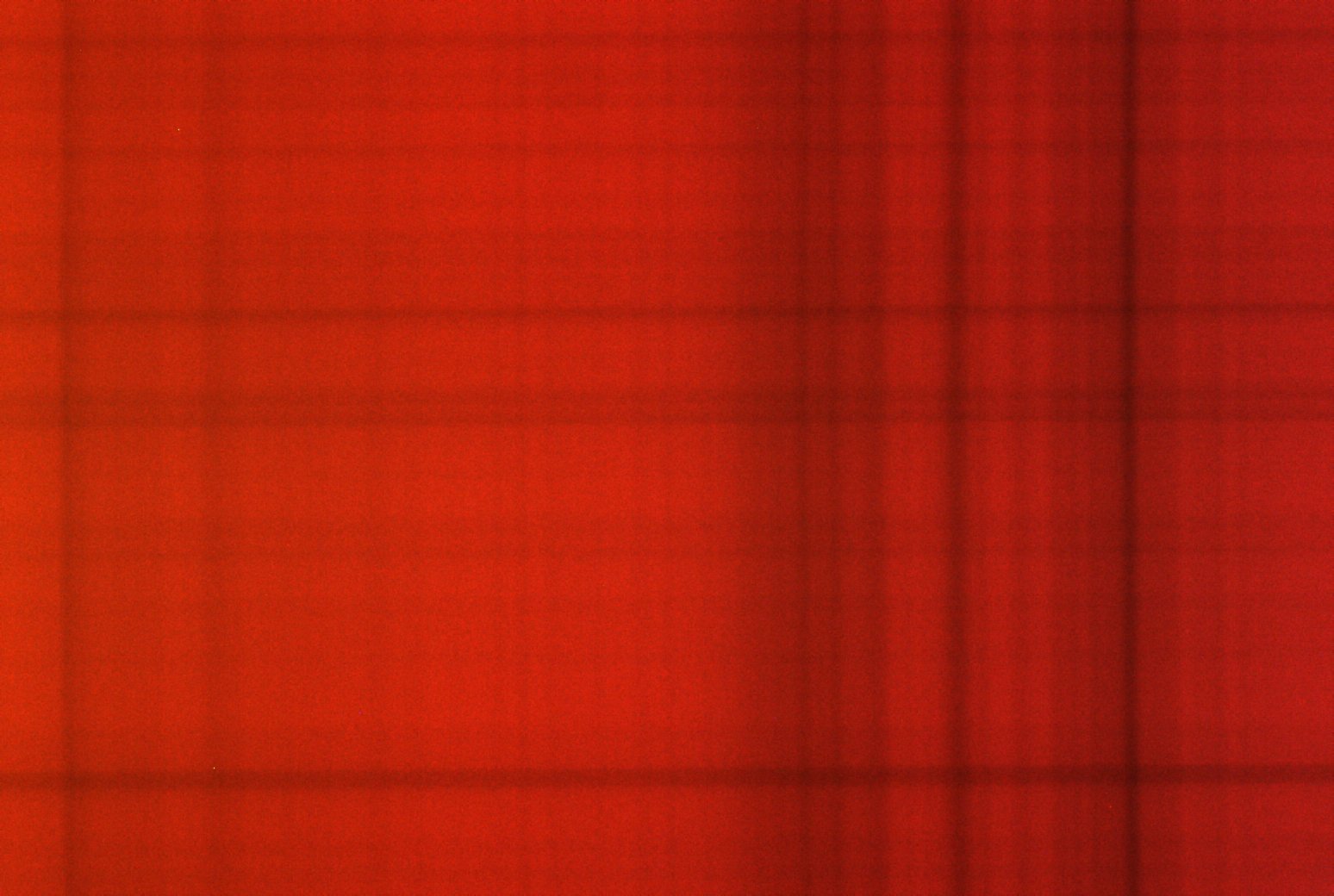}
	\end{subfigure}
	\begin{subfigure}[c]{0.3\linewidth}
		\includegraphics[width=\linewidth]{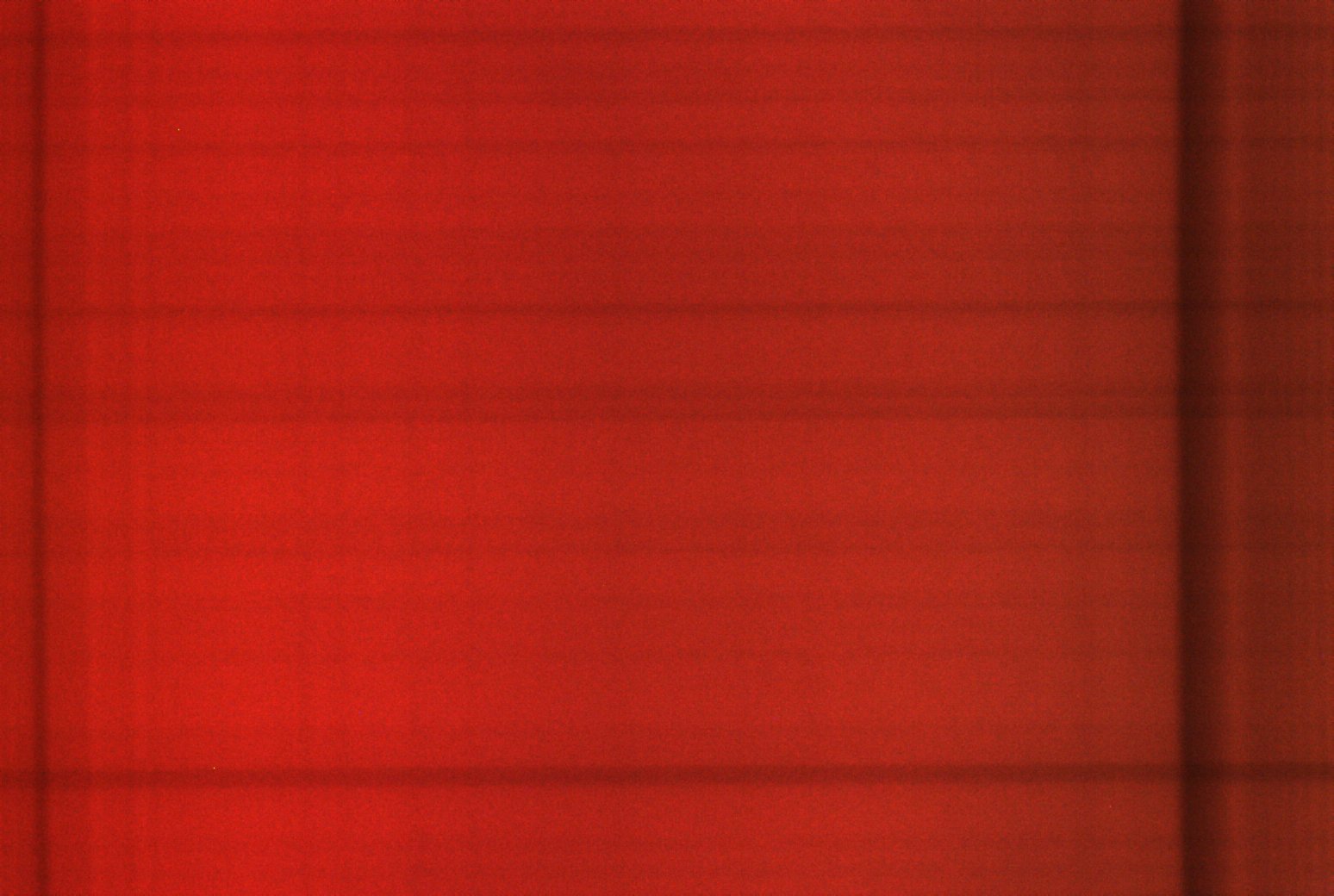}
	\end{subfigure}
	\begin{subfigure}[c]{0.3\linewidth}
		\includegraphics[width=\linewidth]{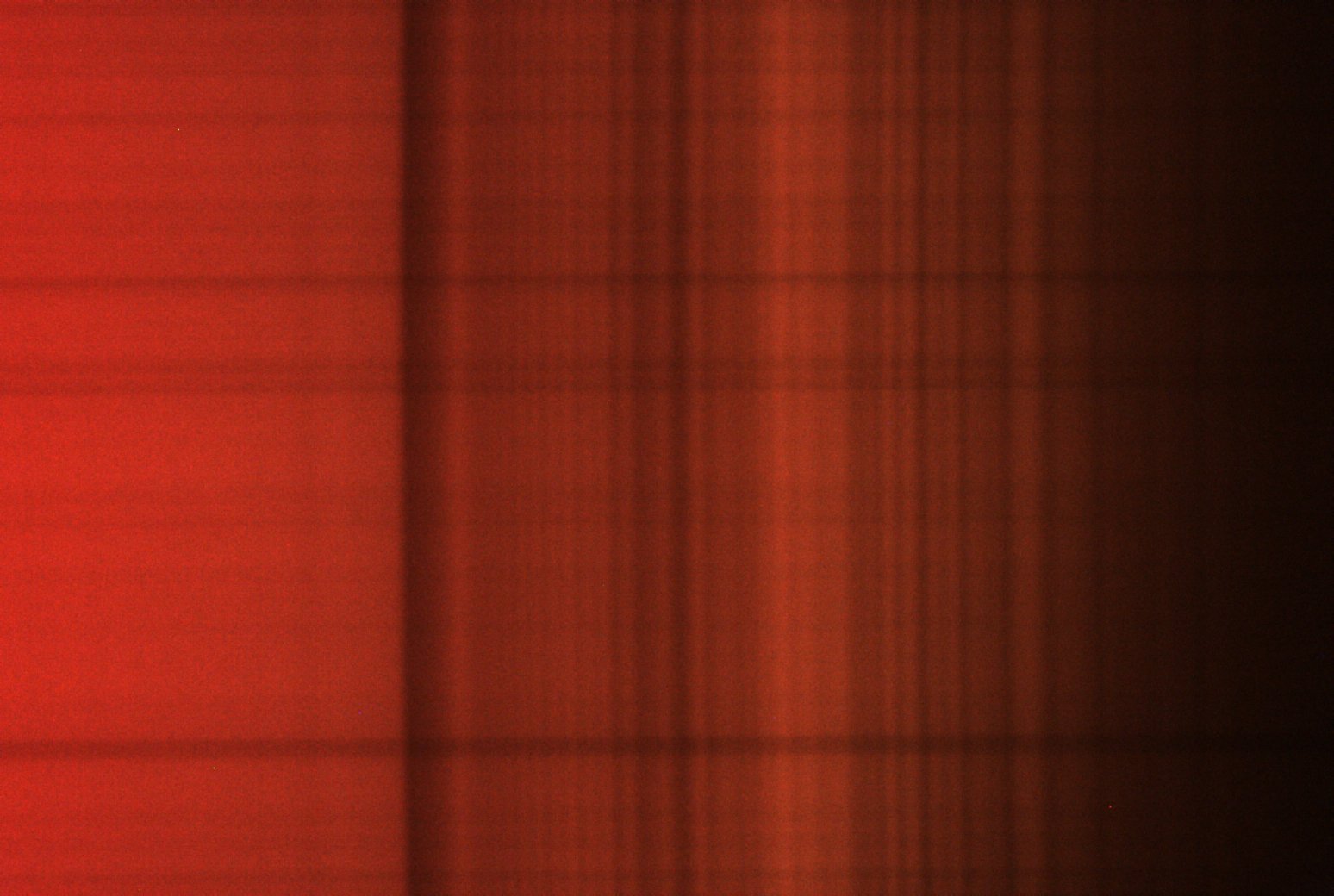}
	\end{subfigure}
	
	\begin{subfigure}[c]{0.3\linewidth}
		\includegraphics[width=\linewidth]{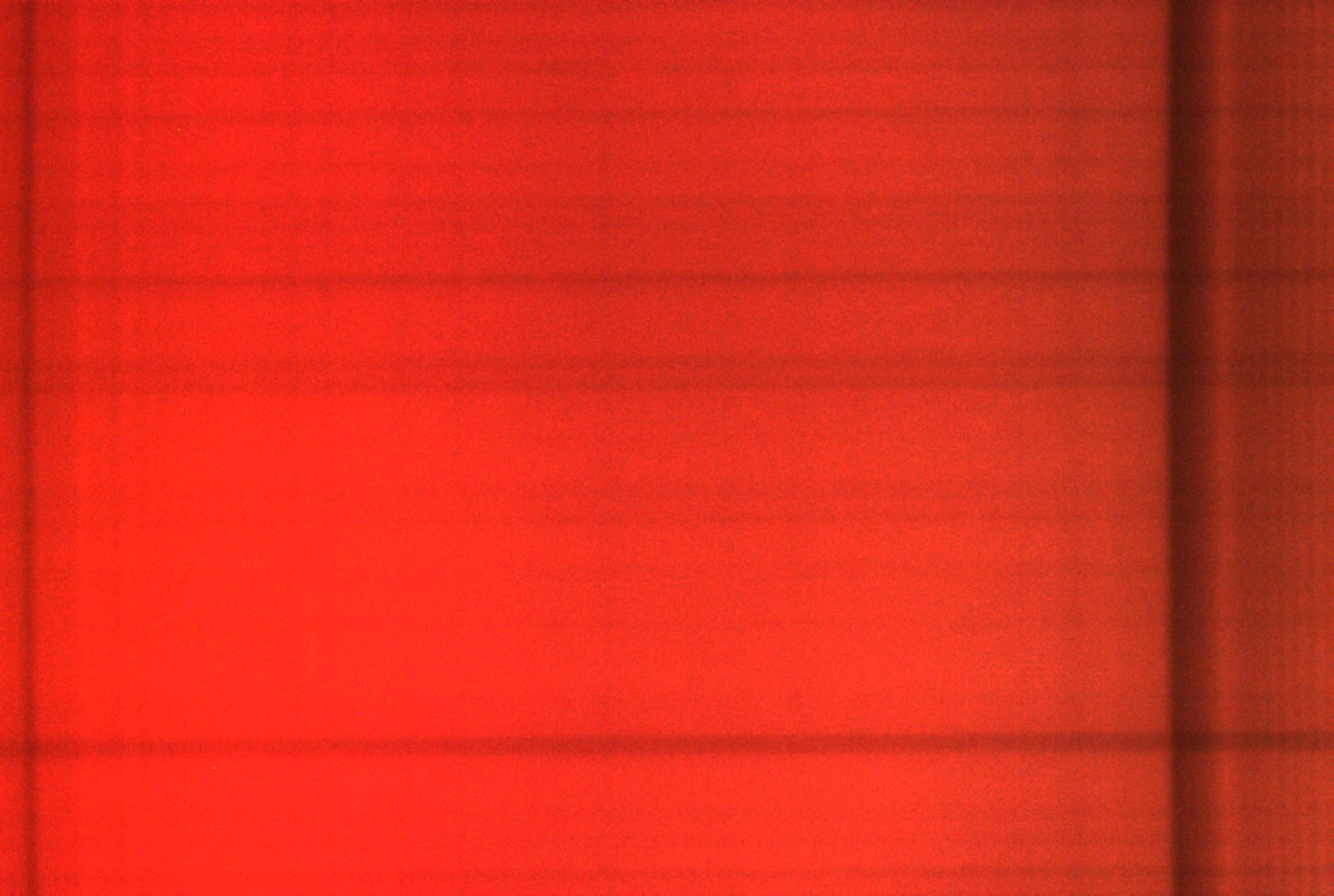}
	\end{subfigure}
	\begin{subfigure}[c]{0.3\linewidth}
		\includegraphics[width=\linewidth]{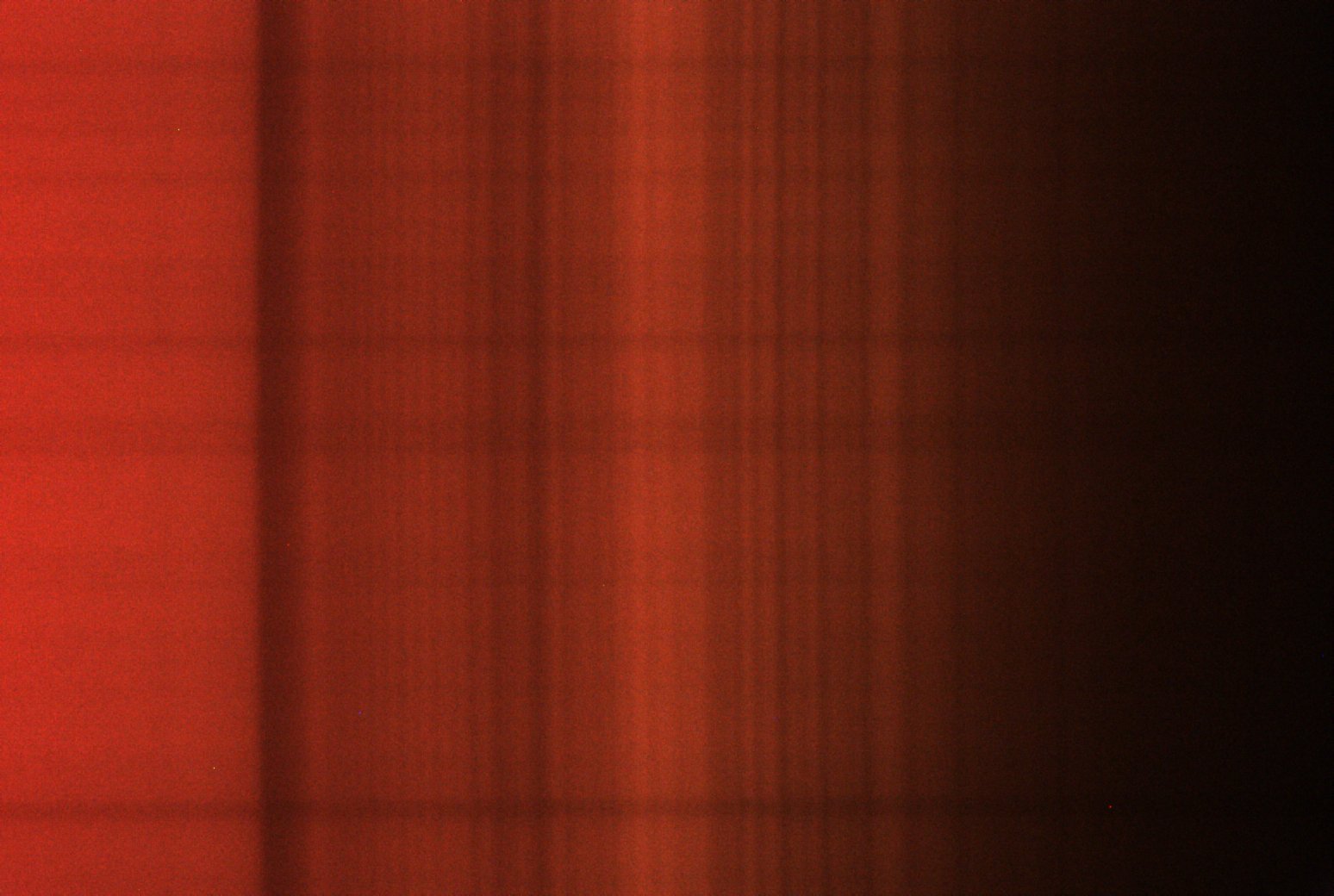}
	\end{subfigure}
\end{figure}

\appendix{Spectrum graphs of the Sun by SPECTRUMMATE}
\label{append:Spectrum graphs of the Sun by SPECTRUMMATE}

\begin{figure}[H]
	\centering
	\includegraphics[width=\linewidth]{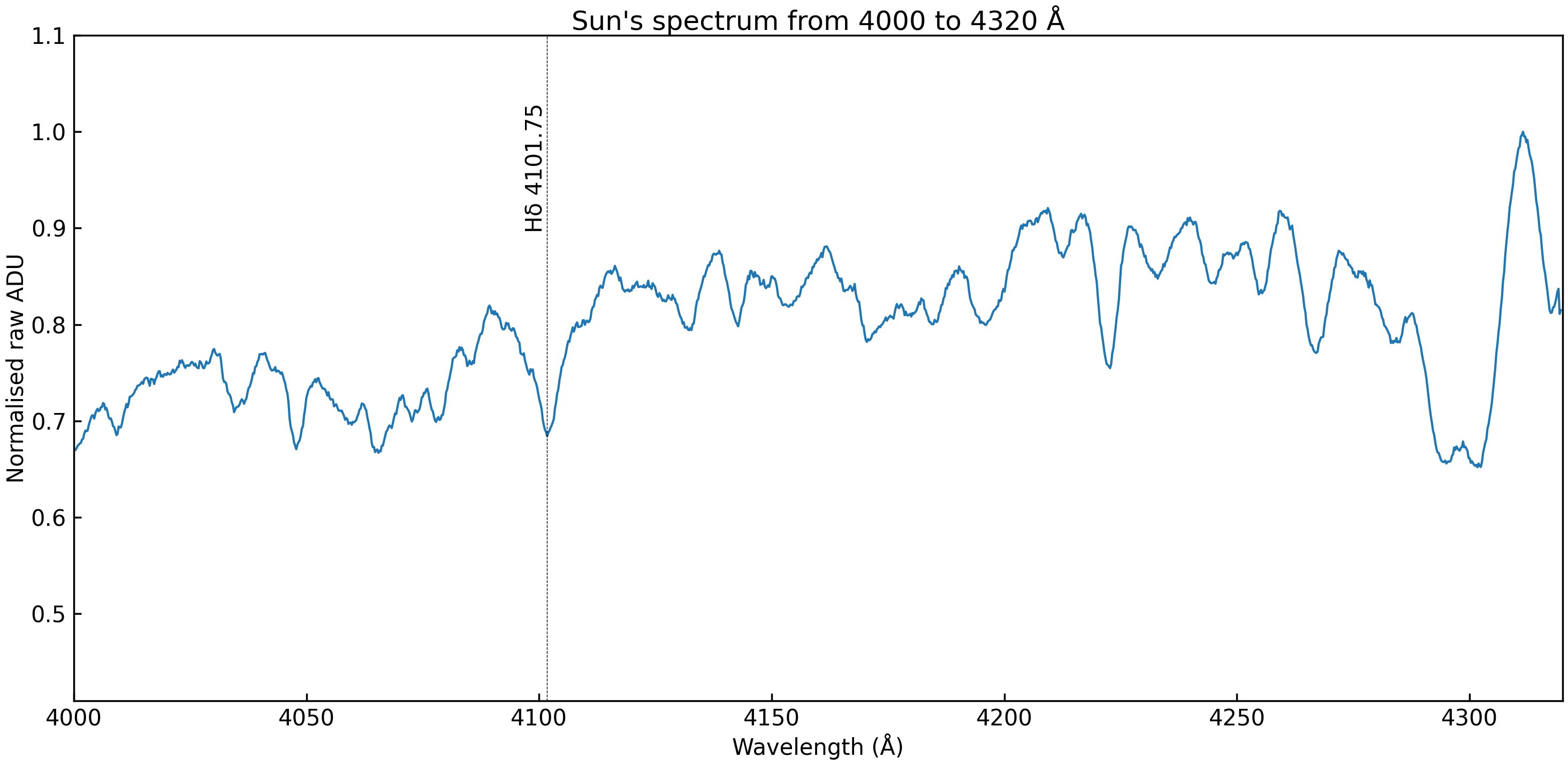}
	\caption{Spectrum of the Sun from 4000 to 4320 \angstrom.}
    \label{fig:Sun4000to4320}
\end{figure}

\begin{figure}[H]
	\centering
	\includegraphics[width=\linewidth]{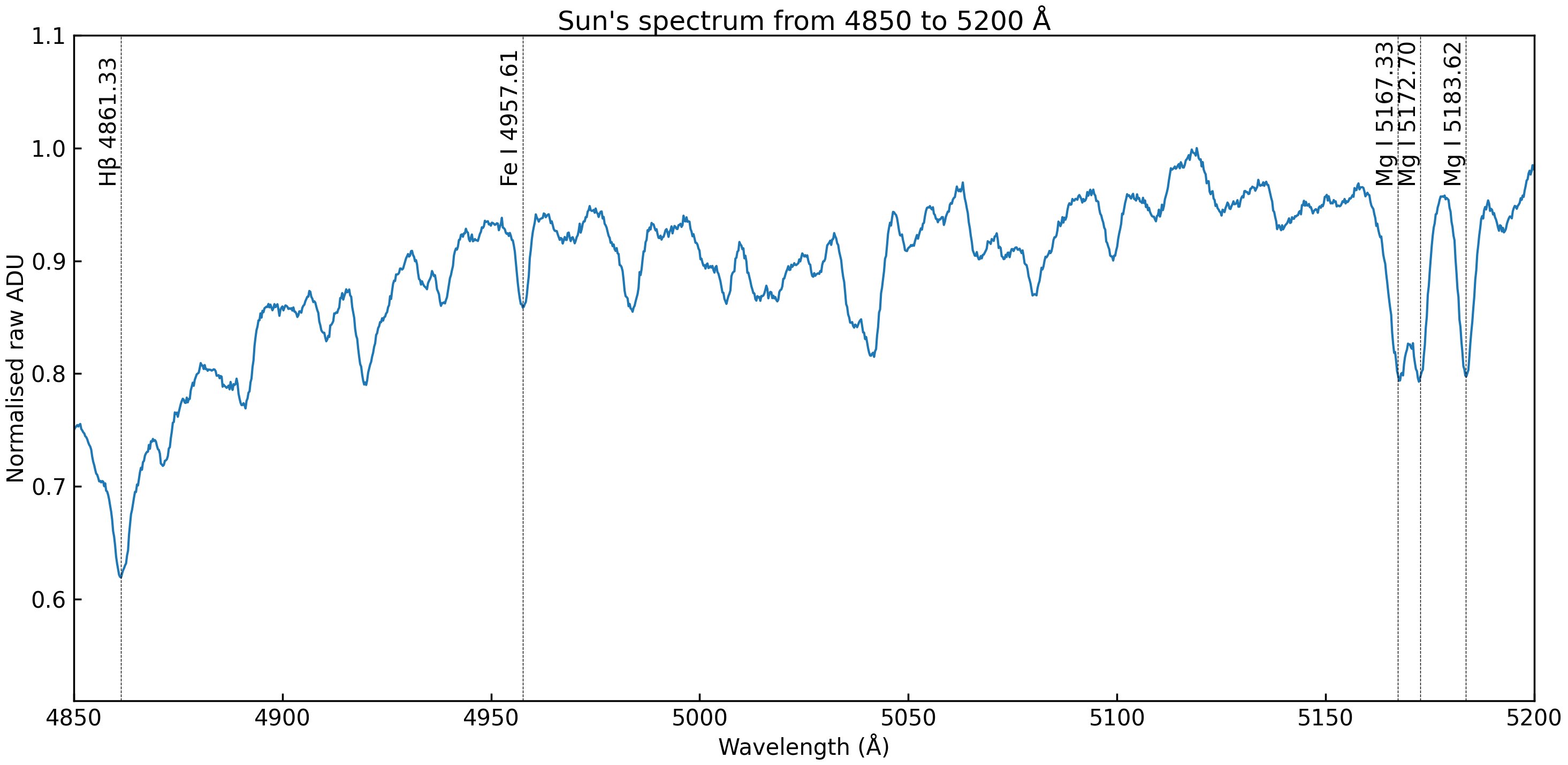}
	\caption{Spectrum of the Sun from 4850 to 5200 \angstrom.}
    \label{fig:Sun4850to5200}
\end{figure}

\begin{figure}[H]
	\centering
	\includegraphics[width=\linewidth]{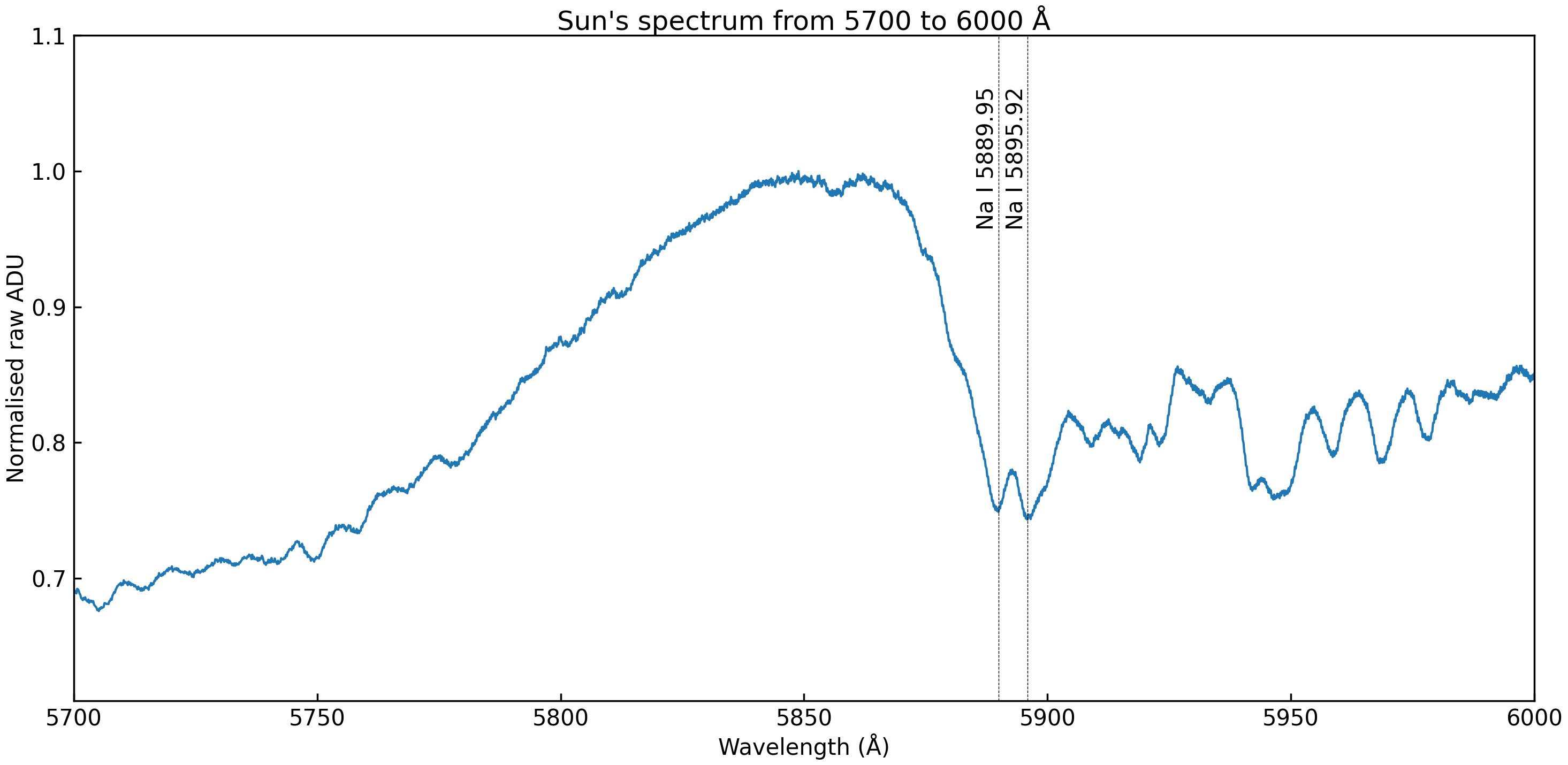}
	\caption{Spectrum of the Sun from 5700 to 6000 \angstrom.}
    \label{fig:Sun5700to6000}
\end{figure}

\begin{figure}[H]
	\centering
	\includegraphics[width=\linewidth]{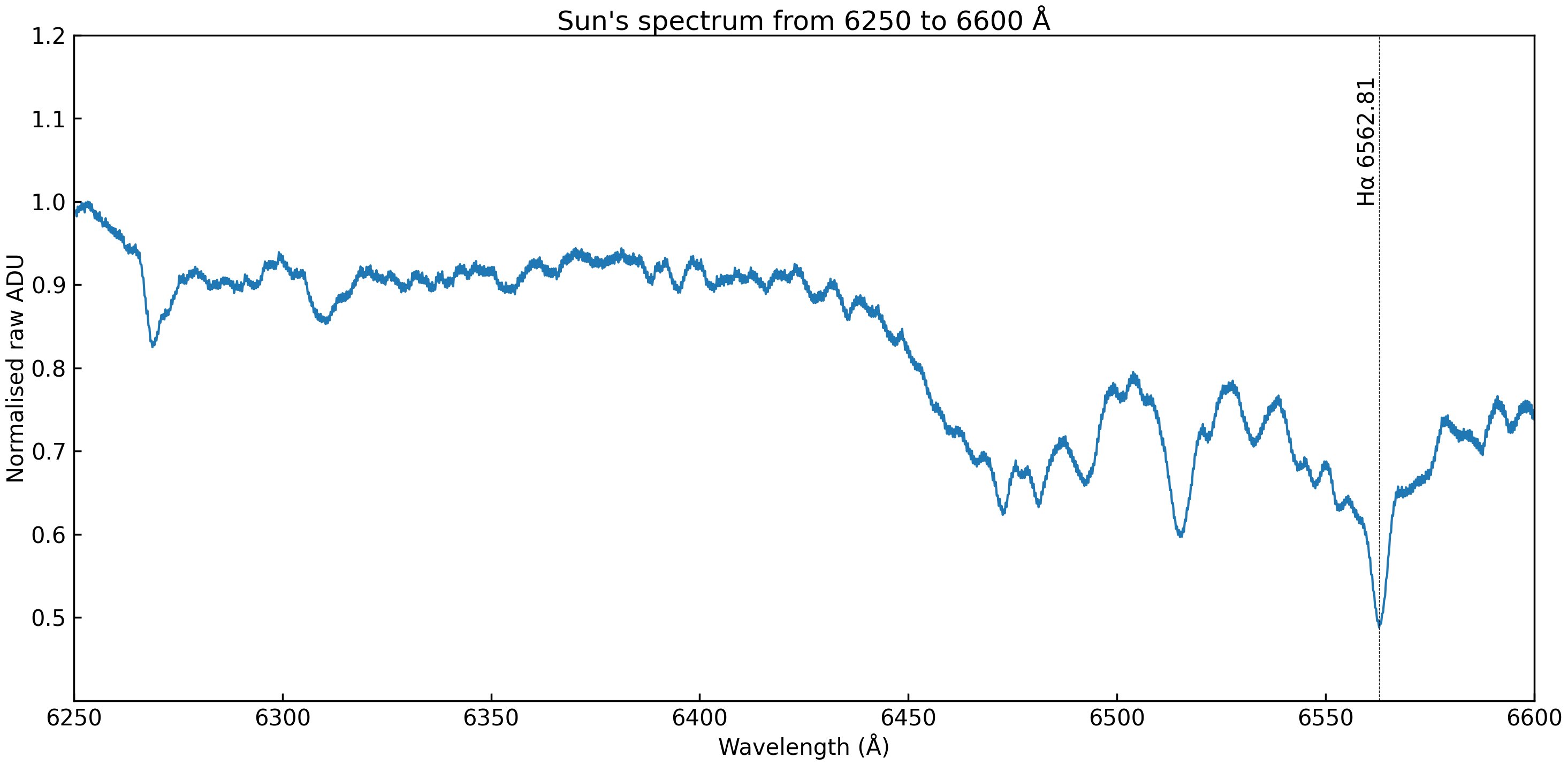}
	\caption{Spectrum of the Sun from 6250 to 6600 \angstrom.}
    \label{fig:Sun6250to6600}
\end{figure}

\bibliographystyle{ws-jai}
\bibliography{mybib}

\end{document}